\documentclass[english,aps,preprint]{revtex4}
\usepackage[T1]{fontenc}
\usepackage[latin9]{inputenc}
\usepackage{textcomp}
\usepackage{amsmath}
\usepackage{amssymb}
\usepackage{graphicx}
\usepackage{esint}

\makeatletter
\@ifundefined{textcolor}{}
{%
 \definecolor{BLACK}{gray}{0}
 \definecolor{WHITE}{gray}{1}
 \definecolor{RED}{rgb}{1,0,0}
 \definecolor{GREEN}{rgb}{0,1,0}
 \definecolor{BLUE}{rgb}{0,0,1}
 \definecolor{CYAN}{cmyk}{1,0,0,0}
 \definecolor{MAGENTA}{cmyk}{0,1,0,0}
 \definecolor{YELLOW}{cmyk}{0,0,1,0}
}

\makeatother

\usepackage{babel}
\begin{document}

\preprint{This line only printed with preprint option}

\title{The principle of the mutual energy }

\author{Shuang-ren Zhao, Kevin Yang, Kang Yang, Xingang Yang and Xintie Yang}
\email{London, Ontario, Canada}

\homepage{http://imrecons.com}

\thanks{.}

\affiliation{Imrecons Inc}
\begin{abstract}
Advanced potential solution of Maxwell equations isn't often accepted.
We have proven if without advanced potential, it is not possible to
satisfy the Maxwell equations. We also shown that it is not the Poynting
vector related energy current transferring energy in the space and
it is the mutual energy really did that. A important result of the
mutual energy theorem is that the advanced potential can suck energy
from the transmitter. This energy is equal to the energy received
at the receiver. Hence a transmitter can not send any energy out without
the receiver. For two remote objects, the energy is transferred only
can by the mutual energy of a retarded potential from the source together
with an advanced potential from the sink. If the sucked energy is
discrete, the summation of mutual energy current of the infinite background
atoms or currents, which can be seen as receivers, is a random process.
This means that the photon energy sent by the transmitter is actually
grabbed by the receiver. Hence the photon from very beginning knows
their destination. This receiver send advanced potential to the transmitter.
This explanation also avoided the wave function collapse. The retarded
potential first reached the receiver, cause the current in the receiver,
the current of receiver send a advanced potential to the transmitter
with a reversed time, in the same time, a photon minus-time-instantly
runs from receiver to transmitter. In our normal feeling, the photon
is still runs from the transmitter to the receiver with a positive
time. How to transfer superluminal signal using advanced potential
is also discussed.

Keywords: Electromagnetic field; Mutual energy; Receive antenna; Transmit
antenna; Speed of light; Retarded potential; Advanced potential; Probability;
Quantum mechanics; Causality; Wave function collapse, superluminal.
\end{abstract}
\maketitle

\part{Mutual energy current in Fourier space}

\section{introduction}

Maxwell equations have two solutions: the retarded potential and the
advanced potential. Many physicists accept the advanced potential\cite{LawrenceMStephenson}.
The absorb theory of D. T. Pegg, Wheeler and Feynman \cite{Wheeler_1,Wheeler_2,Pegg}
also based on the advanced potential. But most antenna engineers or
microwave engineers reject the advanced potential. It is seems without
advanced potential the engineer still can solved all engineering problems
by using the reciprocity theorem\cite{IEEEexample:Lorentz-1,IEEEexample:Carson-1,IEEEexample:Carson-2,IEEEexample:STUART-BALLANTINE,IEEEexample:Rumsey}\cite{IEEEexample:JinAuKong,IEEEexample:J_A_Kong2}. 

The author introduced the concept of mutual energy and the mutual
energy theorem\cite{IEEEexample:shrzhao1,IEEEexample:shrzhao2,IEEEexample:shrzhao3,shuangrenzhaoarxiv}.
There were very closed earlier publications related to the concept
of mutual energy \cite{IEEEexample:Welch}\cite{IEEEexample:Rumsey_VH},
but they did not realize it is the energy and still thought it as
some kind of reciprocity. The author of ref.\cite{IEEEexample:shrzhao1,IEEEexample:shrzhao2,IEEEexample:shrzhao3}
thought this is mutual energy but didn't continue to work with mutual
energy for around 30 years. Only recently the authors have found that
this mutual energy is so important, it is the only way to transfer
energy between two remote objects.

The authors have shown that the reciprocity theorem solution to a
system with one transmitter and a receiver is inadequate. This antenna
system should be explained by using the mutual energy theorem. In
the mutual energy explanation, the receive antenna must send a advanced
potential. From the mutual energy theorem, it shows that the advanced
potential of the receiver sucks the energy from the transmitter. The
transmitter send the retarded potential to the receiver, and then
caused the current of the receiver, the receiver send the advanced
potential to transmitter, the advanced potential and retarded potential
together can transfer the energy form the transmitter to the receiver.

In order to accept the concept of the mutual energy and the advanced
potential solution we have to answer following question, if mutual
energy transfer the energy, what about the energy current calculated
through Poynting vector which is referred as P-current?

If P-current transfer energy also, we found that the advanced potential
cannot be accept. Because even in a one dimensional wave guide, there
is 4 times P-current for a retarded potential together with advanced
potential. Hence we assume that the P-current of the advanced potential
and P-current of the retarded potential all do not exchange energy
with other materials in the environment. The concept of P-current
is similar to the concept of self-energy in quantum physics and the
absorb theory of ref.\cite{Wheeler_1,Wheeler_2,Pegg}.

We have found that in the one dimensional wave guide case, there is
a transmitter and a receiver in each ends of the wave guide, the P-current
is equal to the M-current which is the energy current calculated by
the concept of the mutual energy.

We also shown that in the case of 3D free space, if the the transmitter
situated at the center of the free space, the summation of all M-current
from the background material to the transmitter is equal to the P-current.
Hence in the above two situations the transferred energy still can
be calculated through P-current. In a laser beam, the transferred
energy can also be approximately calculated through P-current. However
in general situation, the calculation with P-current is not the transferred
energy. A example is the line antenna with the cross-section area
close to 0.

After we accept the advanced potential, and accept the advanced potential
can suck the energy from the transmitter, we can discrete the sucking
process, hence the background atoms will send discrete advanced potential
to the transmitter. There are infinite more this background atoms
which can be seen as receivers. Hence the photon energy is randomly
sucked out from the transmitter.

The photon from the very beginning knows exactly where it should to
reach, this avoid the Schrödinger wave function collapse. 

Advanced potential can be seen as ether of the receiver, it can lead
the retarded potential, this can offer the reason why the light speed
is a constant to the receiver. 

The beam width of the mutual energy current is clear different with
the P-current. It is easy to obtain the beam section area for the
mutual energy current, which has the the maximum value in the middle
between the source and sink. This explained the effect referred as
antibunching\cite{Wheeler_1,Wheeler_2,Pegg}. Many results of this
paper agree the absorb theory \cite{Wheeler_1,Wheeler_2,Pegg}, i.e.,
self-energy calculated from P-current can be removed, The spontaneous
emission is caused by remote environment materials. However this article
offer more details how the advanced potential is cooperate with retarded
potential to transfer energy. The retarded potential from transmitter
can only cooperate with a returned advanced potential from the receiver
and synchronize them together. This cause the energy transfer from
transmitter to the receiver.

It is clear this kind of energy transfer can also transfer a electromagnetic
force to remote object, a retarded potential send from the transmitter
to the receiver, the receiver send advanced potential back, the reaction
force on the transmitter is at 0 time. This means at least the electromagnetic
force can be reacted at 0 time. If other forces are transferred also
similar way with a mutual energy cooperated with a retarded potential
and an advanced potential, it can also has a reaction force at 0 time.
This perhaps is the reason of for example what is mass, why the action
and reaction is equal to each other and can be transferred with 0
time? We also discuss how to transfer a superluminal signal.

\section{foundational theory of electromagnetic fields\textbackslash{}\textbackslash{}\label{sec:foundational-theory-ofII}}

\subsection{Maxwell Equations}

Assume the electromagnetic field system is $\zeta=[E,H,J,K,D,B]$,
where $E$, $D$ are electric field and electric displacement field.
$H$, and $B$ are magnetic fields, H-field and B-field. The magnetic
field can be expressed as $\xi=[E,H]$. Assume $J$ and $K$ are current
and magnetic current intensities and can be expressed as $\rho=[J,K]$.
$\epsilon=\epsilon(t)$ and $\mu=\mu(t)$ are permittivity and permeability
which are 3D tensors and a function of time $t$. $\zeta$ is a field
system, but we often referred it also as field. $\zeta$ satisfies
the Maxwell equations,

\begin{equation}
\nabla\times H=J+\partial D\label{eq:0030}
\end{equation}
\begin{equation}
\nabla\times E=-K-\partial B\label{eq:0040}
\end{equation}
where $\partial\equiv\frac{\partial}{\partial t}.$ 
\begin{equation}
D=\epsilon*E,\ \ \ \ \ B=\mu*H\label{eq:0041}
\end{equation}
$\rho=[J,K]$ can be source which sends energy out or sink which receive
energy. The example of the source and sink are transmit antenna and
receive antenna. In case of scattering, $\rho$ can be the source
and sink simultaneously. Another example for $\rho$ is laser sender
and laser receiver. $*$ is time convolution where $f*g=\intop_{-\infty}^{\infty}f(t)(t-\tau)d\tau$.

\subsection{The superimposition of the fields}

Assume in the space there are electromagnetic fields includes advanced
potential and retarded potential. The electromagnetic fields can be
superimposed. The following is the superimposition of two electromagnetic
field systems, 

\[
\zeta=\zeta_{1}+\zeta_{2}
\]
\begin{equation}
=[E_{1}+E_{2},H_{1}+H_{2},J_{1}+J_{2},K_{1}+K_{2},D_{1}+D_{2},B_{1}+B_{2}]\label{eq:0042}
\end{equation}
Normally there is

\begin{equation}
D_{1}+D_{2}=\epsilon*(E_{1}+E_{2})\label{eq:0043}
\end{equation}

\begin{equation}
B_{1}+B_{2}=\mu*(H_{1}+H_{2})\label{eq:0044}
\end{equation}
One important situation for the superimposition is the case where
the two fields one is retarded potential another is advanced potential.
This superimposition is not self-explanatory, if we still do not accept
the advanced potential. The author accept the advanced potential,
and this article we try to make the theory about advanced potential
self-consistent. 

However the above formula can be generalized as 

\begin{equation}
D_{1}+D_{2}=\epsilon_{1}*E_{1}+\epsilon_{2}*E_{2}\label{eq:0046}
\end{equation}

\begin{equation}
B_{1}+B_{2}=\mu_{1}*H_{1}+\mu_{2}*H_{2}\label{eq:0047}
\end{equation}
In the generalized superimposition two fields $\zeta_{1}$ and $\zeta_{2}$
are at different space, each space has different media. The authors
do not claim that this generalized superimposition is a physics field.
The generalized field can be a mathematical field. But if the advanced
potential has different media constant, the above superimposition
is a physical process. Up to now we still assume that the advanced
potential have the same media with retarded potential. 

\subsection{The Poynting theorem}

Poynting theorem is also be referred as Poynting energy conservation
law. There is the Poynting theorem and modified Poynting theorem,

\[
-\nabla\cdot(E\times H)
\]
\begin{equation}
=J\cdot E+K\cdot H+E\cdot\partial D+H\cdot\partial B\label{eq:0048}
\end{equation}
where $\partial\equiv\frac{\partial}{\partial t}.$ If there is only
one media Eq.(\ref{eq:0043},\ref{eq:0044}), the above formula is
referred normal Poynting theorem. If there are different media Eq.(\ref{eq:0046},\ref{eq:0047}),
the above formula is referred as the modified Poynting theorem\cite{shuangrenzhaoarxiv}. 

\subsection{What will happen if there is no advanced potential in 3D free space?}

We have known there is black body which can receive electromagnetic
field but do not send the electromagnetic field out. Here we assume
it do not send retarded potential out. We also know that the black
body received energy by some current. Hence the current in the above
black body cannot vanish.

In this time, we assume there exist no advanced potential in nature.
We also know that a nor-scattering ideal antenna, like a black body,
can not send a retarded field out. Hence a ideal received antenna
has no any field associated to it. No any field retarded or advanced
potentials produced by the current $J_{a},K_{a}$. $J_{a},$$K_{a}$
(which doesn't vanish. Since we know from our experience there is
current in a receive antenna. The subscript $a$ means the current
is a receiver hence ``absorb'' energy.

We have assume there is no any advanced potential in nature, hence,
the electromagnetic field in the space can only be caused by all the
retarded potential of the transmitters or transmit antennas, i.e.,

\begin{equation}
\xi=\sum_{i\in I}\xi_{t_{i}}\label{eq:40000-600}
\end{equation}
assuming $M$ is the Maxwell operator. When Maxwell operator acted
to the field $\xi_{t_{i}}$ will obtain the source, i.e, $M\xi_{t_{i}}=\gamma\rho_{t_{i}}$$\gamma$
is a constant. Hence
\begin{equation}
M\xi=\sum_{i\in I}M\xi_{t_{i}}=\gamma\sum_{i\in I}\rho_{t_{i}}\label{eq:40000-610}
\end{equation}
 This means there is not possible to have

\begin{equation}
M\xi=\gamma\rho_{a}\label{eq:40000-620}
\end{equation}
Here is $\rho_{a}=[J_{a},K_{a}]$ is the current of the receive antenna. 

Hence if there is no advanced potential, in the place of the current
of a receive antenna, it is not possible to satisfy the Maxwell equation. 

This shows ether the Maxwell equation is wrong or there must be the
advanced potential. It is not possible to obtained current of received
antenna from the retarded potential with Maxwell operator. 

Hence, it is better to us to assume that the advanced potential exist
in nature. If the advanced potential can not be easily measured, perhaps
it can be indirect perceived by the result of the theory of the advanced
potential.

\subsection{Retarded and advanced potential}

A retarded and an advanced potential potential,
\begin{equation}
\xi_{r}\propto\iiintop_{V}\frac{\rho(x',t_{r})}{|x-x'|}d^{3}x'\ \ \ \xi_{a}\propto\iiintop_{V}\frac{\rho(x',t_{a})}{|x-x'|}d^{3}x'\label{eq:5000-100}
\end{equation}
Here $x$ and $x'$ are 3D vectors. $x'\in V$.
\begin{equation}
t_{r}=t-\frac{1}{c}|x-x'|,\ \ \ t_{a}=t+\frac{1}{c}|x-x'|\label{eq:5000-110}
\end{equation}
Here $c$ is spead of light. Assume $x$ is restrict to one dimension,
$x>x'$ , $\rho(x',t)=\delta(x')\exp(j\omega t)$, this is oscillator
located at the origin of the coordinates, we have

\begin{equation}
\xi_{r}\propto\frac{\exp(j\omega t_{r})}{x-x'}\ \ \ \ \xi_{a}\propto\frac{\exp(j\omega t_{a})}{x-x'}\label{eq:5000-120}
\end{equation}
here, $x'=0$, considering,

\begin{equation}
\exp(j\omega t_{r})=\exp(j\omega(t-\frac{1}{c}(x-x'))=\exp(j(\omega t-kx))\label{eq:5000-130}
\end{equation}

\begin{equation}
\exp(j\omega t_{a})=\exp(j\omega(t+\frac{1}{c}(x-x'))=\exp(j(\omega t+kx))\label{eq:5000-140}
\end{equation}
where $k=\frac{\omega}{c}$, hence,

\begin{equation}
\xi_{r}\propto\frac{1}{x}\exp(j(\omega t-kx))\ \ \ \ \ \xi_{a}\propto\frac{1}{x}\exp(j(\omega t+kx))\label{eq:5000-150}
\end{equation}
if we do not consider the fact $\frac{1}{x}$, the retarded and advanced
potential can be expressed as,

\begin{equation}
\xi_{r}\propto\exp(j(\omega t-kx))\ \ \ \ \ \ \xi_{a}\propto\exp(j(\omega t+kx))\label{eq:5000-160}
\end{equation}

\subsection{Retarded and advanced potential in loss media}

If there is energy loss, we can assume that $k=k_{0}-ik_{loss}$,
hence there is

\[
|\xi_{r}|\propto|\exp(i(\omega t-(k_{0}-ik_{loss})x))|
\]
\begin{equation}
=|\exp(i(\omega t-k_{0}x))\exp(-k_{loss}x)|=\exp(-k_{loss}x)\label{eq:5000-180}
\end{equation}
For the retarded potential, it is attenuated wave. For the advanced
potential there is

\[
|\xi_{a}|\propto|\exp(i(\omega t+(k_{0}-ik_{loss})x))|
\]

\begin{equation}
=|\exp(i(\omega t+k_{0}x))\exp(+k_{loss}x)|=\exp(+k_{loss}x)\label{eq:5000-190}
\end{equation}
It is a ascending wave.

\section{Review of the theory about the mutual energy theorem\label{sec:Review-of-theIV}}

The theory for the mutual energy theorem includes the following a
few components,

\subsection{The mutual energy theorem formula}

Before the full formula of the mutual energy theorem, there are two
early version of it, which can be seen in\cite{IEEEexample:Welch}\cite{IEEEexample:Rumsey_VH}.
The formula of the mutual energy theorem can be found \cite{IEEEexample:shrzhao1,IEEEexample:shrzhao2,IEEEexample:shrzhao3}.
The formula \cite{IEEEexample:shrzhao1,IEEEexample:shrzhao2,IEEEexample:shrzhao3}
is in Fourier domain. The corresponding time domain mutual energy
theorem can be found in ref.\cite{IEEEexample:Adrianus2}. 

In Fourier domain the modified mutual energy theorem formula can be
written as following\cite{shuangrenzhaoarxiv},

\begin{equation}
(\xi_{1},\xi_{2})_{\varGamma}+(\rho_{1},\xi_{2})_{V}+(\xi_{1},\rho_{2})_{V}=0\label{eq:2000-130-1}
\end{equation}
$V$ is the volume contains the current $\rho_{1}$ and $\rho_{2}$.
$\varGamma$ is the boundary of the volume $V$. $\varGamma$ can
be chosen as infinite big sphere. The media have to meet the condition,
\begin{equation}
\epsilon_{1}^{\dagger}(\omega)=\epsilon_{2}(\omega),\ \ \ \ \mu_{1}^{\dagger}(\omega)=\mu_{2}(\omega)\label{eq:2000-150-1}
\end{equation}
Where ``$\epsilon^{\dagger}=\epsilon^{*T}$. The superscript$*$
expresses the complex conjugate operator, $T$ is matrix transpose
and,

\begin{equation}
(\xi_{1},\xi_{2})_{\varGamma}\equiv\iintop_{\varGamma}\,(E_{1}\times H_{2}^{*}+E_{2}^{*}\times H_{1}^{*})\,\hat{n}dS\label{eq:2000-151}
\end{equation}

\begin{equation}
(\rho_{1},\xi_{2})_{V}\equiv\iiintop_{V}(E_{2}^{*}\cdot J_{1}+H_{2}^{*}\cdot K_{1})\,dV\label{eq:2000-152}
\end{equation}

\begin{equation}
(\xi_{1},\rho_{2})_{V}\equiv\iiintop_{V}(E_{1}\cdot J_{2}^{*}+H_{1}\cdot K_{2}^{*})\,dV\label{eq:2000-153}
\end{equation}
It is possible that the modified mutual energy theorem is not a physical
theorem since the media of the two fields $\zeta_{1}$ and $\zeta_{2}$
can be different or at different spaces. If we assume they are the
same, i.e.,$\epsilon_{1}=\epsilon_{2}$ and $\mu_{1}=\mu_{2}$. There
is

\begin{equation}
\epsilon^{\dagger}(\omega)=\epsilon(\omega),\ \ \ \ \mu^{\dagger}(\omega)=\mu(\omega)\label{eq:2000-154}
\end{equation}
That is lossless condition. Hence in lossless media the mutual energy
theorem (note here, there is no ``modified'') is established. 

\subsection{The mutual energy theorem can be sub-theorem of the Poynting theorem}

We have shown that the modified mutual energy theorem can be derived
from the modified Poyting theorem or the Poyting energy conservation
law. This is also true if the word modified is removed, i.e., the
mutual energy theorem can be also be derived from the Poynting theorem.

It is important that we can show that the mutual energy theorem can
be a direct sub-theorem of Poynting energy conservation law instead
of deriving it from Maxwell equations. In this way, the mutual energy
theorem becomes a energy theorem\cite{shuangrenzhaoarxiv}. 

\subsection{The mutual energy current of a retarded potential and an advanced
potential vanishes in the infinite big sphere}

Assume $\zeta_{1}$ is retarded potential and $\zeta_{2}$ is advanced
potential we can prove that in the free space (where the media is
$\epsilon_{0}$ and $\mu_{0}$) the surface integral of the mutual
energy theorem will vanish at infinite big sphere $\varGamma$.

\begin{equation}
\lim_{r\rightarrow\infty}\iintop_{\varGamma}(E_{1}\times H_{2}^{*}+E_{2}^{*}\times H_{1})\cdot\hat{n}\,dS=0\label{eq:2000-160}
\end{equation}
Here $r$ is the radio of sphere $\varGamma$. Assume 
\begin{equation}
\lim_{r\rightarrow\infty}H_{1}=\frac{1}{Z}\hat{n}\times E_{1}\label{eq:2000-161}
\end{equation}
Here $\hat{n}$ is the direction of wave $\zeta_{1}$ which is retarded
potential.

\begin{equation}
\lim_{r\rightarrow\infty}H_{2}=\frac{1}{Z}(-\hat{n})\times E_{2}\label{eq:2000-162}
\end{equation}
Here $(-\hat{n}$) is the direction of wave $\zeta_{2}$ which is
advanced potential. $Z=\sqrt{\frac{\mu_{0}}{\epsilon_{0}}}$, the
details can be found in \cite{shuangrenzhaoarxiv}.

\subsection{The surface integral in the mutual integral is inner product between
two retarded potentials or two advanced potentials}

The surface integral in the mutual energy formula $(\xi_{1},\xi_{2})_{\varGamma}$
is a inner product \cite{IEEEexample:shrzhao1}. Here we have assumed
that that the two fields $\xi_{1},\xi_{2}$ are two retarded potentials.
i.e. the surface integral satisfies following 3 inner product laws,

I. conjugate symmetry,

\begin{equation}
(\xi_{1},\xi_{2})_{\varGamma}=(\xi_{2},\xi_{1})_{\varGamma}^{*}\label{eq:6010}
\end{equation}

II linear,
\begin{equation}
(\xi_{1}+\xi_{2},\xi_{3})_{\varGamma}=(\xi_{1},\xi_{3})_{\varGamma}+(\xi_{2},\xi_{3})_{\varGamma}\label{eq:6020}
\end{equation}

\begin{equation}
(\alpha\xi_{1},\xi_{2})_{\varGamma}=\alpha(\xi_{1},\xi_{2})_{\varGamma}\label{eq:6030}
\end{equation}

III Positive-definiteness,

\begin{equation}
(\xi,\xi)_{\varGamma}\geq0\label{eq:6040}
\end{equation}

\begin{equation}
(\xi,\xi)_{\varGamma}=0\ \ \ \ \ \ \ iff\ x=0\label{eq:6050}
\end{equation}
Here $iff$ means if and only if. 

If the $\xi$ is advanced potential, the other formulas are all established
except Eq.(\ref{eq:6040}) need to updated to

\begin{equation}
(\xi,\xi)_{\varGamma}\leq0\label{eq:6040-1}
\end{equation}

If we allow the field $\xi_{1}$ and $\xi_{2}$ are retarded potential
and $\xi=\xi_{1}+\xi_{2}$, also advanced potential, the above inner
product laws are still satisfied. Only the formula Eq.(\ref{eq:6050})
does not satisfy. This inner product formula guarantee that we can
use the inner product expression $(\xi_{1},\xi_{2})_{\varGamma}$
it defines the mutual energy current of surface $\varGamma$, 
\begin{equation}
Q_{12}=(\xi_{1},\xi_{2})_{\varGamma}=\iintop_{\varGamma}(E_{1}\times H_{2}^{*}+E_{2}^{*}\times H_{1})\cdot\hat{n}\,dS\label{eq:6052}
\end{equation}
The above discussion tell us the mutual energy current is actually
a very good inner product. We will apply thin inter product to simplify
the formula.

\subsection{The mirror transform of a retarded potential is advanced potential
and vice versa}

The mutual energy theorem often works with magnetic mirror transform,
it is important to know what will happen after the transform for a
retarded potential or a advanced potential. This can be found in \cite{shuangrenzhaoarxiv}.
\begin{equation}
m\zeta=[E^{*},-H^{*}-J^{*},K^{*},\epsilon^{*},\mu^{*}]\label{eq:50000--01}
\end{equation}

\section{The system with transmit antenna and receive antenna\label{sec:The-system-withVII}}

In the case of two coils, the first coil has a current $J$, which
send energy to the free space, if we put the second coil close to
the first coil, the second coil will suck more energy from first coil.
In this situation the second coil has influence to the first coil.
This is the case where the two coils close to each other, for example
a electric transformator which has two coils: the primary coil and
a secondary coil. We know that if we add more load to the secondary
coil of a transformator, it can suck more energy from the primary
coil.

The two coils can be seen as two antennas. If we assume the two antennas
are put very far away, is the second coil (or antenna) still can suck
energy from the first coil (or antenna)? Does the received antenna
only passive receive energy sent from the transmit antenna or it actively
suck the energy from the transmit antenna?, like the the secondary
coil in a transformator? 

Traditionally we thought the receiver antenna, only passive receive
the energy from the the retarded wave of the transmitter. The P-current
carries the energy from the transmitter to the receiver. But why when
the receive antenna close to the receiver it can affect the transmitter,
but when it is far away, this ability is disappeared? According to
my understand, it should not disappeared completely but just decreased!
This section the authors try to answer this question.

\subsection{The traditional way of the explanation of the antenna system using
the reciprocity theorem }

Here we review the traditional way of using the reciprocity theorem
to explain the antenna system. Assume $\zeta_{1}=[E_{1},H_{1},J_{1},K_{1},\epsilon,\mu]$,
$\zeta_{2}=[E_{2},H_{2},J_{2},K_{2},\epsilon,\mu]$ are retarded potentials,
$J_{1},K_{1}$ and $J_{2},K_{2}$ are inside the volume $V_{1}$ and
$V_{2}$ respectively. $V_{1}$ and $V_{2}$ are inside of the volume
$V$, $V$,$V_{1},$$V_{2}$ have the boundary surface $\varGamma$,$\varGamma_{1}$$\varGamma_{2}$.
the reciprocity theorem is
\[
\varoiintop_{\varGamma}(E_{1}\times H_{2}-E_{2}\times H_{1})\cdot\,\hat{n}dS
\]

\begin{equation}
=\iiintop_{V}\,(E_{1}\cdot J_{2}-H_{1}\cdot K_{2})-(E_{2}\cdot J_{1}-H_{2}\cdot K_{1})\,dV\label{eq:40000-10}
\end{equation}
Since $\xi_{1}=[E_{1},H_{2}]$ and $\xi_{2}=[E_{2},H_{2}]$ are retarded
potentials, it can be proven that the surface integral vanishes if
the surface $\varGamma$ is chosen as a infinite big sphere, i.e.,

\begin{equation}
\varoiintop_{\varGamma}(E_{1}\times H_{2}-E_{2}\times H_{1})\cdot\,\hat{n}dS=0\label{eq:40000-20}
\end{equation}
Hence there is 
\begin{equation}
\iiintop_{V_{2}}\,(E_{1}\cdot J_{2}-H_{1}\cdot K_{2})\,dV\label{eq:40000-30}
\end{equation}

\begin{equation}
=\iiintop_{V_{1}}(E_{2}\cdot J_{1}-H_{2}\cdot K_{1})\,dV\label{eq:40000-40}
\end{equation}
For the reciprocity theorem, the media must be symmetric, i.e.,

\begin{equation}
\epsilon^{T}=\epsilon,\ \ \ \ \ \ \mu^{T}=\mu\label{eq:40000-50}
\end{equation}
If we assume there are $K_{1}=K_{2}=0$, there is,

\begin{equation}
\iiintop_{V_{2}}\,E_{1}\cdot J_{2}\,dV=\iiintop_{V_{1}}E_{2}\cdot J_{1}\,dV\label{eq:40000-60}
\end{equation}
Assume the volume $V_{1}$ and $V_{2}$ are cylinders and $V_{1}=L_{1}S_{1}$,
$V_{2}=L_{2}S_{2}$. $L_{1}$ $S_{1}$ are the length and sectional
areas of the cylinders, it is same to $V_{2}$ , hence there is,

\begin{equation}
\intop_{L_{2}}E_{1}dL\iintop_{S_{2}}J_{2}\,dS=\intop_{L_{1}}E_{2}dL\iintop_{S_{1}}J_{1}dS\label{eq:40000-80}
\end{equation}

\begin{equation}
V_{12}I_{2}=V_{21}I_{1}\label{eq:40000-90}
\end{equation}
The reciprocity theorem is derived for two transmit antennas. Where
$V_{12}=\intop_{L_{2}}E_{1}dL$ is potential on antenna 2 which is
produced by current $I_{1}$. $V_{21}=\intop_{L_{1}}E_{2}dL$ is the
potential on antenna 1 which is produced by current $I_{2}$. $I_{1}=\iintop_{S_{1}}J_{1}dS$,
$I_{2}=\iintop_{S_{2}}J_{2}dS$. All books of the electromagnetic
field theory claim that the above derivation also suit to the antenna
systems with a transmit antenna and a receive antenna. However they
did not offer a convinced proof. 

The above explanation seems correct, but it contains a mortal problem.
The authors do not satisfy that. First the current $J_{1}$ and $J_{2}$
must be both sources, they both send retarded potentials $\xi_{1}$,
$\xi_{2}$, (we have known that if and only if $\xi_{1}$ and $\xi_{2}$
are both retarded potential, the surface integral of the reciprocity
theorem vanishes, otherwise the reciprocity theorem of Eq.(\ref{eq:40000-40})
is not established). In the case there is a receive antenna, we can
shown that the reciprocity theorem is wrong in a very simple situation.
In the simple situation we assume that the receive antenna is a black
body, it can only receive energy. Hence $\xi_{2}=[E_{2},H_{2}]$ which
is the field produced by antenna 2 send to space, should vanish (we
have assume it is retarded potential). Hence, there is $\iiintop_{V_{1}}(E_{2}\cdot J_{1}-H_{2}\cdot J_{1})\,dV=0$
and the reciprocity theorem becomes 
\begin{equation}
0=\iiintop_{V_{2}}\,(E_{1}\cdot J_{2}-H_{1}\cdot K_{2})\,dV\label{eq:40000-100}
\end{equation}
If $\iiintop_{V_{2}}\,(E_{1}\cdot J_{2}-H_{1}\cdot K_{2})\,dV$ does
not vanish, the reciprocity theory become 
\begin{equation}
0=something\ not\ zero\label{eq:40000-110}
\end{equation}
which is clear wrong. if $\iiintop_{V_{2}}\,(E_{1}\cdot J_{2}-H_{1}\cdot K_{2})\,dV$
vanishes, the reciprocity theorem tell us 
\begin{equation}
0=0\label{eq:40000-120}
\end{equation}
 which is also meaningless.

The above discussion is also suit to the situation where the antenna
2 is a paraboloid antenna. The energy received by the paraboloid antenna
does not send again to the space, but send to a feed source. If the
scattering energy send from paraboloid antenna to the free space can
be omitted, the paraboloid antenna can be seen as black body. For
this kind idea paraboloid antenna used as receive antenna, the reciprocity
theorem is not suitable.

Second, we have known that there should be the energy current send
from transmit antenna to the receive antenna. The reciprocity theorem
can not offer us any information about the energy between the two
antennas.

Here we do not claim that the reciprocity theorem is all wrong, the
reciprocity theorem is still correct to two transmit antennas, which
is not very useful. In the following, we will explain the antenna
system with the mutual energy theorem.

\subsection{Explanation of a system with a receive antenna using the mutual energy
theorem}

\subsubsection{The result of the mutual energy theorem}

Instead of using reciprocity theorem we will apply the mutual energy
theorem to the system with two antennas, one is transmit antenna and
another is receive antenna. 

Assume there is a source $J_{1},K_{1}$ which is corresponding to
the transmit antenna and there is a sink $J_{2},K_{2}$ which is corresponding
to the receive antenna. Normally there is scatter field send out from
the receive antenna, we assume the scattering field is small and can
be omitted. The corresponding fields to the current $J_{1},K_{1}$
and $J_{2},K_{2}$ are $\xi_{1}$ and $\xi_{2}$ respectively. The
following calculation is in the Fourier domain. According to the modified
mutual energy theorem in Fourier domain there is\cite{shuangrenzhaoarxiv},

\begin{equation}
(\xi_{1},\xi_{2})_{\varGamma}+(\xi_{1},\rho_{2})_{V}+(\rho_{1},\xi_{2})_{V}=0\label{eq:4010}
\end{equation}
\begin{equation}
\epsilon_{2}^{\dagger}=\epsilon_{1},\ \ \ \ \ \ \mu_{2}^{\dagger}=\mu_{1}\label{eq:4020}
\end{equation}
Considering the antenna 1 is a transmit antenna, assume $\xi_{1}=[E_{1},H_{1}]$
is retarded potential. Considering antenna 2 is a receive antenna,
the authors assume $\xi_{2}=[E_{2},H_{2}]$ is advanced potential.
In the above formula $\epsilon^{\dagger}=(\epsilon^{*})^{T}$, $\rho_{1}=[J_{1},K_{1}]$,
$\rho_{2}=[J_{2},K_{2}]$, the superscript $T$ is matrix transpose
and $*$ is complex conjugate operator. $\varGamma$ is spherical
surface located at infinity. For a physical antenna system, we also
require that
\begin{equation}
\epsilon_{2}=\epsilon_{1}=\epsilon,\ \ \ \ \ \mu_{2}=\mu_{1}=\mu\label{eq:4021}
\end{equation}
The above formula means the receive antenna and the transmit antenna
are at the same space with same media $\epsilon,\mu$.

The above two formula from the media can be combined together,

\begin{equation}
\epsilon^{\dagger}=\epsilon,\ \ \ \ \ \ \mu^{\dagger}=\mu\label{eq:4022}
\end{equation}
The above formula indicates that the media must be lossless. With
the above media condition, the word ``modified'' before the word
mutual energy can be removed, hence modified mutual energy theorem
becomes the mutual energy theorem. Hence, the mutual energy theorem
is established in lossless media.

We have proven that in the mutual energy theorem, if $\xi_{1}$ and
$\xi_{2}$ one is retarded potential and another is advanced potential,
the surface integral vanishes, i.e.,
\begin{equation}
(\xi_{1},\xi_{2})_{\varGamma}=\varoiintop_{\varGamma}\,(E_{1}\times H_{2}^{*}+E_{2}^{*}\times H_{1}^{*})\,\hat{n}dS=0\label{eq:4030}
\end{equation}
Here $\varGamma$ is infinite big sphere. $\hat{n}$ is the outward
unit vector of surface $\varGamma$. Hence we have

\begin{equation}
(\xi_{1},\rho_{2})_{V}+(\rho_{1},\xi_{2})_{V}=0\label{eq:4040}
\end{equation}
or

\begin{equation}
\iiintop_{V}((E_{1}\cdot J_{2}^{*}+H_{1}\cdot K_{2}^{*})+(E_{2}^{*}\cdot J_{1}+H_{2}^{*}\cdot K_{1}))\,dV=0\label{eq:4050}
\end{equation}
If we have assumed that the magnetic current does not exist, $K_{1}=0$,
$K_{2}=0$, the above formula becomes,

\begin{equation}
\iiintop_{V}\,(E_{1}\cdot J_{2}^{*}+E_{2}^{*}\cdot J_{1})\,dV=0\label{eq:4060}
\end{equation}
If current $J_{1},K_{1}$ is only inside $V_{1}$ and $J_{2},K_{2}$
is only inside $V_{2}$, we have, 

\begin{equation}
\iiintop_{V_{2}}\,E_{1}\cdot J_{2}^{*}dV+\iiintop_{V_{1}}E_{2}^{*}\cdot J_{1}\,dV=0\label{eq:4060-1}
\end{equation}
or
\begin{equation}
-\iiintop_{V_{2}}\,E_{1}\cdot J_{2}^{*}dV=\iiintop_{V_{1}}E_{2}^{*}\cdot J_{1}\,dV\label{eq:4060-2}
\end{equation}
$\iiintop_{V_{2}}\,E_{1}\cdot J_{2}^{*}dV$ indicate the energy of
the field $E_{1}$ act on the current $J_{2}$. This is the energy
antenna 2 received. This part of energy is negative (assume we have
assumed that the energy contributed to free space is positive, this
energy is received by antenna 2, and hence cannot be sent to the free
space, hence it is negative). Hence $-\iiintop_{V_{2}}\,E_{1}\cdot J_{2}^{*}dV$
is positive.

$\iiintop_{V_{1}}E_{2}^{*}\cdot J_{1}dV$ is the energy sucked by
the field $E_{2}$ from the current $J_{1}$. This also tell us that
the received energy of antenna 2 sent from antenna 1 is same as the
energy sucked from antenna 1 by the advanced potential of the antenna
2. 

If advanced potential exist, it can suck energy from the source. This
sucked energy later is received by the receive antenna. This is the
most important result of the mutual energy theorem to the advanced
potential and received antenna. The authors believe this result is
correct and will apply this to explain more physics result in next
sections.

The above important mutual energy formula Eq.(\ref{eq:4060-2}) is
first obtained in Ref. \cite{IEEEexample:Rumsey_VH} in 1963, can
be converter time-domain, which is

\begin{equation}
-\iiintop_{V_{2}}\,\intop_{-\infty}^{\infty}E_{1}(t+\tau)\cdot J_{2}(\tau)d\tau dV=\iiintop_{V_{1}}\intop_{-\infty}^{\infty}E_{2}(t+\tau)\cdot J_{1}(\tau)\,d\tau dV\label{eq:40000-200}
\end{equation}
This formula is first obtained in ref.\cite{IEEEexample:Adrianus2},
If $t=0$ we can obtain that

\begin{equation}
-\iiintop_{V_{2}}\,\intop_{-\infty}^{\infty}E_{1}(\tau)\cdot J_{2}(\tau)d\tau dV=\iiintop_{V_{1}}\intop_{-\infty}^{\infty}E_{2}(\tau)\cdot J_{1}(\tau)\,d\tau dV\label{eq:40000-210}
\end{equation}
This formula is first obtained by Welch\cite{IEEEexample:Welch} in
1960. From above formula is also clear tell us, that the advanced
potential of antenna 2 sucks time{*}energy from antenna 1 $\iiintop_{V_{1}}\intop_{-\infty}^{\infty}E_{2}(\tau)\cdot J_{1}(\tau)\,d\tau dV$
is equal to the received time{*}energy by antenna 2 $-\iiintop_{V_{2}}\,\intop_{-\infty}^{\infty}E_{1}(\tau)\cdot J_{2}(\tau)d\tau dV$.
Welch knows that $E_{2}$ is advanced potential, but he has not Further
explain the meaning of his formula. Welch call his theorem time-domain
reciprocity theorem instead of some kind energy theorem, this perhaps
obstruct him to thought that the advanced potential can suck energy
from the transmitters. The first author of this article obtained the
mutual energy theorem\cite{IEEEexample:shrzhao1} in 1987, he also
did not notice that actually the advanced potential can suck energy
from transmitters. Perhaps this is because the advanced potential
is a so strange concept, most physicist and engineer reject this concept,
how can think it sucks energy?

Interesting to point out from Welch's formula, that the sucked energy
does not happen in particular time point, for example,

\begin{equation}
-\iiintop_{V_{1}}E_{1}(t)\cdot J_{2}(t)=\iiintop_{V_{1}}E_{2}(t)\cdot J_{1}(t)\,dV\label{eq:40000-220}
\end{equation}
but related a integral with time. Consider the photons has particle
frequency, the formula Eq.(\ref{eq:4060-2}) is in Fourier frequency
domain. The mutual energy transfer is based on wave of particular
frequency, perhaps is related to the fact that the energy particle
which is also based on frequency. 

\subsubsection{Similar result compare to the reciprocity theorem}

Applying magnetic mirror transform $m$ to the the variable with subscript
2, i.e.,

\begin{equation}
\zeta_{2}=m\,\zeta_{2m}\label{eq:4000-230}
\end{equation}
\begin{equation}
=[E_{2m}^{*},-H_{2m}^{*},-J_{m}^{*},K_{m}^{*},\epsilon_{m}^{*},\mu_{m}^{*}]\label{eq:4070}
\end{equation}
to the formula Eq.(\ref{eq:4060-1}), we obtain,
\begin{equation}
\iiintop_{V}\,(E_{1}\cdot(-J_{2m})+E_{2m}\cdot J_{1})=0\label{eq:4080}
\end{equation}
or
\begin{equation}
\iiintop_{V_{2}}\,E_{1}\cdot J_{2m}\,dV=\iiintop_{V_{1}}E_{2m}\cdot J_{1}\,dV\label{eq:4090}
\end{equation}
After the transform the media formula becomes,

\begin{equation}
\epsilon_{2m}^{T}=\epsilon_{1},\ \ \ \ \ \ \mu_{2m}^{T}=\mu_{1}\label{eq:4100}
\end{equation}

\begin{equation}
\epsilon_{2m}^{*}=\epsilon_{1}=\epsilon,\ \ \ \ \ \mu_{2m}^{*}=\mu_{1}=\mu\label{eq:4011}
\end{equation}
For simplification, the subscript $2m$ can be rewritten as 3, the
above formula becomes,

\begin{equation}
\iiintop_{V_{2}}\,E_{1}\cdot J_{3}\,dV=\iiintop_{V_{1}}\,E_{3}J_{1}\,dV\label{eq:4012}
\end{equation}
This formula is close to the the reciprocity theorem Eq.(\ref{eq:4040-1}).
It can be applied to prove that when an antenna is used as the receiver
or as the transmitter, will have the same directivity diagram. The
media formula can be rewritten as 

\begin{equation}
\epsilon_{3}^{T}=\epsilon_{1},\ \ \ \ \ \ \mu_{3}^{T}=\mu_{1}\label{eq:4013}
\end{equation}

\begin{equation}
\epsilon_{3}^{*}=\epsilon_{1}=\epsilon,\ \ \ \ \ \mu_{3}^{*}=\mu_{1}=\mu\label{eq:4014}
\end{equation}

Actually the Eq.(\ref{eq:4012},\ref{eq:4013}) are modified reciprocity
theorem. Eq.(\ref{eq:4014}) is caused by lossless media that we have
used in mutual energy theorem. Substitute Eq.(\ref{eq:4014}) to Eq.(\ref{eq:4013}),
i.e., considering $\epsilon_{1}=\epsilon$, $\epsilon_{3}=\epsilon^{*}$,
$\mu_{1}=\mu$, $\mu_{3}=\mu^{*}$, we still get, 

\begin{equation}
\epsilon^{\dagger}=\epsilon,\ \ \ \ \ \ \mu^{\dagger}=\mu\label{eq:2015}
\end{equation}

\subsection{Explanation of a system with a receive antenna using the mutual energy
theorem in loss media}

\subsubsection{The result of the mutual energy theorem}

Instead of using reciprocity theorem we will apply the mutual energy
theorem to the system with two antennas, one is transmit antenna and
another is receive antenna. For receive antenna, there should have
the wave toward to them, otherwise it can not receiver energy, all
fields send from the transmitter to the receiver is a wave toward
the receiver, but we have proven that cannot meet the Maxwell equation
close to the receiver antenna. Another wave toward to receiver is
its advanced potential. Hence we assume that the transmitter send
the retarded potentail and the receiver send a advanced potential.

We have proven that in the mutual energy theorem, if $\xi_{1}$ and
$\xi_{2}$ one is retarded potential and another is advanced potential,
the surface integral vanishes, i.e.,$(\xi_{1},\xi_{2})_{\varGamma}=0$.
Here $\varGamma$ is infinite big sphere. $\hat{n}$ is the outward
unit vector of surface $\varGamma$. Hence we have
\begin{equation}
Q_{loss}=-(\rho_{1},\xi_{2})_{V}-(\xi_{1},\rho_{2})_{V}\label{eq:4040-2}
\end{equation}
Here $-(\rho_{1},\xi_{2})_{V}$ is the emitted energy of $\rho_{1}$,
$-(\xi_{1},\rho_{2})_{V}$ is the emitted energy of $\rho_{2}$, this
energy output have been used in media, hence it equal to $Q_{loss}$.
If we take out the minus sign, $(\rho_{1},\xi_{2})_{V}$ is the absorbed
energy of $\rho_{1}$, $(\xi_{1},\rho_{2})_{V}$ is the absorbed energ
of $\rho_{2}$.
\begin{equation}
-\iiintop_{V}((E_{1}\cdot J_{2}^{*}+H_{1}\cdot K_{2}^{*})+(E_{2}^{*}\cdot J_{1}+H_{2}^{*}\cdot K_{1}))\,dV=Q_{loss}\label{eq:4050-1}
\end{equation}
If we have assumed that the magnetic current does not exist, $K_{1}=0$,
$K_{2}=0$, the above formula becomes,
\begin{equation}
-\iiintop_{V}\,(E_{1}\cdot J_{2}^{*}+E_{2}^{*}\cdot J_{1})\,dV=Q_{loss}\label{eq:4060-3}
\end{equation}
If current $J_{1},K_{1}$ is only inside $V_{1}$ and $J_{2},K_{2}$
is only inside $V_{2}$, we have, 
\begin{equation}
-\iiintop_{V_{2}}\,E_{1}\cdot J_{2}^{*}dV-\iiintop_{V_{1}}E_{2}^{*}\cdot J_{1}\,dV=Q_{loss}\label{eq:4060-1-1}
\end{equation}
or

\begin{equation}
-\iiintop_{V_{1}}E_{2}^{*}\cdot J_{1}\,dV=\iiintop_{V_{2}}\,E_{1}\cdot J_{2}^{*}dV+Q_{loss}\label{eq:4060-2-2}
\end{equation}
If $Q_{loss}=0$, this is lossless situation, there is, 
\begin{equation}
-\iiintop_{V_{1}}E_{2}^{*}\cdot J_{1}\,dV=\iiintop_{V_{2}}\,E_{1}\cdot J_{2}^{*}dV\label{eq:4060-2-1}
\end{equation}

$-\iiintop_{V_{1}}E_{2}^{*}\cdot J_{1}dV$ is the energy sucked by
the field $E_{2}$ from the current $J_{1}$. This is $J_{1}$emitted
energy.

$\iiintop_{V_{2}}\,E_{1}\cdot J_{2}^{*}dV$ indicate the energy of
the field $E_{1}$ act on the current $J_{2}$. This is the energy
antenna 2 received. 

This also tell us that the received energy of antenna 2 sent from
antenna 1 plus the media loss energy is same as the energy sucked
from antenna 1 by the advanced potential of the antenna 2. 

If advanced potential exist, it can suck energy from the source. This
sucked energy later is received by the receive antenna or is lost
in the media. This is the most important result of the mutual energy
theorem to the advanced potential and the received antenna. The authors
believe this result is correct and will apply this to explain more
physics result in the future.

\subsection{The differences of two methods }

The above mutual energy theorem result is similar to the explanation
of the reciprocity theorem Eq.(\ref{eq:40000-40}), however there
are fundamental differences, see following.

\subsubsection{$J_{3}$ is not a physical current}

1) In the mutual energy theorem, the field $\xi_{2}$ of the receive
antenna is advanced potential. $J_{2},K_{2}$ is sink which receive
wave energy, it is physical current. The current $J_{3}$ is a artificial
current or some kind of the effective current. Even the field of $\xi_{3}$
is retarded potential, but it is also an artificial field, the actually
physical field is still $\xi_{2}$ which is advanced potential. In
the reciprocity theorem the field of the receive antenna $\zeta_{2}$
is retarded potential.

\subsubsection{The are established in different media}

The mutual energy theorem functions in lossless media. This can not
be changed even after we have applied the magnetic mirror transform.
The reciprocity theorem is established in symmetric media. Only if
the media with real values for both mutual energy theorem and reciprocity
theorem can offer similar results. For example the media $\epsilon=i\epsilon_{0}$
(where $\epsilon_{0}$ is the Permittivity in free space) is not lossless
media but it is symmetric media. This This kind of media doesn't suit
to the derivation of the mutual energy theorem, hence the mutual energy
theorem can not apply to the system with transmit antenna and receive
antenna.

\subsubsection{in loss media, mutual energy theorem wins}

The mutual energy theorem (normally when we spoke about mutual energy
theorem, in the formula, there is $Q_{loss}=0$) functions in lossless
media. This can not be changed even after we have applied the magnetic
mirror transform. The reciprocity theorem is established in symmetric
media. Only if the media with real values for both mutual energy theorem
and reciprocity theorem can offer similar results. For example the
media $\epsilon=i\epsilon_{0}$ (where $\epsilon_{0}$ is the Permittivity
in free space) is not lossless media but it is symmetric media. This
kind of media doesn't suit to the mutual energy theorem (in lossless
media). In this situation we can apply the mutual energy with energy
loss.
\begin{equation}
-\iiintop_{V_{1}}E_{2}^{*}\cdot J_{1}\,dV=\iiintop_{V_{2}}\,E_{1}\cdot J_{2}^{*}dV+Q_{loss}\label{eq:4060-2-1-1}
\end{equation}
Assume $Q_{e}\equiv-\iiintop_{V_{1}}E_{2}^{*}\cdot J_{1}\,dV$ is
emission energy, $Q_{a}=\iiintop_{V_{2}}\,E_{1}\cdot J_{2}^{*}dV$
is received or absorbed energy, the above formula can be written as

\begin{equation}
Q_{e}=Q_{a}+Q_{loss}\label{eq:4060-3-1}
\end{equation}
That also means, 

\begin{equation}
Q_{a}<Q_{e}\label{eq:4060-4}
\end{equation}
Now we validate if we can obtained the above formula from the calculation
of with the retarded potential and advanced potential. Assume the
distance from receiver to the transmitter is $L$. $J_{1}$ located
at $0$ and $J_{2}$ located at $L$. We assume $|J_{1}|\sim1$, ``$\sim$''
means ``is close to each other''. Hence 
\begin{equation}
|J_{1}|\sim1\label{eq:5000-200}
\end{equation}

\begin{equation}
|E_{1}(0)|\sim1\label{eq:5000-201}
\end{equation}
$E_{1}(L)$ is retarded potential hence, at distance $L$ the value
decrease,

\begin{equation}
|E_{1}(L)|\sim\exp(-k_{loss}L)\label{eq:5000-210}
\end{equation}
$J_{2}$ is at the place of $L$, and it should influenced by the
field $E_{1}(L)$ hence,

\begin{equation}
|J_{2}|\sim|E_{1}(L)|\sim\exp(-k_{loss}L)\label{eq:5000-220}
\end{equation}
$E_{2}(L)$ is produced by the current $J_{2}$ hence there is,

\begin{equation}
|E_{2}^{*}(L)|\sim|J_{2}|\propto\exp(-k_{loss}L)\label{eq:5000-230}
\end{equation}
$E_{2}$ is advanced potential, when it go from the started point
$L$ to the destination $0$, it should increase the value, because
in the loss media, the advanced potential is ascending wave Eq.(\ref{eq:5000-190}),
hence,

\[
|E_{2}^{*}(0)|\sim E_{2}^{*}(L)\exp(k_{loss}L)
\]

\begin{equation}
\sim\exp(-k_{loss}L)\exp(k_{loss}L)=1\label{eq:5000-240}
\end{equation}

\begin{equation}
|Q_{e}|=|\iiintop_{V_{1}}E_{2}^{*}\cdot J_{1}\,dV|\sim1\label{eq:5000-250}
\end{equation}

\[
Q_{a}|=|\iiintop_{V_{2}}\,E_{1}(L)\cdot J_{2}^{*}dV|
\]
\begin{equation}
\sim\exp(-k_{loss}L)\exp(-k_{loss}L)=\exp(-2k_{loss}L)\label{eq:5000-260}
\end{equation}
We can see, it is really we can obtain that $Q_{a}<Q_{e}$. That also
means the mutual energy theorem with loss media can obtained correct
results.

In this loss media, if we also assume it is symmetric, the reciprocity
theorem still established. Hence we obtain
\begin{equation}
\iiintop_{V_{2}}\,E_{1}\cdot J_{2}\,dV=\iiintop_{V_{1}}E_{2}\cdot J_{1}\,dV\label{eq:500-270}
\end{equation}
That is clear wrong for an antenna system with a transmitter and a
receiver, and the media has energy loss, the reciprocity theorem is
clear wrong. The formula of the reciprocity theorem itself not wrong,
it assume the situation where the two antennas are all transmitters.
If the two antennas are all transmitters, we can also evaluate the
above formula using the similar method.

Since $\xi_{1}$ and $\xi_{2}$ are all retarded potentials, hence

\begin{equation}
|J_{2}|\sim1\ \ \ \ \ \ \ |J_{1}|\sim1\label{eq:5000-300}
\end{equation}

\begin{equation}
|E_{1}(0)|\sim1\ \ \ \ \ E_{2}(L)\sim1\label{eq:5000-310}
\end{equation}

\begin{equation}
|E_{1}(L)|\sim\exp(-k_{loss}L)\ \ \ \ \ \ |E_{2}(0)|\sim\exp(-k_{loss}L)\label{eq:5000-320}
\end{equation}

\begin{equation}
|\iiintop_{V_{2}}\,E_{1}\cdot J_{2}\,dV|\sim|E_{1}(L)J_{2}|\sim\exp(-k_{loss}L)\label{eq:5000-330}
\end{equation}

\begin{equation}
|\iiintop_{V_{2}}\,E_{2}\cdot J_{1}\,dV|\sim|E_{2}(0)J_{1}|\sim\exp(-k_{loss}L)\label{eq:5000-340}
\end{equation}
hence we have

\begin{equation}
|\iiintop_{V_{2}}\,E_{2}\cdot J_{1}\,dV|\sim|\iiintop_{V_{2}}\,E_{1}\cdot J_{2}\,dV|\label{eq:5000-350}
\end{equation}

\subsubsection{The scattering process}

In the derivation of the mutual energy we have assumed that there
is no scattering field sending out from the receive antenna 2. If
there is scattering for antenna 2, We can not obtained Eq.(\ref{eq:4050})
and hence Eq.(\ref{eq:4012}). 

Using these 3 differences we can design experiments to further examine
which is correct, the two theories, the reciprocity theorem or the
mutual energy theorem. We can test the theories in symmetric loss
media in which the reciprocity theorem is established, but the mutual
energy theorem is not established. In this case if we still obtained
the same result of reciprocity for a antenna used as receive antenna
and transmit antenna, then the reciprocity is correct, otherwise the
mutual energy theorem is correct.

There is no any influence of scattering to the reciprocity theorem.
But from derivation of mutual energy theorem we have known that the
scattering play a important role. Since paraboloid antenna has big
difference of scattering when the receive antenna toward to the transmitter
or back against the transmitter. If the directivity pattern of a receive
antenna and a transmit antenna is influenced by the scattering, then
the reciprocity theorem is wrong and the mutual energy theorem is
corrected to the antenna theory.

\subsection{Calculate the mutual energy current}

It is clear there are energy current sent from transmit antenna to
the received antenna. It is not possible to get this kind of energy
current from reciprocity theorem, but it is possible to get this energy
current from the mutual energy theorem.

Since we assume that the receive antenna associate only advanced potential
(assume there is no scattering, and the transmit antenna sent only
retarded potential. We also know that the mutual energy of a retarded
potential and an advanced potential have no pure mutual energy go
to the outside of the infinite big sphere, i.e. the surface integral
of the mutual energy theorem vanishes. Hence all mutual energy current
is stared from transmit antenna to the receive antenna. The authors
believe this energy current is just the energy sending from transmit
antenna to the receive antenna. In the following we offer the formula
to calculate the mutual energy current from the transmit antenna to
the receive antenna.

Assume there is a transmit antenna which has the source $J_{1},K_{1}$
and the corresponding field $\xi_{1}$ and a receive antenna which
has a sink $J_{2},K_{2}$ and corresponding field $\xi_{2}$. Assume
$V_{1}$ is the the volume corresponding to $J_{1},K_{1}$ and $V_{2}$
is the volume corresponding to $J_{2},K_{2}$. The $\varGamma_{1}$
is the boundary of $V_{1}$. We can apply the mutual energy theorem
to the volume $V_{1}$, 
\begin{equation}
(\xi_{1},\xi_{2})_{\varGamma_{1}}+(\rho_{1},\xi_{2})_{V_{1}}=0\label{eq:5010}
\end{equation}
or

\begin{equation}
(\xi_{1},\xi_{2})_{\varGamma_{1}}=-(\rho_{1},\xi_{2})_{V_{1}}\label{eq:5020}
\end{equation}
or
\[
\varoiintop_{\varGamma_{1}}\,(E_{1}\times H_{2}^{*}+E_{2}^{*}\times H_{1})\,\cdot\hat{n}_{1}dS
\]
\begin{equation}
=-\iiintop_{V_{1}}\,(J_{1}\cdot E_{2}^{*}+K_{1}\cdot H_{2}^{*})\,dV\label{eq:5030}
\end{equation}
If we apply the mutual energy theorem to the volume $V_{2}$ there
is
\begin{equation}
(\xi_{1},\xi_{2})_{\varGamma_{2}}+(\xi_{1},\rho_{2})_{V_{2}}=0\label{eq:5040}
\end{equation}
\begin{equation}
-(\xi_{1},\xi_{2})_{\varGamma_{2}}=(\xi_{1},\rho_{2})_{V_{2}}\label{eq:5050}
\end{equation}
or
\[
-\iintop_{\varGamma_{2}}\,(E_{1}\times H_{2}^{*}+E_{2}^{*}\times H_{1})\,\hat{n}_{2}dS
\]
\begin{equation}
=\iiintop_{V_{2}}\,(E_{1}\cdot J_{2}^{*}+H_{1}\cdot K_{2}^{*})\,dV\label{eq:5060}
\end{equation}
From the mutual energy theorem for volume $V$, we know that,
\begin{equation}
(\xi_{1},\rho_{2})_{V}+(\xi_{2},\rho_{1})_{V}=0\label{eq:4040-1}
\end{equation}
or

\[
-\iiintop_{V_{1}}\,(E_{2}^{*}\cdot J_{1}+H_{2}^{*}\cdot K_{1})dV
\]
\begin{equation}
=\iiintop_{V_{2}}\,(E_{1}\cdot J_{2}^{*}+H_{1}\cdot K_{2}^{*})\,dV\label{eq:5070}
\end{equation}
where $V_{1}\subset V$ and $V_{2}\subset V$.

Substitute Eq.(\ref{eq:5030}, \ref{eq:5060}) to Eq.(\ref{eq:5070}),
we obtain,

\[
\iintop_{\varGamma_{1}}\,(E_{1}\times H_{2}^{*}+E_{2}^{*}\times H_{1}^{*})\,\cdot\hat{n_{1}}dS=-\iintop_{\varGamma_{2}}\,(E_{1}\times H_{2}^{*}+E_{2}^{*}\times H_{1})\,\hat{n}_{2}dS
\]

The above formula shows the mutual energy current send from the transmit
antenna $\varGamma_{1}$ is equal to the the energy current received
by the receive antenna on $\varGamma_{2}$. Actually we can choose
any surface between the transmit antenna 1 and the receive antenna
2 to calculate the mutual energy current, which is the energy send
from transmit antenna to the received antenna. The energy current
sending from transmit antenna to the receive antenna can only be calculated
by the mutual energy theorem, it can not be calculated by the reciprocity
theorem or Poynting energy theorem. This is also shows the importance
of the concept of the mutual energy theorem,

The left side of Eq.(\ref{eq:5030}) and Eq.(\ref{eq:5060}) are the
mutual energy current go through transmit antenna 1 to the receive
antenna 2. The right side of the above formula is the receive energy
of the receive antenna. 

We know the real Poynting vector is defined as

\[
S=Re\{E\times H^{*}\}
\]
\begin{equation}
=\frac{1}{2}(E\times H^{*}+E^{*}\times H)\label{eq:40000-300}
\end{equation}

Constant $\frac{1}{2}$ can be omitted, hence we can also write
\begin{equation}
S=E\times H^{*}+E^{*}\times H\label{eq:40000-310}
\end{equation}

Similar to the above definition of Poynting vector, we can define
a mutual energy vector, 
\begin{equation}
S_{12}=E_{1}\times H_{2}^{*}+E_{2}^{*}\times H_{1}\label{eq:400-330}
\end{equation}

The energy current from the transmit antenna to the receive antenna
are 
\begin{equation}
Q_{12}\equiv\iintop_{\varGamma}S_{12}\cdot\hat{n}d\varGamma=\iintop_{\varGamma}(E_{1}\times H_{2}^{*}+E_{2}^{*}\times H_{1})\,\cdot\hat{n}d\varGamma\label{eq:40000-450}
\end{equation}
Where $\varGamma$ is any surface between antenna 1 and antenna 2.

\section{Is advanced potential just the retarded potential}

Lawrence M. Stephenson\cite{LawrenceMStephenson} point out there
is one situation there advanced potential can not avoid, that is one
dimensional antenna system. For example there is a wave guide, in
the left end there is transmitter and in the right end there is a
receiver. Assume there is no any energy loss. Hence the retarded potential
of the transmitter is also the advanced potential of the receiver.

In the beginning the author thought this is a good idea, i.e. the
advanced potential perhaps is the retarded potential of combination
of some retarded potential. The authors built the following antenna
theory.

\subsection{Simple antenna system}

\subsubsection{One dimension antenna system}

For one dimension antenna system the antenna 1 send wave energy out
and the antenna 2 receive the wave energy. Assume there is no any
energy loss. Hence all energy send from antenna 1 has been received
by antenna 2. It is clear that the field between antenna 1 and antenna
2 is retarded potential for antenna 1 and advanced potential for antenna
2.

\subsubsection{Simple 3 dimension antenna system}

Consider another simple antenna system, there is a transmit antenna,
which send wave from the source current $\rho_{1}=[J_{1},K_{1}]$
and a receive antenna which is at big sphere and the receiver can
receive all electromagnetic waves. The current on the antenna 2 is
$[J_{s2},K_{s2}]$ which is surface current. Assume antenna 2 is supper
conductor and super magnetic conductor, hence on the surface of antenna
2, the electric field and magnetic field both vanish. the current
$[J_{s2},K_{s2}]$ can be found from the boundary conditions as,

\begin{equation}
J_{s2}=\hat{n}\times(H_{1}^{out}-H_{1}^{in})\label{eq:4200}
\end{equation}
\begin{equation}
K_{s2}=-\hat{n}\times(E_{1}^{out}-E_{1}^{in})\label{eq:4210}
\end{equation}
where $E_{1}^{out}$ is the electric field at outside of the big sphere
surface. $E_{1}^{in}$ is the electric file at inside of the infinite
sphere surface. $H_{1}^{out}$ and $H_{1}^{in}$ have also the similar
meaning. Since the antenna 2 are super conductor and super magnetic
conductor, $E_{1}^{out}=H_{1}^{out}=0$. For this kind of antenna
system, the field send by antenna 1 is retarded potential $\zeta_{1}$,
but it is also the advanced potential $\zeta_{2}$ for antenna 2.
$\hat{n}$ is the normal vector of the sphere

\begin{equation}
\zeta_{1}=\zeta_{2}\label{eq:4230}
\end{equation}

\subsection{The problem of this antenna theory}

There is the antenna theory\cite{LawrenceMStephenson}, it supported
advanced potential, but claim the advanced potential is just part
of the retarded potential.

Lawrence M. Stephenson\cite{LawrenceMStephenson} point out, for the
simple antenna system, the field between the transmit antenna and
the receive antenna can be seen as retarded potential for transmit
antenna and received potential for the receive antenna. Every thing
is fine. It is seems this example support the existing advanced potential.
Lawrence M. Stephenson thought this is an example to support the advanced
potential.

But after reflectingly thought, the authors can not agree. The author
thought this example actually do not support the existing the advanced
potential, if the receive antenna has associated the same field as
the transmit antenna.

If the advanced potential exists, even for this simple situation the
received antenna should has the field of its own. Assume $\rho_{1}=[J_{1},K_{1}]$
is the transmit antenna, $\rho_{2}=[J_{2},K_{2}]$ is the receive
antenna. The field $\xi_{1}$ is the field produced by antenna 1.
$\xi_{2}$ is the field of received antenna 2 which is advanced potential.

The author believe that the total field of this system should be
\begin{equation}
\xi=\xi_{1}+\xi_{2}\label{eq:50000-10}
\end{equation}
That means the field should be possible to be superimposed. Even for
this simple situation, the fields transmit antenna should be possible
to be superimposed. 

We have mentioned if the field of received antenna can be obtained
by a combination of other transmit field, i.e.,

\begin{equation}
\xi_{a}=\sum_{i=1}^{N}a_{i}\xi_{t_{i}}\label{eq:50000-20}
\end{equation}

\[
M\xi_{a}=\sum_{i=1}^{N}a_{i}\xi_{t_{i}}
\]

\[
\sum_{i=1}^{N}a_{i}\text{\ensuremath{M}}\xi_{t_{i}}
\]

\begin{equation}
=\sum_{i=1}^{N}a_{i}\gamma\rho_{t_{i}}\label{eq:50000-30}
\end{equation}
$\xi_{a}$ is the absorbed or received. 

Where $\gamma$ is a constant. The above formula is clear conflict
to the fact, we need in the place close to the receive antenna, there
should be 

\begin{equation}
M\xi_{a}=\gamma\rho_{a}\label{eq:50000-32}
\end{equation}
Where $\rho_{a}$ is the current of receiver(or absorber). Hence this
kind of antenna theory the field can not satisfy the Maxwell equations. 

\subsection{The correction to the above antenna theory}

\subsubsection{The superimposition, should be insistent, i.e.,}

Even there are retarded potential $\xi_{1}$ and advanced potential
$\xi_{2},$ which is produced by transmit antenna source $\rho_{1}$
and the receive antenna sink $\rho_{2}$. The field should be the
superimposition of the two field
\begin{equation}
\xi=\xi_{1}+\xi_{2}\label{eq:50000-400}
\end{equation}
But we also knows that for the simple situation there is

\begin{equation}
\xi=\xi_{1}=\xi_{2}\label{eq:50000-410}
\end{equation}
How can we make the above two different formulas self-consistent? 

\subsubsection{The mutual energy current is important}

The mutual energy current is the energy transfer between the two antenna.
The Poynting vector of related energy current can not transfer energy
between two antennas

Even in one dimensional wave guide system the above 2 points should
also be true. Assume the transmit antenna is at $x=a$, and the receive
antenna is at $x=b$, here $a<b$, 

\begin{equation}
\xi_{2}=\begin{cases}
-\xi_{1} & \ \ \ \ x<a\\
\xi_{1} & \ \ \ \ x\geq a,\,x\leq b\\
-\xi_{1} & \ \ \ \ x>b
\end{cases}\label{eq:50000-430}
\end{equation}
If we superimpose the field according the traditional way there is

\begin{equation}
\xi=\xi_{1}+\xi_{2}\label{eq:5000-440}
\end{equation}

\begin{equation}
=\begin{cases}
0 & x<a\\
2\xi_{1} & x\geq a,\,x\leq b\\
0 & x>b
\end{cases}\label{eq:50000-450}
\end{equation}
Hence, the energy transfer between two antenna is

\[
Q=\iintop_{\varGamma}S\,\hat{n}d\varGamma
\]

\[
=\iintop_{\varGamma}(E_{1}+E_{2})\times(H_{1}^{*}+H_{2}^{*})+(H_{1}+H_{2})\times(E_{1}^{*}+E_{2}^{*})\,\hat{n}d\varGamma
\]

\[
=4\iintop_{\varGamma}(E_{1}\times H_{1}^{*}+H_{1}^{*}\times E_{1})\,\hat{n}d\varGamma
\]

\begin{equation}
=4\iintop_{\varGamma}S_{11}\,\hat{n}d\varGamma\label{eq:50000-460}
\end{equation}

Here $S$ is the Poynting vector. $S_{11}=E_{1}\times H_{1}^{*}+H_{1}^{*}\times E_{1}$.
This can be referred the energy current 4 times difficulty. That means
if the advanced potential is normal field, in this one dimension antenna
system the transferred energy will be 4 times big than the energy
transferring energy calculated with the Poynting vector.

The above result is clear not correct. The explanation for this is
that the advanced potential is not a normal field, it is the field
perhaps in other implicit space. Hence the above field energy calculation
is meaning less. The only important between two antennas is the mutual
energy which is the transferred energy. the advanced potential can
act to the current of a retarded potential. The retarded potential
can act on the current of an advanced field. Hence the formula of
mutual energy
\begin{equation}
(\xi_{1},\rho_{2})_{V_{2}}+(\rho_{1},\xi_{2})_{V_{1}}=0\label{eq:50000-470}
\end{equation}
is still established. In the above formula the first item $(\xi_{1},\rho_{2})_{V_{2}}$
is the retarded potential offered energy to current $\rho_{2}$. The
second item $-(\rho_{1},\xi_{2})_{V_{1}}$ is corresponding to the
sucked energy by the advanced potential from the source current $\rho_{1}.$
\begin{equation}
(\xi_{1},\xi_{2})_{\varGamma_{1}}+(\rho_{1},\xi_{2})_{V_{1}}=0\label{eq:50000-480}
\end{equation}
The first item is the mutual current send out to the transmitter antenna,
second item $-(\rho_{1},\xi_{2})_{V_{1}}$ is sucked energy from the
transmitter antenna. And we have, 

\begin{equation}
(\xi_{1},\rho_{2})_{V_{2}}=-(\xi_{1},\xi_{2})_{\varGamma_{2}}\label{eq:50000-490}
\end{equation}
or

The right of the above formula is the mutual energy send to the receive
antenna. The left is the energy current cent to the receive antenna.

The energy transferred from transmit antenna to the received antenna
is still the mutual energy current,
\[
Q_{12}=(\xi_{1},\xi_{2})_{\varGamma}
\]
\[
=\iintop_{\varGamma}(E_{1}\times H_{2}^{*}+E_{2}^{*}\times E_{1})\,\hat{n}d\varGamma
\]
\begin{equation}
=\iintop_{\varGamma}S_{12}\,\hat{n}d\varGamma\label{eq:50000-500}
\end{equation}
Hence, we can define mutual energy current vector,

\begin{equation}
S_{12}=E_{1}\times H_{2}^{*}+E_{2}^{*}\times E_{1}\label{eq:50000-510}
\end{equation}
Here, if $\varGamma$ is chosen between the two ends, since $\xi_{1}=$$\xi_{2}$
(notice, only the value is same, actually one is retarded potential
and one is advanced potential, they are different things), the mutual
energy 

\[
Q_{12}=\iintop_{\varGamma}(E_{1}\times H_{1}^{*}+E_{1}^{*}\times E_{1})\,\hat{n}d\varGamma
\]

\begin{equation}
=\iintop_{\varGamma}S_{11}\,\hat{n}d\varGamma=Q_{11}\label{eq:5000-520}
\end{equation}
Here, $S_{11}$is the Poynting vector corresponding the retarded potential
$\xi_{1}$. It is same there is 
\begin{equation}
Q_{12}=Q_{22}\label{eq:50000-530}
\end{equation}
Hence, there is

\begin{equation}
Q_{11}=Q_{12}=Q_{22}\label{eq:50000-540}
\end{equation}
In the above the mutual energy $Q_{12}$ is the real energy transferred
from the transmit antenna to the receive antenna. This energy in this
situation just equal to the energy send out from the transmitter $Q_{11}$
and also the energy received by the receive antenna $Q_{22}$.

If $\varGamma$ is chosen less the left end, $\xi_{1}$ is the wave
toward out side $\xi_{2}$ is the wave toward inside, it is clear
the mutual energy of
\begin{equation}
Q_{12}(x<a)=0\label{eq:50000-550}
\end{equation}

If $\varGamma$ is chosen at outside right end, $\xi_{1}$ is the
wave toward out side $\xi_{2}$ is the wave toward inside, it is clear
the mutual energy of
\begin{equation}
Q_{12}(x>b)=0\label{eq:50000-570}
\end{equation}

From above discussion, in this situation the transferred energy is
equal to $Q_{12}$, the value of $Q_{12}$ is equal to $Q_{11}$or
$Q_{22}$ accidently. Here we speak about ``accidently'' is because
$Q_{12}=Q_{11}=Q_{22}$ is only in this simple one dimension wave
guide antenna system. In general, $Q_{12}\neq Q_{11}\neq Q_{22}$

\subsection{Summary}

If we assume there is no advanced potential in 3D space, the Maxwell
equation cannot be satisfied at the place of the receive antenna.
Hence there should exist the advanced potential.

If advanced field is the normal field, in the one dimensional wave
guide situation, the field between the transmitter and the receiver
receiver will be doubled according to the fields superimposition.
If the superimposed field is doubled, the energy current will 4 times
bigger than the energy current calculated through Poynting vector.
This means also that some thing is wrong. Hence we have to assume
the advanced potential is not a normal field. Hence the calculation
result about the superimposition of advanced field and a retarded
field is also not normal. We can assume the advanced potential is
at some other space outside our space. However it can act on to current
which created the retarded potential. 

The advanced potential can suck energy from the current sours (or
transmit antenna or transmitter). The sucked energy is equal to the
energy of the retarded potential applied to the sink (the received
antenna or the receiver). The energy transfer in the space is through
the mutual current. 

The important thing about energy transfer to an antenna system is
the mutual energy current, which is the energy transferred from the
transmitter antenna to the receiver antenna.

The energy current calculated through Poynting vector does not offer
the energy transferring between transmit antenna and the receive antenna.
This is clear if considered a section area of receive antenna wire
is close to 0. This kind antenna obtained 0 energy from energy current
related to the Poyning vector. However we know that the antenna received
energy is not depending to the section area of the wire. Even the
wire is as thin as possible, it still can receive nearly same energy
compared to a wire with no zero section area.

The current calculated by Poynting vector is also not the energy transfer
between the transmitter and the receiver in the case of one dimensional
wave guide. However accidently the mutual energy current is just equal
to the energy current calculated by the Poynting vector.

We also answered the question whether the advanced potential can be
a combination of other retarded potential, the answer is negative.
The advanced potential can not be any combination of the retarded
potential, it is the potential ``sent'' out by the sink current.
Hence ``sent'' is very important, the advanced potential is caused
by the sink!

In case of one dimensional wave guide, or 3D situation the background
is seen as receiver antenna, the advanced potential is equal to the
retarded potential in mathematics values, but they are still different
things. But in general situation, for example there is a transmit
antenna and a receiver antenna in free space, the advanced potential
is not equal to the retarded potential, because the retarded potential
is sent by transmitter and the advanced potential is sent from the
receiver.

\section{Re-explanation of the energy calculated from Poynting vector \label{sec:Re-explanation-of-The}}

\subsection{The problems of Poynting vector}

We have few times met the problems with the energy current calculated
from Poynting vector. In the following it is referred as P-current,
defined as
\begin{equation}
Q=\iintop_{\varGamma}S\cdot\hat{n}d\varGamma\label{eq:60000-10}
\end{equation}
where
\begin{equation}
S=E\times H^{*}+E^{*}\times H=2Re\{E\times H^{*}\}\label{eq:60000-20}
\end{equation}

\begin{equation}
Q=(\xi,\xi)=\iintop_{\varGamma}S\cdot\hat{n}d\varGamma\label{eq:60000-30}
\end{equation}
In our definition of Poynting vector is a real value. We have added
a constant 2 to make it has similar definition with mutual energy
current density vector, which is defined as

\begin{equation}
S_{12}=E_{1}\times H_{2}^{*}+E_{2}^{*}\times H_{1}\label{eq:60000-50}
\end{equation}
We can also define the mutual energy current (or for simple M-current)
as

\begin{equation}
Q_{12}=(\xi_{1},\xi_{2})_{\Gamma}=\iintop_{\varGamma}S_{12}\cdot\hat{n}d\varGamma\label{eq:60000-60}
\end{equation}

\subsubsection{Receive antenna with thin wire }

First we have known that the received energy from a received antenna
is not related to P-current. When the section area of wires become
as small as possible the antenna can still receive enough energy.
Here we can assume this wire is produced by superconductor material,
hence the current on the thin wire can be big enough. This problem
is known for long time, it has been also noticed by Ref.\cite{LawrenceMStephenson}.
The electronic engineer just overlook this phenomenon. 

\subsubsection{4-times of the energy current.}

Second in the last section there is so called 4-times difficulty.
In one dimensional wave guide or 3D space the transmitter situated
at the center of a sphere, the receive antenna is at the sphere, and
assume there is no energy loss in the system. In this situation the
retarded potential and advanced potential is equal in values. The
P-current of superimposed field of two fields in which one is retarded
and the other one is advanced, obtains 4-times of the P-current compare
to the P-current of the transmit antenna or the receive antenna. This
means if we accept the advanced potential, we have to face the 4-time
energy current difficulty. Most of us meet this difficulty perhaps
will reject the advanced potential. But the authors strongly insist
the advanced potential, hence the wrong should be in the side of Poynting
vector.

\subsubsection{An advanced potential in 3D free space}

We assume there is only a current situated in the center of space,
there are no other material. Assume this a current (sink) has advanced
potential, the P-current is clear no zero, that means there is an
energy current always toward to the current. We have said the advanced
potential only can be used to receive energy. Now it seems even without
other source, a sink can still obtained energy, this is very strange.
This is also another reason perhaps why normally the people reject
the advanced potential. 

The authors strongly believe the advanced potential. In order to solve
these problems, the authors have to assume that it is not the P-current
transfer the energy from a transmitter to a receiver. P-current has
no any power to transfer the energy, it is the self-energy current
belong the current itself. It does not communicate with other world.
The energy current transferring from transmitter to a receiver can
only be done with the mutual energy current, i.e., M-current.

In case of one dimensional wave guide situation discussed in last
section we have found that the P-current is equal to the M-current
accidentally. 

In another case, a transmitter antenna sits at the center of free
space. The background of free space can can be seen as infinite more
receiver. Each receivers will send an advanced potential to the transmit
antenna. The summation of this all mutual energy current can be calculated
by assume the background as big receive antenna, we can show this
summation of the M-current from the transmit antenna to the background
receiver is just same to the P-current.

We can assume the background as a big antenna situated at the infinite
big sphere, this antenna receive all energy from the transmit antenna.
Hence it send the advanced field which is same as the field of transmit
antenna, between the transmit antenna and the infinite big sphere.
That means,

\begin{equation}
\xi_{r}=\xi_{a}\label{eq:60000-100}
\end{equation}
The P-current is
\begin{equation}
(\xi_{r},\xi_{r})=\iintop_{\varGamma}(E_{r}\times H_{r}^{*}+E_{r}^{*}\times H_{r})\,\hat{n}d\varGamma\label{eq:60000-110}
\end{equation}

\begin{equation}
(\xi_{r},\xi_{a})=\iintop_{\varGamma}(E_{r}\times H_{a}^{*}+E_{r}^{*}\times H_{a})\,\hat{n}d\varGamma\label{eq:60000-130}
\end{equation}
Hence we have 

\begin{equation}
(\xi_{r},\xi_{a})=(\xi_{r},\xi_{r})\label{eq:60000-150}
\end{equation}
After this re-explanation for P-current, we can avoid the 4-times
difficulty mentioned before and also can explain the reason why a
receive antenna received energy is nothing to do with P-current. 

After this re-explanation for P-currents, we also do not need to assume
the advanced potential is at other space instead of our space, which
is also too strange. But any way the retarded potential and advanced
potential are still different and they need different method to measure
them.

After this re-explanation we also solved the problem of conflict between
assumptions of following two formula,
\begin{equation}
\xi=\xi_{1}+\xi_{2}\label{eq:60000-170}
\end{equation}
\begin{equation}
\xi=\xi_{1}=\xi_{2}\label{eq:60000-180}
\end{equation}
where $\xi_{1}$ is retarded potential and $\xi_{2}$ is an advanced
potential. The two formula are all correct. The first is correct in
general and $\xi$ is superimposed field, this field perhaps is not
a normal field. The second formula Eq.(\ref{eq:60000-180}) is correct
in case of one dimensional wave guide, or in a free space the background
can be seen as receive antennas, without energy loss and reflection,
and here $\xi$ here is a normal field we can detected with normal
technology. In the one dimensional wave guide, here $\xi_{1}=\xi_{2}$
is only on mathematical value, $\xi_{1}$, $\xi_{2}$ are still two
different physical thing $\xi_{1}$ is retarded potential, $\xi_{2}$
is advanced potential. $\xi=\xi_{1}$ is tell us in the wave guide
the normal field just the retarded potential.

In this re-explanation the P-current of advanced potential do not
cause any energy current, hence advanced potential can be existed
to arbitrary current. If there is no source or transmit antenna, there
is no energy current of advanced potential go towards the current
of the receiver. We know that From Maxwell equations we can obtain
two solutions one is retarded potential one is advanced potential.
Now we know that the advanced potential exists to any current even
it is a transmit antenna. The P-current of this advanced potential
do not cause an energy toward the current source or sink. 

The retarded potential send out the energy, however we have shown
it is not the retarded potential's P-current send energy out, it is
the back ground which can be seen as infinite small antennas, which
have sucked the energy from the transmit antenna through their advanced
potential. The summation of all mutual energy current, i.e. M-current
just equal to the P-current of the retarded potential.

\subsection{The mutual energy of two retarded potential}

Assume there are two current sources $\rho_{1}=[J_{1},K_{1}]$, $\rho_{2}=[J_{2},K_{2}]$,
we have produced two retarded potential, $\xi_{1}$ $\xi_{2}$,
\begin{equation}
\text{\ensuremath{\xi}=\ensuremath{\xi_{1}+\xi_{2}}}\label{eq:60000-190}
\end{equation}
is also retarded potential, the Poyning theorem can be obtained by
\[
Q=(\xi,\xi)_{\varGamma}
\]
\[
=(\xi_{1},\xi_{1})_{\varGamma}+(\xi_{1},\xi_{2})_{\varGamma}+(\xi_{2},\xi_{1})_{\varGamma}+(\xi_{2},\xi_{2})_{\varGamma}
\]

\begin{equation}
=Q_{11}+Q_{12}+Q_{21}+Q_{22}\label{eq:60000-200}
\end{equation}

We have known that $Q_{11}$ and $Q_{22}$ and $Q$ are the P-current
of $\xi_{1}$,$\xi_{2}$ and $\xi$ respectively, we have know that
this current belong self energy current, it has no any contribution
to outside world. $Q_{11}+Q_{12}$ is some part of $\xi$. It should
also no any contribution to the energy change. The energy change of
$Q_{12}$ and $Q_{21}$ are also retarded energy current, it will
have no contribution to the energy current. 

\subsection{Summary about P-current}

\subsubsection{For a current in free space}

The authors assume, any currents have advanced potential and retarded
potential. P-current of the advanced potential and the retarded potential
do not transfer energy. We can also think that the retarded potential
it transfer energy to infinite far away and then this energy change
to advanced potential and has been received by this current, so there
is no any pure exchange of energy to other current. But this need
in the infinite remote distance there should have a tunnel which has
the power to connect the future to the past. 

Since we cannot detector this kind of P-current, that means not only
the P-current of advanced potential but also the P-current of the
retarded potential all can not be detected.

\subsubsection{M-current}

The authors assume, it is only the mutual energy current M-current,
which is responsible for the transferring the energy from transmit
antenna to the receive antenna. Here the mutual energy of a retarded
potential and an advanced potential. The advanced mutual energy must
send from the receiver and cause the the current in the receiver changed,
and the change a current in receiver send a back a advanced potential
to the receiver. Since this this two potential can be synchronized.
The mutual energy can send from the transmitter to the receiver in
a retarded meaning.

It is also possible the receiver send a advanced potential to the
transmitter and the transmitter caused a current change which in turn
produced a retarded potential, that potential is send bake to the
receiver. In this case the energy is also retarded from the transmitter
transferred to the receiver in retarded meaning.

\subsubsection{One dimensional wave guide situation}

In case of one dimensional wave guide or the case the transmit antenna
in the center of a big sphere, the receive antenna is at the big sphere,
the M-current is just equal to the P-current of receive antenna or
the P-current of the transmit antenna. Hence we still can calculate
the transferred energy by use P-current for this special situation.
This waym the calculation result is correct, but the method is wrong. 

\subsubsection{In case there are of laser beam}

In the case of laser beam if we can calculate the P-current of the
receiver, which is

\begin{equation}
Q=(\xi_{1},\xi_{1})=\iintop_{\varGamma_{S}}(E_{1}\times H_{1}+E_{1}^{*}\times H_{1})\cdot\hat{n}d\varGamma\label{eq:60000-210}
\end{equation}
and we can also calculate the total M-current, Here $\varGamma_{M}$
is the beam area of M-current,

\begin{equation}
Q_{12}=(\xi_{1},\xi_{2})=\iintop_{\varGamma_{M}}(E_{1}\times H_{2}+E_{2}^{*}\times H_{1})\cdot\hat{n}d\varGamma\label{eq:60000-230}
\end{equation}

\begin{figure}
\includegraphics{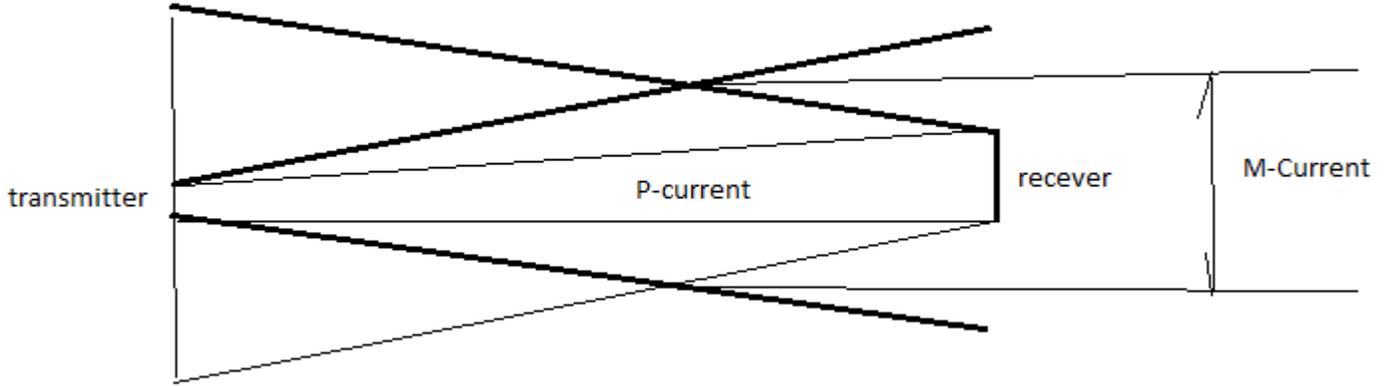}

\caption{Laser beam situation.\label{fig:9-01}}
\end{figure}

Here $\varGamma_{M}$ is the beam area of M-Current. The beam shape
for P-current and M-current can been seen in the figure \ref{fig:9-01}.

I believe that the M-current of the receiver is little bigger than
the P-current of the transmitter (please see, not only part of P-current
has been received by the receiver), because the beam area of M-current
is bigger than that of P-current, see Figure \ref{fig:9-01}. 

This is also clear for us, we know that line antenna have much bigger
effect area than the actual section area of the antenna. The plate
antenna has also larger effective receive area than the plate is.
The reason is also because energy is transferred by M-current instead
of P-current. This also explained as the antibunching\cite{Wheeler_1,Wheeler_2,Pegg}
phenomenon.

\section{The principle of the mutual energy}

Normally we speak that Poynting theorem is the energy conservation
law for electromagnetic field theory. However summary of this article
are much different. The Poynting vector loss some of importance. The
concept mutual energy win more importance. The following can be seen
as mutual energy principle (because we have not proven it completely,
it is accept as principle or law). Assume there is a system with currents
(sours and sink) separate remotely.

\subsection{Energy is not transferred by Poynting vector}

Assume we have two current $\rho_{1}$ and $\rho_{2}$ not very close
to the origin of the infinite bigger sphere. The current $\rho_{1}$,
$\rho_{2}$ can be seen as remote to each other, they send advanced
potential and retarded potential, $\xi_{1a}$, $\xi_{1r}$, $\xi_{2a},$$\xi_{2r}$ 

The energy current of infinite big sphere $\varGamma$ is calculated,
we found the following energy current do not transfer energy:

\subsubsection{The energy current calculated from Poynting vector of the retarded
potential, 
\[
con(\xi_{1r},\xi_{1r})_{\varGamma}=con(\xi_{2r},\xi_{2r})_{\varGamma}=0
\]
}

\subsubsection{The energy current calculated from Poynting vector of the advanced
potential 
\[
con(\xi_{1a},\xi_{1a})_{\varGamma}=con(\xi_{2a},\xi_{2a})_{\varGamma}=0
\]
}

\subsubsection{The mutual energy of two retarded potentials, 
\[
con(\xi_{1r},\xi_{2r})_{\varGamma}=0
\]
}

\subsubsection{The mutual energy of two advanced potentials 
\[
con(\xi_{1a},\xi_{2a})_{\varGamma}=0
\]
}

In the above the symbol $con$ means contributed energy to the outside
of the surface.

These energy current can be seen as some kind of self-energy. Self
energy do not exchange energy with other materials. It belong to the
current itself.

Actually the in the first 4 principles, only the first two are required.
We can prove the third and the fourth can be derived from the first
two. 

Proof: 

Assume in the beginning there is only a current $\rho_{_{1}}=[J_{1},K_{1}]$
in the empty space (their are no any background materials). The corresponding
field is $\xi_{1}=[E_{1},H_{1}]$. We can calculate the Poynting vector
as
\[
Q_{1}=(\xi_{1},\xi_{1})=\varoiintop_{\varGamma}(E_{1}\times H_{1}^{*}+E_{1}^{*}\times H_{1})\,\hat{n}d\varGamma
\]

Assume now we put a second current to the space. The field in the
space is $\xi=[E_{2},H_{2}]$, the superimposed current is $\rho_{2}=[J_{2},K_{2}]$
is
\[
\xi=\xi_{1}+\xi_{2}
\]
We have
\[
Q=(\xi,\xi)=(\xi_{1},\xi_{1})+(\xi_{1},\xi_{2})+(\xi_{2},\xi_{1})+(\xi_{2},\xi_{2})
\]

\[
=Q_{1}+Q_{12}+Q_{21}+Q_{2}
\]
In the above we find $Q_{1}$ is still same as there is no current
$\rho_{2}$. Hence $Q_{1}$ is no thing to do with $\rho_{2}=[J_{2},K_{2}]$.
$\rho_{2}=[J_{2},K_{2}]$ can be the whole environment materials,
hence this part of energy do not change after, $\rho_{2}=[J_{2},K_{2}]$
appear. Assume $con(Q)$ is the contributed energy current of $Q$
to the energy can be measured.

\[
con(Q)=con(Q_{1})+con(Q_{2})+con(Q_{12}+Q_{21})
\]
This part of energy current cannot be received, if we define the energy
current as energy current can be received, which is $con(Q)$ 

\[
con(Q_{1})=0
\]
Even $Q$ is not zero, but we only care the part of energy which can
be received by other material. This part of energy is 0 for $Q_{1}$.

In the same way, there is 

\[
con(Q_{2})=0
\]

Since $\xi$ is the retarded potential of $[J_{1}+J_{2},K_{1}+K_{2}${]},
the corresponding contribution of P-current should also vanish, i.e.,
\[
con(Q)=0
\]
Assume $\xi_{1}$ and $\xi_{2}$ are retarded potential, the above
discussion tell us that we have $con(Q)=0$, $con(Q_{1})=0$ and $con(Q_{2})=0$,
hence we have
\[
con(Q_{12}+Q_{21})=0
\]

\[
con(Q_{12}+Q_{21})
\]

\[
=con((\xi_{1},\xi_{2})+(\xi_{2},\xi_{1}))
\]
\[
=(\xi_{1},\xi_{2})+(\xi_{1},\xi_{2})^{*}
\]
\[
=2Re\{(\xi_{1},\xi_{2})\}
\]

\[
=2con(Q_{12})
\]
or

\[
con(Q_{12})=0
\]
This tell us the contribution of the mutual energy current of the
two retarded potential is 0. Similar this should be also the contribution
of the mutual energy current of the two advanced potential is 0.

\subsection{Only mutual energy between two object transfer energy}

\subsubsection{The electromagnetic field energy is transferred through only a special
kind of the mutual energy current, which is the current of one retarded
potential and one advanced potential.}

We have add a condition for above principle, the currents are separated
remotely. Since if they are too close their should be near fields,
we do not clear how the near field transfer energy. 
\[
(\xi_{1a},\xi_{2r})_{\varGamma_{1}}\neq0
\]
\[
(\xi_{1a},\xi_{2r})_{\varGamma_{2}}\neq0
\]
\[
(\xi_{1a},\xi_{2r})_{\varGamma_{3}}\neq0
\]
Here $\varGamma_{1}$,$\varGamma_{2}$ are the surfaces which enclosed
the current $\rho_{1}$ and $\rho_{2}$, $\varGamma_{3}$ is a plane
between $\rho_{1}$ and $\rho_{2}$. This principle is depended to
the above principle. If P-current can not transfer energy, the only
possibility of energy transfer is the mutual energy of a retarded
potential and a advanced potential. 

\subsection{The mutual energy of a retarded potential and an advanced potential.}

Assume $\rho_{1}$ is close to $\rho_{2}$ and close to the origin
of a infinite big sphere, the corresponding field $\xi_{1}$ is retarded
potential, $\xi_{2}$ is advanced potential. Here $\varGamma$ is
infinite big sphere surface.

\paragraph{
\[
Q_{12}=(\xi_{1a},\xi_{2r})_{\varGamma}=0
\]
}

means that the mutual energy can only send energy between the transmitter
to the receiver but not send any energy to the out space. We have
proved this in Fourier domain. But this is also very easy to understand
from time domain, the two retarded potential one is end to the future,
one is send to the past, this two energy current can not be synchronized
in the big sphere.

\section{important result to physics\label{sec:important-result-toVIII}}

\subsection{The advanced potential}

The advanced potential cannot be vanish for receive antenna like a
black body. We know this antenna has current $J,K$ which does not
vanish. We know black body do not send retarded potential out. Hence
if there is no advanced potential, the field of $J,K$ can only obtained
from other transmit antennas. This kind field can not satisfy Maxwell
equation. This means that we can not avoid the advanced potential.
The author assume there exist the advanced potential to the any current
$\rho=[J,K]$. The advanced potential is only depended to the current
$\rho=[J,K]$ and it cannot be composed with other retarded potential.
The advanced potential same as retarded potential always associated
to the current $\rho$, even this $\rho$ send also a retarded potential.

\subsection{The P-current}

If the energy current calculate through Poynting vector expresses
a energy current, P-current of the advanced potential will bring a
continuous energy to the current, even there is no any transmitter
send energy out. This is very strange. Hence the author assumed that
the P-current of the advanced potential do not carry any energy current.
To make consistent, the author also assume the P-current of the retarded
potential do not carry current energy. This assumption, also solved
the problem of so called 4-times difficulty and the difficulty by
thin wire antenna. Hence in free space there is no any receiver, a
current $\rho=[J,K]$ will produce a P-current for it's retarded potential
and a P-current for it's advanced potential. This current belong to
the current $\rho$ itself, it do not exchange any energy with other
current in the environment. The contribution to energy current of
the P-current is $0$.

\subsection{Mutual energy}

From above discussion we have known that the energy sent from transmit
antenna to the receive antenna is the mutual energy current or M-current.
If we calculate the field from the source, and calculate the energy
current send out from the source using Poynting vector, i.e. P-current,
which has no energy contribute to the receive antenna. In a few special
situation the P-current can equal to the summation of all M-current
between the transmitter and all the background environment, which
can be seen as infinite small receivers. But in general for example
in the case of thin wire antenna situation, the energy current can
only be calculated through the mutual energy current 
\begin{equation}
Q_{12}=(\xi_{1},\xi_{2})\label{eq:10000-10}
\end{equation}
where $\xi_{1}$ is retarded potential, and $\xi_{2}$ is the advanced
potential. 

\subsection{Advanced potential sucks the energy from the transmitter}

The energy received by receive antenna is equal to the energy sucked
by the advanced potential of the receive antenna on the transmit antenna.
The receive antenna play a important role in sending the energy out
from the transmit antenna. This can lead very important physics consequence
of the following. The sucked energy is transferred from the transmitter
to the receiver is by the M-current or mutual energy current. The
mutual current is a inner product of the advanced potential and the
retarded potential.

People think advanced potential can not transfer energy, actually
this is correct. Advanced wave itself really donot transfer energy.
But actually the retarded potential itself also donot transfer energy.
People thought retarded potential can transfer energy, this mistake
obstruct us to understand the advanced potential. The energy transfer
is by the inner product of the advanced potential and a retarded potential.
This inner product is just the mutual energy current of the retarded
potential and the advanced potential. This mutual energy current will
transfer the energy in the retarded meaning. That is the retarded
potential and advanced potential together created the retarded energy
current. Without the advanced potential, only with the retarded potential
can not transfer any energy! 

\subsection{Reflection}

Assume there is a mirror, the retarded potential of a source can be
received by the mirror, in the mirror a current is caused by this
retarded potential, the current will send an advanced potential to
the transmitter and also send a retarded potential to surround materials.
The advanced potential will cooperate with the retarded potential
of the source to transfer energy through the mutual energy to the
mirror. The retarded potential of the mirror will be sent to the materials
surrounded the mirror. If the materials received the retarded potential
and reacted it by sending an advanced potential back, the mutual energy
current of the retarded potential and advanced potential can be transferred
from the mirror to the materials. 

\subsection{The probability explanation of the quantum physics}

We know that there is probability explanation for quantum physics.
But why the light souse send the photo randomly according to the probability?
Einstein said God do not play dice. Many physicist do not agree the
probability explanation of quantum physics, included Einstein and
Schrödinger. 

The authors believe the reason of the probability is because of the
advanced potential and the mutual energy. A source send energy to
the space can not be done by the source itself. It need the advanced
potential of other receiver to suck the energy out from the source.
If this sucking process is discrete, all the material surround the
transmit antenna can be seen as many small receivers, which will randomly
sucks the energy from the source or the transmitter. Hence the source
can only send the photo randomly according to some probability.

This explanation can also avoid the concept of wave function collapsed.
There is no the wave function collapsed after a receiver received
a photon or a particle like electron. If the advanced potential belong
to a receiver, in this case perhaps it is a atom, in this situation
the photon send out from the the transmitter, it is clear it will
be received by which atom, that is the atom applied the advanced potential
to the transmitter. It is not the wave function collapsed to the atoms
(which can be seen as receivers). The photon is decided from the very
beginning which atom it should reach. This like the atom (playing
the role of the receiver) has stretched its hand to the transmitter
and taken a photo back to the receiver. Hence the photon knows where
to go. Hence the wave function collapsed can be avoid.

If electron wave transfer energy similar to photon by some kind of
mutual energy and advanced potential and retarded potentila, then,
the electron will also similar to photon, the eplantation for photon
can be applied to electron. In this situation, The concept about Schrödinger
wave function collapse can be avoid too.

If photo through their mutual energy cooperate the retarded potential
and advanced potential to transfer energy, perhaps other particle
do the same. In that case all problebility explanation is because
of the mutual energy.

This looks very strange if the distant between the receiver and the
sender is very big for example many light year. The photo knows which
atom is their destination. However if we accept the mutual energy
and the advanced potential we have to accept also this very strange
result.

\subsection{In the space there is no any background material}

In absorb theory\cite{Wheeler_1,Wheeler_2,Pegg}, it said current
send retarded P-current toward outside and received, and received
P-current toward to itself. Hence the current have no gain and loss
any energy. 

In the case there is no any background material, if the current $J,K$
situated on the center of space, it will send a retarded potential
and a advanced potential continuously. The P-current of the retarded
potential bring energy out from the transmitter and the P-current
of advanced potential received energy current from out space. The
current has no any pure energy loss, it can stay in space stably.

It is looks fine. But actually there is still problem, the advanced
potential is send to past, and the retarded potential is sent to the
future. If the current change is a short impulse. There should two
impulse current send to the future and to the past. How can we measure
this two impulse? 

We have successfully explain the energy transfer by a retarded potential
and advanced potential. Only with retarded potential or only with
advanced potential seems can only transfer energy mathematically and
but not physically. It is better do not accept energy can be transferred
with P-current. 

Since we have assume their is no any material surround the transmitter,
hence these two potentials both do not have any energy transferring
to other material. Their retarded potential and advanced potential
do not cause any energy toward the transmitter or toward the infinite
big sphere.

\subsection{Photon}

The authors have to thought about what is photon. First if the receiver
do not receive the retarded wave, is the receiver can send a advanced
potential? That is sames not possible. Hence the authors assume in
the time the retarded potential reached the receiver that caused the
current change in the receiver. However this current change send the
the advanced potential, because it is advanced potential it has the
ability to handshake with a past current in the transmitter, even
that is a current belong to many many years ago. That means the photo
is transferred between the two currents the sink and the source instantly.
The word instantly is not enough to discernible how fast it is, because
it use a minus time! It is better to create a word minus-time-instantly.
So we can say that the photo runs from the the receiver to the transmitter
minus-time-instantly. Here we do not speak about the photo runs from
the transmitter to the receiver, instead it runs from the receiver
to the transmitter. We have said that the receiver has the power to
grab the energy from the transmitter. That is only our imagination.
Actually, more accurate statement is the receiver direct send a photo
to the transmitter through the advanced potential. The advanced potential
is caused by the current in receiver, the current of receiver is caused
by the retarded potential, but since the advanced potential runs so
fast, minus-time-instantly, it still can combine with retarded potential
to be come the mutual energy current. We have known that the mutual
energy current is $(\xi_{1},\xi_{2})_{\varGamma}$, where $\xi_{1}$
is retarded potential, $\xi_{2}$ is advanced potential. This mutual
energy current is just the photo's energy current.

We know that the 1-D retarded potential is
\begin{equation}
\phi_{r}(x,t)=\intop_{-l}^{l}f(t-\frac{|x-x'|}{c})dx'\label{eq:10000-30}
\end{equation}
advanced potential is

\begin{equation}
\phi_{a}(x,t)=\intop_{-l}^{l}f(t+\frac{|x-x'|}{c})dx'\label{eq:10000-40}
\end{equation}
for simplicity, assume $l\rightarrow0$, $x'=0$

\begin{equation}
\phi_{r}(x,t)=\delta(t-\frac{|x|}{c})\label{eq:10000-50}
\end{equation}

\begin{equation}
\phi_{a}(x,t)=\delta(t+\frac{|x|}{c})\label{eq:10000-60}
\end{equation}
assume $x>0$,

\begin{equation}
\phi_{r}(x,t)=\delta(t-\frac{x}{c})\label{eq:10000-70}
\end{equation}

\begin{equation}
\phi_{a}(x,t)=\delta(t+\frac{x}{c})\label{eq:10000-80}
\end{equation}
or

\begin{equation}
\phi_{r}(x,t)=\delta(-\frac{1}{c}(x-ct))\label{eq:10000-90}
\end{equation}

\begin{equation}
\phi_{a}(x,t)=\delta(\frac{1}{c}(x+ct))\label{eq:10000-100}
\end{equation}
or, assume at $t=T$, the retarded potential reached the destination,
it become, assume in time of $t=0$ the retarded potential is sent
out, the shape is following,

\begin{equation}
\phi_{r}(x,t=0)=\delta(-\frac{1}{c}(x))\label{eq:10000-110}
\end{equation}
Assume in time $t$=T, the retarded wave reached to the receiver,
\begin{equation}
\phi_{r}(x,t=T)=\delta(-\frac{1}{c}(x-cT))\label{eq:10000-120}
\end{equation}
In this time the receiver send an advanced potential back to transmitter,
the shape of the wave in this time should same as the reached retarded
potential, i.e., 

\begin{equation}
\phi_{a}(x,t=0)=\delta(-\frac{1}{c}(x-cT))\label{eq:10000-130}
\end{equation}
Hence the advanced potential is 

\begin{equation}
\phi_{a}(x,t)=\delta(-\frac{1}{c}(x-cT+ct))\label{eq:10000-150}
\end{equation}
after a time $t=T$, the advanced potential travel to the following
place,

\begin{equation}
\phi_{a}(x,t=T)=\delta(-\frac{1}{c}(x-cT+cT))\label{eq:10000-160}
\end{equation}

\begin{equation}
=\delta(-\frac{1}{c}(x))\label{eq:10000-170}
\end{equation}
We can see the advanced potential after a time $t=T$ it has the same
shape with the the retarded potential. 

Since the time $t$ here is for the advanced potential, it is actually
means a time in the past, this means that even the advanced potential
is caused by the retarded potential, it still can synchronized with
the retarded potential.

This means the light wave and light particle are different concept.
The wave like property is because of the retarded potential. When
the retarded potential from the transmitter reaches the receiver,
there is still no energy particle send out, hence light is only waves.
This is reason why light can have wave interference pattern. Later
the current in the receiver begin to change, this change caused advanced
potential, it can also produced their retarded potential which is
scattering, here we do not like to talk about scattering. The receiver
send the advanced potential out, this advanced potential reached a
current in the transmitter of long time ago, and immediately grab
the photon from transmitter to the receiver. In other words the photo
runs from the receiver to transmitter minus-time-instantly. In this
time a photon come from the transmitter to the receiver. Hence the
photon process is always happens at the later time comparing the retarded
potential. 

Hence light wave and light particle are two concepts. Light wave is
still go as wave in the space, but photon is only appear when the
light wave reached the receiver. After this the current change on
the receiver caused an advanced potential, the advanced potential
together with retarded potential produced a mutual energy current,
this mutual energy current's energy is the photon.

Why energy is discrete? Advanced potential and retarded potential
has to be synchronous. If the energy particle is too big, the advanced
potential can become out of the phase with the retarded potential.
Hence the energy particle cannot too big. 

If the frequency of advanced potential has a little bit difference
with the retarded potential or they have a very small speed difference,
the two wave some where will cancel each other, some where will increase
their magnitude, this can make the energy discrete. 

Hence it is better the advanced potential has the very closed speed
to the retarded potential. In this case the minus-time-instantly handshake
between the receiver and the transmitter can happen. This is the photo.
The photo is nothing else but the mutual energy current. 

Energy particle (photon) is also possible caused by the change of
the current in the receiver which must discrete. That is the energy
particle only exist in the receive or transmit process. It is not
exist in the space. In the space it is still the waves: the mutual
energy of the retarded potential and advanced potential.

When we guess it is the Poynting vector transfer the energy, the photon
is very difficult to understand, because the Poynting current is a
diverge to the whole space, even for a light beam, the beam width
is always increase. Why latter the retarded wave can collapse to one
atom? Now we know the mutual energy transfer the energy, the beam
of mutual energy of a transmitter to a receiver is not always diverge,
it first diverge and then converge to the receiver or just a atom
(if it is the receiver). Hence the mutual energy beam do not need
to collapse to one point.

\subsection{Spontaneous emission}

From above we assume that the source is first send the retarded potential.
However that is possible the receiver first send the advanced potential
and that cause the source to have the spontaneous emission. Hence
the principle of the mutual energy supported the absorb theory\cite{Wheeler_1,Wheeler_2,Pegg},
which claim that the spontaneous emission is caused by the remote
environment atoms in the future. The mutual energy solution offers
a clear picture to this phenomena and tell us why this can happen.

\subsection{Stimulated radiation}

If the source produced a retarded potential $\xi_{1}$, that will
cause in the receiver has a current $\rho=[J,K]$, if this current
also produced a retarded potential $\xi_{2}$, the mutual energy of
these two retarded potential will produce a mutual energy current.
We have known that the mutual energy current cannot vanish in the
big sphere for the two retarded potentials. This also mean that there
will a mutual energy current send out to the free space. Corresponding
to the quantum theory this phenomenon is stimulate radiation.

From the principle of mutual energy, looks this stimulate radiation
is much different compare to the other above mentioned phenomenon
which need a transmitter and a receiver and a advanced potential and
a retarded potential. 

Two retarded potential produced the mutual energy should be also belong
to the P-current, which can not send energy current. It need advanced
potential to receive their energy. 

\subsection{Speed of light}

\subsubsection{One possibility}

From above discussion we have known that the energy current send from
the transmitter to the receiver is the mutual energy current. The
calculation of the mutual energy current needs the advanced potential
of the received antenna. 

The antenna system needs a advanced potential for receive antenna
means that the speed of electromagnetic wave is depended on the speed
of the advanced potential. The speed of advanced potential should
be calculated from the time-space coordinates of the receive antenna. 

Assume the the advanced potential has the speed of $v_{a}$ and the
retarded potential has the speed of $v_{r}$ the speed of the mutual
energy current should be some kind average of the two speeds, i.e.,
\[
c=average(v_{a},v_{r})
\]
Here $c$ is light speed. The speed of mutual energy current offers
us some method to guess what should be about the speed of light.

We have said that the advanced potential grab the energy from the
transmitter, it perhaps also leads the retarded potential, hence the
speed of retarded potential is the same to the advanced potential.
This means the advanced potential is possible to be the ether of the
retarded potential, hence the retarded potential has the same speed
to the advanced potential, in this case we obtain,

\[
c=average(v_{a},v_{a})=v_{a}
\]
Hence the speed of light $c$ is the same with that of the advanced
potential. This will grantee the speed of light is a constant to the
receiver. 

We have also said the condition of the mutual energy current can transfer
energy is that the retarded potential must synchronized with advanced
potential. Hence the retarded potential must have same speed with
the advanced potential. That is 
\[
v_{r}=v_{a}
\]
Hence the retarded light speed must equal to the advanced potential.
Hence the light speed is same with the receiver instead of the transmitter.

\subsubsection{Second possibility}

In the last sub-section we have said there is handshake between the
receiver and the transmitter, this also ask that the retarded potential
perhaps transferring in the same speed of the advanced potential.

However even if the retarded potential has different speed with advanced
wave, the handshake process is only depended to the advanced potential.
Hence we can allow that the retarded potential has the speed different
with the advanced potential. But the light speed is depend on the
advanced potential only, because the handshake process is controlled
by the advanced potential. 

Even the photo is instantly from transmitter go to the receiver, but
because actually it is a handshake between a receiver current and
a current of transmitter in perhaps many years ago, hence we feel
that the photo needs a time to go from the transmitter to the receiver.
The speed of this photon is the speed of advanced potential. This
is way the light has the speed same as the advanced potential. 

We can say that the speed of advanced potential is the speed of grab
process or handshake speed which is also the photo's speed.

This way we have allowed the retarded potential has a different speed
as the advanced potential. Assume the transmitter move toward the
receiver in a speed $v_{t}$, assume the speed of retarded potential
have a speed $c+v_{t}$ toward the receiver, assume the transmitter
send a short impulse, when the retarded potential reached the receiver,
the receiver send its advanced potential, which runs (with a reversed
time) with a speed of $c$ to the transmitter. In that time, the transmitter
have move a distance from the original place, the short impulse has
gone since the impulse has very short time. The handshake can not
happen. No energy can transfer from transmitter to the receiver. This
is not what we have seen.

This become very complicate if the retard potential has different
speed with the advanced potential. It is better that we still assume
the retarded potential has the same speed with the advanced potential.

Assume the retarded potential has different speed in the beginning,
but when the advanced potential is returned from the receiver to transmitter,
the retarded potential has to synchronized to the advanced potential,
it has to be adjust to the same speed of the advanced potential. 

\subsubsection{third possibility}

A difference of the mutual energy theory with special relativity is
that the speed of light. According to the special relativity that
the speed of light is same at any inertial frame. According to the
mutual energy theorem, the speed of light is same to the receiver.
It is the receiver send the advanced potential which is same as ether,
the retarded potential must synchronized to the advanced potential,
hence even in the beginning their speed can be any thing for example
dependent to the speed of it's source, but after the receiver send
back any advanced potential, the retarded potential must take the
same speed as the advanced potential. If it's speed is different to
the advanced potential, it can not synchronized and hence can not
transfer any energy. We can assume the transmitter can send the retarded
potential in a speed a little bit differ with light speed $c$ and
only the retarded potential with same speed of the corresponding advanced
potential survived. 

Hence there is the difference between the special relativity theorem
and the point of view of the mutual energy theorem. According to the
authors point view we can build a system receiver the signal fast
than the speed of light. For example, assume there is a transmitter
which send the light signal. Assume I sit down at another place with
a receiver, if i receive the signal, it's speed is c, that is clear.
Assume I am sit down on big wheel, the wheel is rotated around me.
In the rim of wheel there are detect which can receive the light signal.
Hence this detector has different speed to me. The light received
by this detector have also different speed. Even in my inertial frame,
the light speed is different depending which detector receive it.
Some will have speed larger than light speed $c$ and some will have
the speed smaller than the light speed c. Assume a detector received
the signal speed fast than light, then it send me their signal, this
do not caused any time. Hence this way we should get the signal a
lit earlier than that received by myself. According special relativity
all speed of light even it belong to different receiver. Hence detector
at the rim of wheel cannot offer more fast to get the signal or less
fast to get the the signal. They all get the same speed of signal.

According to mutual energy current(it is referred as M-current) theory
the all light particle have the speed depending their receiver.

\subsubsection{Forth possibility}

Since the energy transferred by together of advanced potential and
the retarded potential. Only when the two potential synchronized it
can transferring energy. Hence the retarded potential must transfer
in the same steed as the advanced potential. If all the receiver has
the different speeds, the energy sucked in the source should has different
speed corresponding to different receiver, hence the light speed is
different corresponding the receiver. The advanced potential of receiver
be come the ether of the light, any retarded potential with different
speed with the advanced potential cannot synchronized and hence can
not carry the energy. The speeds of photon in the transmitter is different
depend their receiver.

Perhaps in the beginning the retarded potential has the same speed
with the light speed in the transmitter's coordinates. But after the
receiver's current changed, it send the advanced potential, the retarded
potential immediately correct their speed.

\subsection{Who is first, the retarded potential or the advanced potential}

There is two possibility, first the transmitter has a current change,
it cause a retarded potential, this retarded potential cause a current
change in the remote receiver, the receive send a advanced potential
back. In this case the retarded action is first, and the advanced
reaction in the latter. 

But in other situation the receiver has a current change it send advanced
potential, this retarded potential reached the transmitter, the transmitter
send a retarded potential back. Hence first have an advanced action,
and then a retarded reaction,

The mutual energy formula do not forbid any of them, my feeling is
the first has a little bit better for causality. That means the action
is retarded is easy to understand. But from the absorb theory\cite{Wheeler_1,Wheeler_2,Pegg},
the atom spontaneous decay is caused by the remote environment of
future. This means the advanced action is also possible. 

Any current should allowed to have a retarded potential and a advanced
potential associated to it. Hence if there is any current change,
it will produced two actions, one is retarded action direct to the
future and an advanced potential directed to the past. That means
the environment should be possible to receive advanced potential fir
also. 

\subsection{Superluminal signal}

Assume we have a wave guide, there is a transmitter in the left end
and a receiver in the right end. The energy transferred from the transmitter
to the receiver is because the mutual energy of the retarded potential
and the advanced potential. If the voltage of the transmitter changed,
the current in the transmitter is changed, the retarded potential
is changed first, and this cause the the current of receiver change,
this change in turn cause an advanced potential. The advanced potential
cooperate with the retarded potential to become the mutual energy
current, the mutual energy current send the energy from transmitter
to the receiver in the retarded meaning. That means current in the
transmitter take place first. The current in the receiver change later. 

However if the current in the receiver is changed first, the advanced
potential is sent from the receiver to the transmitter first. The
transmitter current will change before the current change of in the
receiver. This change in turn produced a retarded potential, the retarded
potential cooperated with the advanced potential to become the mutual
energy current, sent form the receiver to the transmitter. The mutual
energy current is send from the transmitter to the receiver in the
retarded meaning.

However the signal change is taken place at receiver first, but the
change in the transmitter will take place at a time before the receiver
change the current.

The current on the receiver can be changed by change their the resistance
or impedance impedance of the receiver.

Their will be perhaps also a retarded signal have been return to the
transmitter. We can find method absorb that return retarded wave.
Hence this system only receive wave, if the resistance/impedance of
receiver is changed, the transmitter must have a change before the
change happens in the receiver. Hence the signal transmitted from
the receiver with time t= -T.

In the space we can also by change the current in the receiver to
send a signal to the transmitter. This signal will reach the transmitter
in -T time. 

The experiment also can be done through two antenna located not to
far away. Assume they the receive antenna is ideal, i.e., that it
do not scatter energy. Assume their is no energy loss, that the energy
send from the transmitter antenna can be received by the receiver
antenna hundred percent. 

Now we transfer some wave for example microwave between them, every
thing is fine suddenly we change the load resistance of the the receiver
in the one end of the wave guide, that cause the current in the receiver
is changed, this change will send an advanced potential to the transmitter,
transmitter also adjust their current. The current change in the transmitter
will happen before the resistance change in the receiver.

If we think the above is wrong, the receiver cannot send the advanced
potential, we have known that this is ideal receiver antenna it can
also not send the retarded potential out. Hence the transmitter cannot
fell there is a load change in receive antenna, it will still send
same energy to the receiver, but the receiver cannot absorb this energy.
This is not possible. Hence if we assume the load in the receiver
can affect the transmitter, it should send advanced potential to tell
the transmitter its change! In this way we can transfer a sign with
minus time. This is superluminal signal.

\subsection{Causality}

From the above discussion we found that even the advanced potential
is runs backward or reverse with time, the photo, I mean the M-current,
is still runs from past to the future. 

However we can transfer the signal and energy from present to the
past to a remote object, if we use the above superluminal method.
If we make a circle wave guide the begin and end is in the same place,
the signal can send to the same place but in the past. Hence the causality
is possible to be violate.

\subsection{Spin and polarization}

First the polarization, we have been taught that the circle polarization
is combined by two fields and between the two fields there is a 90
degree delay in phase. If photon is only related to one retarded potential
it is very difficult to think about why in side the photo there have
two fields, and the two fields have phase delay 90 degree. Hence we
can only say that the photon has this spin property. This property
is intrinsic. But this actually did not tell us any thing.

According to the authors M-current theory there are two field, $\xi_{1}$
and $\xi_{2}$, one is retarded and one is advanced, but the two fields
have been nearly synchronized in phase. If the the transmitter send
retarded potential $\xi_{1}=[E_{1},H_{1}]$, wen the receiver received
the field $\xi_{1}$, it send a advanced potential, $\zeta_{2}=[E_{2},H_{2}]$,
if this two potentials synchronized completely, it be come a line
polarization. If the receiver/transmitter send a advanced wave have
90 degree delay, then it become a circle polarization. 

If the receiver first send advanced potential $\xi_{2}$, when the
transmitter received this potential, it send retarded potential $\xi_{1}$,
$\xi_{1}$ will have 90 degree delay in phase. If the transmitter
first send $\xi_{1}$ as retarded potential, when the receiver received
it, and send $\xi_{2}$ as advanced potential, $\xi_{2}$ will have
a 90 degree delay compare to $\xi_{1}$.

We can assume the above transmitter and receiver are the two atoms,
a emission atom and an absorb atom. Hence the photon or M-current
has circle polarization is because the above described re-sending
process of the potentials at two atoms. The existence of left and
right polarization tell us, both receiver and transmitter can be the
first one send the potentials.

Hence M-current theory can offer us a very good explanation about
circle polarization of light energy. This explanation is much nature
than tell us there is a spin for photon.

\subsection{Why is this special atom react to the retarded potential}

A transmitter send a retarded potential to the environment, there
is infinite atoms received this potential, why that special atom send
back a advanced potential not others? We can assume the atom in the
receiver can be awake only in a very short time to receive energy
from the retarded potential, for example $10^{-10}$second. Other
time it just sleep do not react to the retarded potential. The time
in which a atom can receive energy become a random variable. In special
time only a few atom can receive energy. This short time window is
also the reason why the photo is also a short time energy particle. 

Transferring through the mutual energy with two potential retarded
and advanced can make the the energy focused to a special atom, The
above time window make the energy also focused to a short time. This
too functions create a photon particle with a local place in a short
time. That amount energy can be seen as photon. Hence photon is not
a particle in space, it is only a particle in the time it is created
and it is received. In other time it is retarded potential and advanced
potential. 

\subsection{Action with a remote object with 0 time }

We have known the current change in the transmitter cause a retarded
potential, the retarded potential applied an action to the receiver.
The receiver give a reaction back. This reaction perhaps is the photon,
the receiver sent to the transmitter minus-time-instantly. What is
the time the reaction applied to the transmitter? Since the the retarded
potential transfer from transmitter need some time. The advanced potential
apply the reaction to transmitter need the time which is negative.
Together the time is just zero. 

This means for transmitter it can fell the object in any remote distance
with a 0 time. This means the receiver can apply a reaction to the
transmitter immediately with no time. The transmitter can feel a force
from the receiver, the reaction is immediately. 

If the gravitational field has the similar functions like electromagnetic
field, also have the advanced potential, the force reaction between
stars are also with 0 time. It is seems it is correct that we normally
believe there is no need of time to apply a gravitational force from
a star to another. If it is true that means the gravitation should
also have the advanced potential for the gravitational waves. And
the force reaction can be with 0 time between far away stars.

The concept of mutual energy and mutual energy theorem, the principle
of mutual energy support many results of absorb theory\cite{Wheeler_1,Wheeler_2,Pegg},
but give more details what the mutual energy can be calculated. How
the energy is transferred from atoms. And also give the picture why
only retarded potential alone and advanced potential alone do not
work, but their combination works.

The concept the energy is transferred through the mutual energy of
a retarded potential of transmitter and advanced potential of the
receiver allow us to obtain the action and reaction only take 0 time
means that the electromagnetic force can be done with 0 time. If the
gravitation force has same principle with electromagnetic field, it
should also can action with remote object with 0 times.

\subsection{Mass}

We have known there is two kind of mass $m_{a}$, the mass because
of acceleration and the mass of the gravitation $m_{g}$, experiment
shows this two mass are equivalent, that means 
\begin{equation}
m_{a}\propto m_{g}\label{eq:10000-20}
\end{equation}
From above discussion we have known that the object can apply a electromagnetic
action to the remote environment, and this action with remote environment
do not need time. This means it can linked with the environment with
this fast action. The environment can apply a reaction to the object
with 0 time. Hence if we try to accelerate a object, we will fell
a force to us. This force is the reaction of far away environment.

If gravitation is also cased by action and reaction between two objects.
The action and the reaction transferred perhaps same with a electromagnetic
field, mutual energy of the retarded potential and advanced potential.
In this case the two mass can be equivalent to each other, i.e., there
is Eq.(\ref{eq:10000-20}).

\subsection{Quantum entanglement}

When we talk about quantum entanglement, we often said that there
is no any method to link the two electrons, to let one tell the other
what is their state of spin. Know at least we know there is superluminal
signal can connect the two electrons. Even we do not know exactly
what happens for entanglement but their signal perhaps is the 0 time
signal like the force reaction or superluminal signal of electromagnetic
signal.

\section{New explanation of the wave function for the quantum physics}

\subsection{Electron}

Electron should also had its advanced potential. Electron should also
some kind of mutual energy theorem for electron, that means there
perhaps is another kind of more accurate wave function than current
Schrödinger or Dirac equations, so that the electron also have the
retarded potential and the advanced potential, this make the electron
transmitter can communicate with another electron receiver. The electron
transmitter an atom. The electron receiver is another atom. The electron
itself is wave like light wave. the electron's mutual energy beam
should similar to the mutual energy of the light beam, it is very
narrow in both transmitter point and at the receiver point. The electron
M-current beam should be very wide in the place between the transmitter
and the receiver.

Time window of the electron receiver or transmitter can react should
be very narrow. Hence there is only one receiver to react the retarded
potential send out from the transmitter. This particular transmitter
atom and the particular receiver atom have a handshake or marriage.
All energy of a electron is send from this particular transmitter
atom to another particular receiver atom. 

\subsection{Wave function}

In the quantum physics, assume $\psi$ is the wave function, then
$|\psi|^{2}$ is explained as probability. However the authors thought
that is because the lack of the knowledge of M-current for light.
If 90 years ago Schrödinger and Dirac knew the above theory about
light they will build their quantum theory looks like light. Here
the author means the theory which explain light as M-current. The
M-current is composed of retarded potential and an advanced potential.

After we have the new understand above about light, we know that,
the situation of quantum physics should be similar to electromagnetic
wave or light wave. If we accept the advanced potential in electromagnetic
field and light, if we accept the light is just the mutual energy
current of the two potentials, one is retarded potential, another
is advanced potential, we can immediately thought in the quantum physics
perhaps there is also the advanced potential. Assume $\psi_{1},\psi_{2}$
are two potentials, we can define the the M-current (mutual energy
current) of quantum physics as

\[
Q_{12}=(\psi_{1},\psi_{2})=\psi_{1}\text{\ensuremath{\psi_{2}^{*}}}
\]
It is possible that $\psi_{1},\psi_{2}$ are vectors like in the electromagnetic
field situation, in that situation $(\psi_{1},\psi_{2})$ will define
the mutual energy current. In quantum physics $\psi_{1},\psi_{2}$
are scales (the scale is possible only a simplified version of vector
field, just light in optics we can use scale value to describe the
electromagnetic field, which actually is a vector field). In case
$\psi_{1},\psi_{2}$ are scales, $(\psi_{1},\psi_{2})=\psi_{1}\text{\ensuremath{\psi_{2}^{*}}}$.

Assume in quantum physics, $\psi_{1}$ is retarded potential which
send out from a transmitter atom, $\psi_{2}$ is advanced potential
which is send out from a receiver atom. 

When the electron is in side the orbit of the atom, the advanced potential
can be completely synchronized with the retarded potential. In this
case $\psi_{2}=\psi_{1}$, Assume the field of electron can be superposed

\[
\psi=\psi_{1}+\psi_{2}
\]
$\psi$ is the electron field inside the orbit. Hence 

\[
\psi_{1}=\psi_{2}=\frac{1}{2}\psi
\]

\[
(\psi,\psi)=(\psi_{1}+\psi_{2})(\psi_{1}+\psi_{2})^{*}
\]

\[
=\psi_{1}\psi_{1}^{*}+\psi_{1}\psi_{2}^{*}+\psi_{2}\psi_{1}^{*}+\psi_{2}\psi_{2}^{*}
\]

$\psi_{1}\psi_{1}^{*}$ is the retarded potential's self-energy current. 

$\psi_{2}\psi_{2}^{*}$ is the advanced potential's self-energy current.

$\psi_{1}\psi_{2}^{*}+\psi_{2}\psi_{1}^{*}$ is the mutual energy
current.

In this situation since $\psi_{1}=\text{\ensuremath{\psi_{2}}}$,
the calculation only with retarded potential, ans assume that is $\psi$
will obtain the same result. This is why if we do not introduce the
concept of advanced potential and the M-current, quantum physics is
still obtain corrected results in the situation of electron inside
orbit. 

\subsection{Election in the free space}

The self-energy current $\psi_{1}\psi_{1}^{*}$, $\psi_{2}\psi_{2}^{*}$
have no contribution to the transmitter atom and the receiver atom,
this is similar to the situation of light. $\psi_{1}\psi_{1}^{*}$
is a beam diverged from the transmitter, when it reach the receiver
atom, since the the section area of the receiver atom is too small,
the receiver atom received energy from the $\psi_{1}\psi_{1}^{*}$
can be omitted. $\psi_{2}\psi_{2}^{*}$ is diverged from receiver,
when it reached to the transmitter atom, since the section area of
the transmitter is too small, this part of transferred energy can
be omitted. In this situation only the mutual energy is important,
which is

\[
\psi_{1}\psi_{2}^{*}+\psi_{2}\psi_{1}^{*}=2Re\{\psi_{1}\psi_{2}^{*}\}
\]

The mutual energy current similar to the situation of light the beam
is that the electron beam first diverged from the transmitter and
then converged to the receiver. Here the transmitter and the receiver
are two atoms which can send or absorb the electrons. Here since the
beam of M-current can focus to a small point, it does not need the
concept of wave function collapse. 

The wave function collapse is because we do not know there is also
the advanced potential. So We calculate $\psi_{1}\psi_{1}^{*}$ which
is a diverged beam. At the place of the wave is received, the beam
of the energy current $\psi_{1}\psi_{1}^{*}$ become very widely spread
out. When we $\psi_{1}\psi_{1}^{*}$ is the result of quantum physics,
we have to face the wave function suddenly collapsed to a point. After
we explain electron as M-current, the property of first diverge and
converge can thoroughly avoid the wave function collapse. 

The probability explanation for the wave function is only because
of we only calculated the retarded potential, $\psi_{1}\psi_{1}^{*}$.

The author don't clear why this particular transmitter atom married
to another particular receiver atom. We have said it is because perhaps
because just in the time the retarded potential reached the receiver
atom, their time window matched together. But this is only one possibility,
that is also possible the transmitter send retarded potential includes
a special cryptograph code, which can be understand only one receiver
atom. It is also has some positive feedback between the transmitter
and the receiver that makes the connection of one pairs of atoms become
strong than others. Finally they become marry together. The electron
is send out from the transmitter atom to the receiver atom.

\subsection{Spin}

In the traditional quantum physics, there is only one wave function,
when we measured some thing rotated, it is difficult to give an explanation,
hence we call it spin. However in the authors' new quantum explanation,
there are two wave functions, one $\psi_{1}$ is retarded and the
other $\psi_{2}$ is advanced. The two waves are nearly synchronized.
But there is allow they have small phase difference. The spin is also
similar to the situation of light. If we assume $\psi_{2}$ has 90
degree phase difference compare to $\psi_{1}$, there is a circle
polarization. Here we assume $\psi_{1}$ and $\psi_{2}$ are transverse
field. This circle polarization is the so called spin. 

In the explanation of the mutual energy current, the spin just two
waves have a 90 degree phase difference. This phase difference is
caused by the receiver atom or the transmitter atom there is a reaction
delay to their wave re-sending process.

\subsection{The Schrödinger equation considered the advanced potential}

The original Schrödinger equation which is corresponding to the retarded
potential

\[
ih\partial\psi(x,t)=[-\frac{h^{2}}{2\mu}\nabla+V(x,t)]\psi(x,t)
\]
Corresponding to the advanced potential, there is

\[
-ih\partial\psi(x,t)=[-\frac{h^{2}}{2\mu}\nabla+V(x,t)]\psi(x,t)
\]
The above is only one example to create the advanced potential, we
also can created the advanced potential using Klein-Gordon equation
or Dirac equation, or any other equation still not found, but that
is beyond the discussion of here.

\subsection{Summary}

For a free electron, we should calculate M-current which $\psi_{1}\psi_{2}^{*}$.
M-current is a beam first diverge and then converge, for this kind
wave beam, the concept of wave function collapse can be avoid. 

When the electron inside orbit, the two potential can synchronized
completely, and hence the retarded potential and advanced potential
is equal to each other. In this situation, the calculation of $\psi_{1}\psi_{1}^{*}$,
the P-current (we can referred it as P-current similar to the light
situation) is same as M-current $\psi_{1}\psi_{2}^{*}$. Even there
still exists the advanced potential, but the result is same when we
only use retarded potential to calculate, $\psi_{1}\psi_{1}^{*}=\psi_{1}\psi_{2}^{*}$.
It is same to the wave guide, in the orbit, the energy transferred
half by P-current $\psi_{1}\psi_{1}^{*}+\psi_{2}\psi_{2}^{*}$ and
half by M-current $\psi_{1}\psi_{2}^{*}+\psi_{2}\psi_{1}^{*}$ 

I the free space the contribution of P-current can be omitted completely.
Only M-current is left. Hence electron is also M-current, which is
composed of two waves retarded potential and advanced potential.

Even the above new explanation has not change the calculation of the
quantum field. However because it abandons the concept of the wave
function collapse, the probability, thing become easy to understand.
Electron in the free space is nothing else, it is just M-current.
The M-current is composed of two waves one is retarded the other is
advanced. The two wave are nearly synchronized. There is 90 degree
phase difference which can be seen as the behind scene of spin. 

In the authors new explanation the square absolute value of wave function
$\psi_{1}\psi_{1}^{*}$ is the P-current which is only a approximation
to the mutual energy $\psi_{1}\psi_{2}^{*}$. Since $\psi_{2}^{*}$
is difficult to obtain, there is hundreds and thousands $\psi_{2}$
in the environment corresponding to each atom which can receive the
electron, we still only calculate $\psi_{1}$. In this situation we
have to use the probability to explain the result $\psi_{1}\psi_{2}^{*}$. 

\section{The problem is still not solved}

Since this new theory, there are lot of thing need to do. I list some
of that as following,

\subsection{Why action and reaction is is same?}

Newton has the third law, that the action force is same to the reaction
force. 

We can assume the force is caused by a action try to accelerate the
object, when a object accelerate it send a retarded potential to the
remote environment, the remote environment send an advanced potential
back to the object. this become the reaction, the reaction happens
with $0$ time because the advanced potential travel minus-time-instantly.
But why this reaction is same to the action?

\subsection{Planck constant?}

Why has this Planck constant? Is this constant can be explained with
the concept of the mutual energy?

\subsection{In the loss media the mutual energy theorem is not established, what
should be the mutual energy theorem in that special situation?}

If the advanced potential has different transfer character, for example
the media for the advanced potential is not the same as with retarded
potential, how energy will transfer? For example if we let the advanced
potential go a way different with the retarded potential and then
put it back to the same way as before, is it possible to transfer
energy?

If the energy is transferred by M-current when the retarded potential
and the advanced potential is separated, is the energy can be transferred?
If it can still transfer the energy. Dose this mutual energy current
still has the meaning?

We know the fields can still transfer, hence in this situation should
be still possible to transfer the energy, then what is wrong with
the mutual energy current?

If the advanced potential can be separated with retarded potential
and the energy still can be transferred, that means the mutual energy
is also not the one transferring energy, the energy is transferred
only with the retarded potential and the advanced potential.

This problem I still did clear.

\subsection{What about polarization to the advanced wave?}

Is the polarization property belong to retarded potential or advanced
potential or both? If it is the property of the retarded potential,
is the advanced potential has the same property of polarization? 

\subsection{Optics to advanced wave}

For example what is the Huygens principle to the advanced potential? 

\subsection{The theory for the whole microwave network}

If the theory about mutual energy and advanced potential is correct,
the whole electric network theory need to be updated. For example
only for coaxial cable, originally we can tell students that the transferred
energy can be calculated using Poynting vector, which can be calculated
with only retarded potential. Now we must tell that is wrong, actually
the energy is transferred by the retarded wave and advanced wave together,
it is the mutual energy transfer the energy. But in this situation
the result calculated by the mutual energy is same as the results
calculated the energy current with Poynting vector.

\section{Conclusion\label{sec:Conclusion_IX}}

We have shown that the traditional way to explain the system with
a transmit antenna and a receive antenna using the reciprocity theorem
is inadequate, which is only correct to the system with two transmit
antennas. Instead it is need a new explanation. In the new explanation
the mutual energy and advanced potential is involved. 

We have shown that the energy current related to the Poynting vector
does not transfer any energy. It is the mutual energy current that
is responsible to transfer the energy from the transmitter to the
receiver.

We have also shown that the advanced potential can suck energy from
the transmitter. This sucked energy is just equal to the energy received
by the receiver.

The process of advanced potential sucking the energy should be also
discrete. This discrete energy is just the photon. 

The background material can be seen as countless receivers, these
receivers will randomly apply their advanced potential to the transmitter,
hence the transmitter randomly send their energy or photon out. This
is the reason that the quantum mechanics result has to be explained
with a probability. We also discussion the reason of speed of the
light. We also discuss how to transfer a superluminal signal.

\part{Photon model in Time domain}

... It will be added later.

\part{The Principle of Mutual energy}

\section{The problem of the superimposition of the field}

\subsection{History of field theory}

In the time Newton, there is action and reaction. action and the reaction
is equal. This is Newton's third law. This is no any problem, for
newton because that time the action can be seen with infinite fast
speed. But today we know the action of electromagnetic field have
the speed $c$ it is not infinity any more. Hence the action at distance
change to two viewpoint. (1) the emitter send action to the field,
the field propagate to the absorber, then the absorber receive the
action from the absorber. (2) the action is still same as newton said,
action is still equal's the reaction. The action is only take place
between the charges. 

The first view of point (1) is very successful in the history of classical
electromagnetic field theory. Later become the Maxwell equation and
theory. This we all clear. The second view of point (2) is not well
know, it is try to solve the problem of the first one. We all know
the Maxwell electromagnetic field theory has some problem which lead
the self action infinity. We can not calculate a energy of a charge
in a point. That create a infinity. Quantum physics has to do the
process of re-normalization to eliminate the infinity. In quantum
physics Maxwell theory tell us the wave have send to all direction
through Poynting vector, however we always receive a energy package
as photon, hence we have to speak about wave collapse. However no
one tell us what equation satisfy by this collapse process. Schwarzchild
first offers the results for the second view of point. His theory
has been referred as direct interaction theory. Tetrode and Fokker
further developed this theory and offers the same principle but suitable
to relative theory. In this direct interaction theory, the action
is still only between the two charges. There is no field exist without
any one of the charge.

The second view of point continually developed by Dirac and later
Wheeler and Feynman\cite{Wheeler_1} as absorber theory and adjunct
field theory. The adjunct field that means it is only adjunct to the
action of the two particle, it is not a independent field. Hence the
real calculation still need to use Maxwell equation. Wheeler and Feynman
tell us

``(1) There is no such concept as ``the'' field, an independent
entity with degrees of freedom of its own.''

``(2) There is no action of an elementary charge upon itself and
consequently no problem of an infinity in the energy of the electromagnetic
field.''

``(3) The symmetry between past and future in the prescription for
the fields is not a mere logical possibility, as in the usual theory,
but postulation of requirement''

When I read the above, my heart spring out. When I now still blind
try to solve the self-action problem, our forefathers has loon clear
tell us the results. Wheeler and Feynman claim they can derive the
Maxwell equation from this adjunct field theory. Actually they derive
the Maxwell equation for a charge.

Wheeler and Faynman offers the the principle what should to solve
the problem but not really gives a new electromagnetic theory which
can replace the Maxwell equation or correct the Maxwell equations.
This will be our task.

\subsection{Poynting theorem is equivalent to Maxwell equation in principle }

We can derive the Poynting theorem from Maxwell equation. We do not
know how to derive Maxwell equation from Poynting theorem. However
we can derive all reciprocity theorem (also the mutual energy theorem)
from Poynting theorem{[}XXXX{]}. We also know the Green function theory
can be derived from reciprocity theorem. Maxwell equation can be solved
through green functions. From all solutions of Maxwell equations,
the Maxwell equation can derived by induction. Our proof is not very
stringent. But physics is not mathematics, this kind of proof is good
enough. 

In the following if we cannot easy found some thing wrong in Maxwell
equation, we can try to stack the Poynting theorem. If Poynting theorem
can be corrected, in the same wave we actually correct the Maxwell's
theory.

\subsection{Is field can be superimposed?}

There two version of superimposition of the field, for example there
are $N$ charges. The first one is the all contribution of the chargesa
\begin{equation}
\overrightarrow{E}(x)=\sum_{j=1}^{N}\overrightarrow{E}(x_{j},x)\label{eq:P3-1-10}
\end{equation}
This is our traditional way define the field. Maxwell follows this
way. However this definition is only suitable to the situation the
field is calculated not on the position of charges.

\begin{equation}
\overrightarrow{E}(x_{i})=\sum_{j=1,j\neq i}^{N}\overrightarrow{E}(x_{j},x_{i})+\overrightarrow{E}(x_{i},x_{i})\label{eq:P3-1-20}
\end{equation}
but 

\begin{equation}
\lim\overrightarrow{E}(x_{i},x_{i})=\infty\label{eq:P3-1-30}
\end{equation}
Hence we change to the following definition 
\begin{equation}
\overrightarrow{E}(x)=\begin{cases}
\sum_{j=1}^{N}\overrightarrow{E}(x_{j},x) & x\notin I\\
\sum_{j=1,j\neq i}^{N}\overrightarrow{E}_{j} & x\in I
\end{cases}\label{eq:P3-1-40}
\end{equation}
$I=1,\cdots i\cdots,$ it is the sets of position of the charges.
The above section definition is also not satisfy. Many will ask is
this correct that the field is extended to the any position without
a charge?

According to the adjunct field theory of Wheeler and Feynman, that
the field can only defined on the position of charges, because the
force is only defined on the position of the charges.

\begin{equation}
\overrightarrow{F}(x_{i})=\sum_{j=1,j\neq i}^{N}\overrightarrow{F}(x_{j},x_{i})\ \ \ \ \ x\in I\label{eq:P3-1-50}
\end{equation}
Hence

\[
\overrightarrow{E}(x_{i})=\frac{1}{q_{i}}\overrightarrow{F}(x_{i})\ \ \ \ \ x\in I
\]
\[
=\sum_{j=1,j\neq i}^{N}\frac{1}{q_{i}}\overrightarrow{F}(x_{j},x_{i})\ \ \ \ \ x\in I
\]
\begin{equation}
=\sum_{j=1,j\neq i}^{N}\overrightarrow{E}(x_{j},x_{i})\ \ \ \ \ x\in I\label{eq:P3-1-60}
\end{equation}

Since the force only defined at the position charge, the field should
also only defined at the position of the charge. 

We believe the theory of Wheeler and Feynman is correct, because if
in the empty space there are only two charges, if the action of the
first can produce a field and action with the field, send the energy
to the field, the most energy will go outside the system of this two
charges. The second charge can only get a very small part of energy
from the first charge. The energy is not conserved for this two particle
system. 

But if the field cannot be defined on the space other than the position
of charges, it become not very useful. Here we at least has 3 different
field definitions. Hence field is a very confused concepts.

\subsection{Energy or Power}

From last subsection we know the field is very confused concept, if
we cannot correctly define the field, let us see if we still can apply
the concept of energy or power.

If the charge move and has the speed $\overrightarrow{v}_{i}$, we
can define the Power which is

\begin{equation}
P(x_{i})=\overrightarrow{F}(x_{i})\cdot\overrightarrow{\overrightarrow{v}}_{i}\label{eq:P3-1-70}
\end{equation}

We believe this still correct. Hence the power of the whole system
with $N$ charges will be,

\begin{equation}
P=\sum_{i=1}^{N}P_{i}=\sum_{i=1}^{N}q_{i}\overrightarrow{E}(x_{i})\cdot\overrightarrow{v}_{i}=\sum_{i=1}^{N}\sum_{j=1,j\neq i}^{N}\overrightarrow{E}(x_{j},x_{i})\cdot(q_{i}\overrightarrow{v}_{i})\label{eq:P3-1-80}
\end{equation}
We can write

\begin{equation}
\overrightarrow{J}_{i}=q_{i}\overrightarrow{v}_{i}\label{eq:P3-1-90}
\end{equation}
where $\overrightarrow{J}_{i}$ is the current of charge $q_{i}$.
Hence we have the Power of the whole system are
\begin{equation}
P=\sum_{i=1}^{N}\ \sum_{j=1,j\neq i}^{N}\overrightarrow{E}(x_{j},x_{i})\cdot\overrightarrow{J}_{i}\label{eq:P3-1-100}
\end{equation}
We find when we calculate Power, we have used the following summation.
\begin{equation}
\sum_{i=1}^{N}\ \sum_{j=1,j\neq i}^{N}\label{eq:P3-1-110}
\end{equation}
The above power calculation is no disputed, why we do not started
from this to redefine what is field? In the above we have offers 3
version of field definition no one is satisfied all situations.

For the whole system
\begin{equation}
P=Const\label{eq:P3-1-120}
\end{equation}
The above is energy conserved law, the whole system energy 

Hence we define field as following,
\begin{equation}
\overrightarrow{E}(x)=[\overrightarrow{E}(x_{j},x),\cdots E(x_{j},x)\cdots]\label{eq:P3-1-130}
\end{equation}
or

\begin{equation}
\overrightarrow{E}(x)=[\cdots\overrightarrow{E}_{j}\cdots]\label{eq:P3-1-140}
\end{equation}
In the above definition, we can define a field to any place where
is not at the position of the charge. However we give up the superimposition
principle. We do not clear how to ``add'' the fields of many particles.
What we know is the Power of the whole system still can be given. 

\subsection{The Poynting theorem of $N$ charges}

According to the traditional electromagnetic field theory, The Poynting
theorem is give as following,

\[
-\nabla\cdot(\overrightarrow{E}\times\overrightarrow{H})
\]
\[
=\overrightarrow{E}\cdot\overrightarrow{J}
\]
\begin{equation}
(\overrightarrow{E}\cdot\partial\overrightarrow{D}+\overrightarrow{H}\cdot\partial\overrightarrow{B})\label{eq:p3-1-150}
\end{equation}
According to the Traditional definition, Here we still not apply our
new field definition.
\begin{equation}
\overrightarrow{E}=\overrightarrow{E}_{1}+\cdots+\overrightarrow{E}_{i}+\cdots+\overrightarrow{E}_{N}\label{eq:P3-1-160}
\end{equation}

\begin{equation}
\overrightarrow{H}=\overrightarrow{H}_{1}+\cdots+\overrightarrow{H}_{i}+\cdots+\overrightarrow{H}_{N}\label{eq:P3-1-170}
\end{equation}
We have

\[
-\nabla\cdot(\sum_{j}\sum_{i}\overrightarrow{E}_{i}\times\overrightarrow{H}_{j})
\]
\[
=\sum_{j}\sum_{i}(\overrightarrow{E}_{i}\cdot\overrightarrow{J}_{j})
\]

\begin{equation}
+\sum_{j}\sum_{i}(\overrightarrow{E}_{i}\cdot\partial\overrightarrow{D}_{j}+\overrightarrow{H}_{i}\cdot\partial\overrightarrow{B}_{j})\label{eq:P3-1-190}
\end{equation}
In last subsection we see the summation

\begin{equation}
\sum_{i=1}^{N}\ \sum_{j=1,j\neq i}^{N}\label{eq:P3-1-200}
\end{equation}
But it is not used in $N$ system Poynting theorem. We use this new
summation to replace the original summation, we obtain,

\[
-\nabla\cdot(\sum_{j}\sum_{i\neq j}\overrightarrow{E}_{i}\times\overrightarrow{H}_{j})
\]
\[
=\sum_{j}\sum_{i\neq j}(\overrightarrow{E}_{i}\cdot\overrightarrow{J}_{j})
\]

\begin{equation}
+\sum_{j}\sum_{i\neq j}(\overrightarrow{E}_{i}\cdot\partial\overrightarrow{D}_{j}+\overrightarrow{H}_{i}\cdot\partial\overrightarrow{B}_{j})\label{eq:P3-1-210}
\end{equation}
or

\[
-\varoiintop_{\Gamma}(\sum_{j}\sum_{i\neq j}\overrightarrow{E}_{i}\times\overrightarrow{H}_{j})\cdot\hat{n}d\Gamma
\]
\[
=\iiintop_{V}(\sum_{j}\sum_{i\neq j}(\overrightarrow{E}_{i}\cdot\overrightarrow{J}_{j})dV
\]

\begin{equation}
\iiintop_{V}(\sum_{j}\sum_{i\neq j}(\overrightarrow{E}_{i}\cdot\partial\overrightarrow{D}_{j}+\overrightarrow{H}_{i}\cdot\partial\overrightarrow{B}_{j})dV\label{eq:P3-1-220}
\end{equation}

It is the rest items of Poynting theorem if all self items as following 

\[
-\varoiintop_{\Gamma}(\sum_{i}\overrightarrow{E}_{i}\times\overrightarrow{H}_{i})\cdot\hat{n}d\Gamma
\]
\[
=\iiintop_{V}(\sum_{i}(\overrightarrow{E}_{i}\cdot\overrightarrow{J}_{i})dV
\]

\begin{equation}
\iiintop_{V}(\sum_{j}(\overrightarrow{E}_{i}\cdot\partial\overrightarrow{D}_{i}+\overrightarrow{H}_{i}\cdot\partial\overrightarrow{B}_{i})dV\label{eq:P3-1-240}
\end{equation}
are taken away. This is the the mutual energy theorems{[}XXXXXXXXX{]}.
This formula is correct in two ways. (1), if Maxwell equation is correct
this formula is also correct, it is easy to prove this. Because we
take away all self items all satisfy Poynting theorem for a single
charge. From the Poynting theorem of $N$ charges take away all corresponding
Poynting theorem for single charge this guarantees the rest part still
satisfies Maxwell equations, since Poynting theorem can be derived
from Maxwell equations. (2) The second way to show this formula is
correct is it satisfies also the ``direct interaction'' theory.
This formula actually offers a correct definition of the adjunct field
of the Wheeler and Feynman. In the following we show this formula
satisfy the direct interaction theory. In this formula the left side
is
\begin{equation}
P=\iiintop_{V}(\sum_{j}\sum_{i\neq j}(\overrightarrow{E}_{i}\cdot\overrightarrow{J}_{j})dV\label{eq:P3-1-250}
\end{equation}
which clear is the whole power of the system with all charges same
as last section. The first term of right side is
\begin{equation}
-\varoiintop_{\Gamma}(\sum_{j}\sum_{i\neq j}\overrightarrow{E}_{i}\times\overrightarrow{H}_{j})\cdot\hat{n}d\Gamma\label{eq:P3-1-260}
\end{equation}
is the power send to outside space, it is the energy current send
to outside of system. If there is only $N$ charge in a empty space,
there should no energy current go outside. We have know from the mutual
energy theorem if photon's field either retarded field for the emitter
or advanced field from absorber, the mutual energy current vanishes.
Hence this term is $0.$ 

The following is

\begin{equation}
\iiintop_{V}(\sum_{j}\sum_{i\neq j}(\overrightarrow{E}_{i}\cdot\partial\overrightarrow{D}_{j}+\overrightarrow{H}_{i}\cdot\partial\overrightarrow{B}_{j})dV\label{eq:P3-1-270}
\end{equation}
is the system energy in the space. If started from some time there
are no action or reaction to a end time there are also no action and
reaction. The integral of this energy is vanishes, i.e.,

\begin{equation}
\intop_{t=-\infty}^{\infty}\iiintop_{V}(\sum_{j}\sum_{i\neq j}(\overrightarrow{E}_{i}\cdot\partial\overrightarrow{D}_{j}+\overrightarrow{H}_{i}\cdot\partial\overrightarrow{B}_{j})dVdt=0\label{eq:P3-1-280}
\end{equation}
Hence we have the last term 

\begin{equation}
\intop_{t=-\infty}^{\infty}\iiintop_{V}(\sum_{j}\sum_{i\neq j}(\overrightarrow{E}_{i}\cdot\overrightarrow{J}_{j})dVdt=0\label{eq:P3-1-290}
\end{equation}
This term also vanishes. The above formula tell us system all energy
is conserved. Hence this corrected formula. It is much meaning full
compare to the the original formula of Poynting theorem. This formula
actually is the mutual energy theorem we also call it mutual energy
principle (in the past we call it principle is because we believe
it also can be established on quantum physics for example the theory
of electrons instead photons. In that time we still not decided to
replace Maxwell equation with this mutual energy theorem). 

Compare to Poynting theorem, it have

\begin{equation}
\iiintop_{V}(\overrightarrow{E}_{i}\cdot\overrightarrow{J}_{i})dV=\infty\label{eq:P3-1-300}
\end{equation}
If the charge is a point.
\begin{equation}
-\varoiintop_{\Gamma}(\overrightarrow{E}_{i}\times\overrightarrow{H}_{i})\cdot\hat{n}d\Gamma\neq0\label{eq:P3-1-310}
\end{equation}
The system always some energy go to outside even there is empty without
other charges, which we do not understand. It is very strange. It
is clear Poynting theorem conflicts with direct interaction principle.
But the mutual energy theorem doesn't.

\subsection{Looking for new principle to replace the Maxwell equation}

In the above we have prove the Poynting theorem is problem as physic
theorem. This wrong also lead the Maxwell equations fail. We have
said the Poynting theorem is nearly equivalent to Maxwell equations.

We have shown that mutual energy theorem is correct and satisfies
energy conservation and direct interaction principles. Why we do not
started from mutual energy theorem and build the whole electromagnetic
field theory? 

\subsection{Upgraded the mutual energy theorem as mutual energy principle}

We have tried a lot times to prove the self energy items vanishes
in the Maxwell theory, it seems successful with a return process,
that means the field is end to infinity some reason it can return
back. The return wave satisfy time-reversed Maxwell equations.

However there is always something doesn't very self consistent. Its
time for us to solve this problem thoroughly. We believe the problem
is at the Maxwell equations. Hence we will replace the the Maxwell
equations. 

Now we speak about the mutual energy principle instead of mutual energy
theorem. This is because we would like to replace Maxwell equations
by the mutual energy theorem. After the replace the mutual energy
theorem upgraded to as the mutual energy principle.

\subsection{What is the conditions for a principle}

Wheeler and Feynman offers the 3 conditions to started the electromagnetic
theory which are,

``(1) Well defined. (2) economical in postulates and (3) in agreement
with experience.''

We know that Maxwell equation and the mutual energy theorem are all
well defined. In the above sections we have clear shows that the mutual
energy theorem is agree with energy conservation and direct interaction
principle. Poynting theorem and hence Maxwell equation doesn't satisfy
(3).

Now we look the condition (2), economical in postulates.

In the above when we speak the Maxwell equation has problem, we mean
it is problem for a system with $N$ charges. In case there is only
two charges, one is emitter and another is absorber, Maxwell equation
still possible to be correct in the following way.

If we assume take mutual energy theorem as principle, we need to solve
the equation of the mutual theorem equation which for only two charges
are following,

\[
-\nabla\cdot(\overrightarrow{E}_{1}\times\overrightarrow{H}_{2}+\overrightarrow{E}_{2}\times\overrightarrow{H}_{1})
\]
\[
=\overrightarrow{E}_{2}\cdot\overrightarrow{J}_{1}+\overrightarrow{E}_{1}\cdot\overrightarrow{J}_{2}
\]

\begin{equation}
+\overrightarrow{E}_{1}\cdot\partial\overrightarrow{D}_{2}+\overrightarrow{E}_{2}\cdot\partial\overrightarrow{D}_{1}+\overrightarrow{H}_{1}\cdot\partial\overrightarrow{B}_{2}+\overrightarrow{H}_{2}\cdot\partial\overrightarrow{B}_{1}\label{eq:P3-1-330}
\end{equation}
This can rewritten as

\[
-(\nabla\times\overrightarrow{E}_{1}\cdot\overrightarrow{H}_{2}-\overrightarrow{E}_{1}\cdot\nabla\times\overrightarrow{H}_{2}+\nabla\times\overrightarrow{E}_{2}\cdot\overrightarrow{H}_{1}-\overrightarrow{E}_{2}\cdot\nabla\times\overrightarrow{H}_{1})
\]

\[
=\overrightarrow{E}_{2}\cdot\overrightarrow{J}_{1}+\overrightarrow{E}_{1}\cdot\overrightarrow{J}_{2}
\]

\begin{equation}
+\overrightarrow{E}_{1}\cdot\partial\overrightarrow{D}_{2}+\overrightarrow{E}_{2}\cdot\partial\overrightarrow{D}_{1}+\overrightarrow{H}_{1}\cdot\partial\overrightarrow{B}_{2}+\overrightarrow{H}_{2}\cdot\partial\overrightarrow{B}_{1}\label{eq:P3-1-340}
\end{equation}
or
\[
\overrightarrow{E}_{1}\cdot(\nabla\times\overrightarrow{H}_{2}-\partial\overrightarrow{D}_{2}-\overrightarrow{J}_{2})+(-\nabla\times\overrightarrow{E}_{2}-\partial\overrightarrow{B}_{2})\cdot\overrightarrow{H}_{1}
\]

\[
+\overrightarrow{E}_{2}\cdot(\nabla\times\overrightarrow{H}_{1}-\partial\overrightarrow{D}_{1}-\overrightarrow{J}_{1})+(-\nabla\times\overrightarrow{E}_{1}-\partial\overrightarrow{B}_{1})\cdot\overrightarrow{H}_{2}
\]
\begin{equation}
=0\label{eq:P3-1-350}
\end{equation}
from this we derive that, if 
\begin{equation}
\nabla\times\overrightarrow{H}_{2}-\partial\overrightarrow{D}_{2}-\overrightarrow{J}_{2}=0\label{eq:P3-1-360}
\end{equation}
\begin{equation}
-\nabla\times\overrightarrow{E}_{2}-\partial\overrightarrow{B}_{2}=0\label{eq:P3-1-370}
\end{equation}

\begin{equation}
\nabla\times\overrightarrow{H}_{1}-\partial\overrightarrow{D}_{1}-\overrightarrow{J}_{1}=0\label{eq:P3-1-380}
\end{equation}
\begin{equation}
-\nabla\times\overrightarrow{E}_{1}-\partial\overrightarrow{B}_{1}=0\label{eq:P3-1-390}
\end{equation}
The mutual energy theorem is satisfied. The nonzero solution of the
mutual energy theorem ask the field $[\overrightarrow{E}_{1},\overrightarrow{H}_{1}]$
and $[\overrightarrow{E}_{1},\overrightarrow{H}_{1}]$ nonzero in
the same time. The above formula are Maxwell equation for singular
charges,
\begin{equation}
\begin{cases}
\nabla\times\overrightarrow{H}_{1}=\overrightarrow{J}_{1}+\partial\overrightarrow{D}_{1}\\
\nabla\times\overrightarrow{E}_{1}=\partial\overrightarrow{B}_{1}
\end{cases}\label{eq:P3-1-400}
\end{equation}
and

\begin{equation}
\begin{cases}
\nabla\times\overrightarrow{H}_{2}=\overrightarrow{J}_{2}+\partial\overrightarrow{D}_{2}\\
\nabla\times\overrightarrow{E}_{2}=\partial\overrightarrow{B}_{2}
\end{cases}\label{eq:P3-1-410}
\end{equation}
Hence if we choose the mutual energy theorem as the principle, the
Maxwell equation still can be obtained as sufficient conditions of
the mutual energy theorem.

In other hand if we choose Maxwell equation as principle, the above
4 formula is still not enough, we have to add a special condition
ask the two fields $[\overrightarrow{E}_{1},\overrightarrow{H}_{1}]$
and $[\overrightarrow{E}_{1},\overrightarrow{H}_{1}]$ nonzero in
the same time. This even not enough, we still have to explain any
place in the space if one of filed vanish $[\overrightarrow{E}_{1},\overrightarrow{H}_{1}]$,
the other field loss its meaning$[\overrightarrow{E}_{2},\overrightarrow{H}_{2}]$.

For example if the emitter and absorber is at two sides of the metal
plate. In the metal plate there is a hole. the fields of the emitter
and the absorber become cone beam fields. The two cone beams has a
overlap region. The field can only nonzero in this overlap region
simultaneously. In other place there is only one field ether $[\overrightarrow{E}_{1},\overrightarrow{H}_{1}]$
and $[\overrightarrow{E}_{2},\overrightarrow{H}_{2}]$ is nonzero.
Hence in the region which is not the overlap region, the field is
not meaning any more even there still one field not vanish according
to the Maxwell equation. 

If we use the mutual energy theorem as principle, this is clear, we
only care the overlap place where the mutual energy is meaningful.
Hence if we use mutual energy theorem as principle we need only one
formula. If we use Maxwell equations we need 4 formula and plus a
very strangle conditions. This means if we use the mutual energy theorem
as principle, it is much more economy than use of the Maxwell equations. 

\subsection{The necessary condition of the Mutual energy theorem}

We just have said that the Maxwell equations plus the two fields are
nonzero simultaneously are the sufficient conditions. It is not the
necessary condition. There is the possibility that the Maxwell equation
doesn't satisfy but the mutual energy theorem still satisfied. If
we found this kind solution in the real world. The Maxwell equations
will thoroughly loss its power as a principle. 

\subsection{Maxwell equations in Macrocosm}

We know that we never can prove Maxwell equations in microcosm (the
Maxwell equation for only one charge) from the Maxwell equation in
Macrocosm (The Maxwell equation for many many charges). Hence traditionally
we started with Maxwell equation in microcosm plus the principle that
field can be superimposed, and derive the Maxwell equations of macrocosm.
However from above discussion we have now clear that the superimposition
principle has problem. We have give up the field can be superimposed
this concept. Hence even we have the Maxwell equations in microcosm
can be principle, they still cannot derive the Maxwell equation in
macrocosm. Without a Maxwell equation in macrocosm, one with a Maxwell
equation in microcosm we still cannot build the whole theory of electromagnetic
field theory. In other hand for mutual energy theorem, the macrocosm
mutual energy theorem just the summation of all the mutual energy
theorem for a pair charges. If the action and reaction of a pair charges
defined by the mutual energy theorem, the whole mutual energy is just
found all connection pairs and put all this together. Hence from microcosm
to the macrocosm for the mutual energy theorem is easy to do.

\subsection{The field can only be defined as a approximate concept in case $N\rightarrow\infty$}

We assume in electromagnetic field, Maxwell equations are not exactly
correct. The field superimposed principle is also not exactly satisfied.
That means we cannot have
\begin{equation}
\overrightarrow{E}=\sum_{i=1}^{N}\overrightarrow{E}_{i}\label{eq:P3-1-440}
\end{equation}
What we have is the only mutual energy principle. In a two electrons'
system the mutual energy principle tell us

\[
-\nabla\cdot(\overrightarrow{E}_{1}\times\overrightarrow{H}_{2}+\overrightarrow{E}_{2}\times\overrightarrow{H}_{1})
\]
\[
=\overrightarrow{E}_{2}\cdot\overrightarrow{J}_{1}+\overrightarrow{E}_{1}\cdot\overrightarrow{J}_{2}
\]

\begin{equation}
+\overrightarrow{E}_{1}\cdot\partial\overrightarrow{D}_{2}+\overrightarrow{E}_{2}\cdot\partial\overrightarrow{D}_{1}+\overrightarrow{H}_{1}\cdot\partial\overrightarrow{B}_{2}+\overrightarrow{H}_{2}\cdot\partial\overrightarrow{B}_{1}\label{eq:P3-1-450}
\end{equation}
This is the starting point. This is the principle of the whole electromagnetic
field theory. If there are more charges, the mutual energy principle
becomes, 

\[
-\nabla\cdot(\sum_{j}\sum_{i\neq j}\overrightarrow{E}_{i}\times\overrightarrow{H}_{j})
\]
\[
=\sum_{j}\sum_{i\neq j}(\overrightarrow{E}_{i}\cdot\overrightarrow{J}_{j})
\]

\begin{equation}
+\sum_{j}\sum_{i\neq j}(\overrightarrow{E}_{i}\cdot\partial\overrightarrow{D}_{j}+\overrightarrow{H}_{i}\cdot\partial\overrightarrow{B}_{j})\label{eq:P3-1-460}
\end{equation}
where
\begin{equation}
i,j=1,2\cdots N\label{eq:P3-1-470}
\end{equation}
When $N\rightarrow\infty$, the following formula is close to the
above one, hence is also correct approximatively,

\[
-\nabla\cdot(\sum_{j}\sum_{i}\overrightarrow{E}_{i}\times\overrightarrow{H}_{j})
\]
\[
=\sum_{j}\sum_{i}(\overrightarrow{E}_{i}\cdot\overrightarrow{J}_{j})
\]

\begin{equation}
+\sum_{j}\sum_{i}(\overrightarrow{E}_{i}\cdot\partial\overrightarrow{D}_{j}+\overrightarrow{H}_{i}\cdot\partial\overrightarrow{B}_{j})\label{eq:P3-1-480}
\end{equation}
or

\[
-\nabla\cdot(\sum_{i}\overrightarrow{E}_{i}\times\sum_{j}\overrightarrow{H}_{j})
\]
\[
=(\sum_{i}\overrightarrow{E}_{i}\cdot\sum_{j}\overrightarrow{J}_{j})
\]

\begin{equation}
+(\sum_{i}\overrightarrow{E}_{i}\cdot\partial\sum_{j}\overrightarrow{D}_{j}+\sum_{i}\overrightarrow{H}_{i}\cdot\partial\sum_{j}\overrightarrow{B}_{j})\label{eq:P3-1-490}
\end{equation}
or

\begin{equation}
-\nabla\cdot(\overrightarrow{E}\times\overrightarrow{H})=\overrightarrow{E}\cdot\overrightarrow{J}+\overrightarrow{E}\cdot\partial\overrightarrow{D}+\overrightarrow{H}\cdot\partial\overrightarrow{B}\label{eq:P3-1-500}
\end{equation}
This is the Poynting theorem. We have written 
\begin{equation}
\overrightarrow{E}=\sum_{i}\overrightarrow{E}_{i}\label{eq:P3-1-510}
\end{equation}
, 
\begin{equation}
\overrightarrow{H}=\sum_{i}\overrightarrow{H}_{i}\label{eq:P3-1-520}
\end{equation}
, 
\begin{equation}
\overrightarrow{J}=\sum_{i}\overrightarrow{J}_{i}\label{eq:P3-1-530}
\end{equation}
 and 
\begin{equation}
\overrightarrow{D}=\sum_{i}\overrightarrow{D}_{i}\label{eq:P3-1-540}
\end{equation}
, 
\begin{equation}
\overrightarrow{B}=\sum_{i}\overrightarrow{B}_{i}\label{eq:P3-1-550}
\end{equation}
Hence we obtained the results if $N\rightarrow\infty$ we have the
Poynting theorem. 

Important to know that here Poynting theorem is not a exact correct
physics formula, but only when $N\rightarrow\infty$ it is correct
approximatively. From Poynting theorem we can have
\begin{equation}
0=\nabla\times\overrightarrow{E}\cdot\overrightarrow{H}-\nabla\times\overrightarrow{H}\cdot\overrightarrow{E}+\overrightarrow{E}\cdot\overrightarrow{J}+\overrightarrow{E}\cdot\partial\overrightarrow{D}+\overrightarrow{H}\cdot\partial\overrightarrow{B}\label{eq:P3-1-560}
\end{equation}
 or
\begin{equation}
(\nabla\times\overrightarrow{E}+\partial\overrightarrow{B})\cdot\overrightarrow{H}+(-\nabla\times\overrightarrow{H}+\overrightarrow{J}+\partial\overrightarrow{D})\cdot\overrightarrow{E}=0\label{eq:P3-1-570}
\end{equation}
We obtained the sufficient conditions 

\begin{equation}
\nabla\times\overrightarrow{E}+\partial\overrightarrow{B}=0\label{eq:P3-1-580}
\end{equation}
\begin{equation}
-\nabla\times\overrightarrow{H}+\overrightarrow{J}+\partial\overrightarrow{D}=0\label{eq:P1-580}
\end{equation}
or

\begin{equation}
\nabla\times\overrightarrow{E}=-\partial\overrightarrow{B}\label{eq:P3-1-590}
\end{equation}
\begin{equation}
\nabla\times\overrightarrow{H}=\overrightarrow{J}+\partial\overrightarrow{D}\label{eq:P3-1-600}
\end{equation}
These are Maxwell equations. Here we did not derived the Maxwell equation
as the sufficient and necessary conditions. However since Poynting
theorem can derive all reciprocity theorems, from the reciprocity
theorems we can get all Green function solution of Maxwell equations.
If we have obtained all solution of Maxwell equations we can obtained
Maxwell equations by induction. Hence we finally obtained the Maxwell
equations.

Please keep in mind that from the authors' above theory, that the
Poynting theorem and Maxwell equations all are only approximatively
correct. It only correct when $N\rightarrow\infty$.

In this way the Maxwell equations are simplified to only with one
formula. Two formula be come one mutual energy principle formula.
However we are not only try to simplify the Maxwell equations. We
found that Maxwell equation is wrong. It need to be correct. It is
only correct at $N\rightarrow\infty$. It is possible has huge problem
even $N\rightarrow\infty$. That is problem the infinity problem the
quantum physics meet, there a re-normalization process is required. 

We achieved this result by many deep thought. The following we will
offers how we finally achieved the above result. In the following
we will shown in the beginning we did not know the problem is at Maxwell
equations.

\subsection{The other two Maxwell equations}

The other two equations of Maxwell equations are,
\begin{equation}
\nabla\cdot\overrightarrow{B}=0\label{eq:P3-1-610}
\end{equation}
\begin{equation}
\nabla\cdot\overrightarrow{D}=\rho\label{eq:P3-1-620}
\end{equation}
or in the integral formula,

\begin{equation}
\varoiintop_{\Gamma}\overrightarrow{B}\cdot\hat{n}d\Gamma=0\label{eq:P3-1-630}
\end{equation}
or
\begin{equation}
\varoiintop_{\Gamma}\overrightarrow{D}\cdot\hat{n}d\Gamma=\iiintop_{V}\rho dV\label{eq:P3-1-640}
\end{equation}
The last formula is derive from Coulomb's law which is

\begin{equation}
\overrightarrow{F}_{ji}=\frac{q_{i}q_{j}}{4\pi\epsilon_{0}}\frac{||\overrightarrow{x}_{i}-\overrightarrow{x}_{j}||}{||\overrightarrow{x}_{i}-\overrightarrow{x}_{j}||^{3}}\label{eq:P3-1-660}
\end{equation}

\begin{equation}
\overrightarrow{E}_{ji}=\frac{\overrightarrow{F}_{ji}}{q_{i}}=\frac{q_{i}q_{j}}{4\pi\epsilon_{0}}\frac{||\overrightarrow{x}_{i}-\overrightarrow{x}_{j}||}{||\overrightarrow{x}_{i}-\overrightarrow{x}_{j}||^{3}}\label{eq:P3-1-670}
\end{equation}
or $\overrightarrow{E}_{ji}$ is the charge $q_{j}$ produced field
at the position $\overrightarrow{x}_{i}$. The total force on the
charge $q_{i}$ is

\begin{equation}
F_{i}=\sum_{j=1,j\neq i}^{j=N}\overrightarrow{F}_{ji}=\sum_{j=1,j\neq i}^{j=N}\overrightarrow{E}_{ji}q_{i}\label{eq:P3-1-680}
\end{equation}
The total power is
\begin{equation}
P=\sum_{i=1}^{N}F_{i}v_{i}=\sum_{i=1}^{N}\sum_{j=1,j\neq i}^{j=N}\overrightarrow{E}_{ji}q_{i}v_{i}\label{eq:P3-1-690}
\end{equation}
This power should be vanishes

\begin{equation}
\sum_{i=1}^{N}\sum_{j=1,j\neq i}^{j=N}\overrightarrow{E}_{ji}\overrightarrow{J}_{i}=0\label{eq:P3-1-700}
\end{equation}
where $J=q_{i}v_{i}$ we find this is also the mutual energy theorem
in static situation. In the case there is no radiation filed hence
in the mutual energy theorem the term

\begin{equation}
-\nabla\cdot(\sum_{i}\sum_{j\neq i}\overrightarrow{E}_{i}\times\overrightarrow{H}_{j})\label{eq:P3-1-710}
\end{equation}
and 

\begin{equation}
+\sum_{i}\sum_{j\neq i}(\overrightarrow{E}_{i}\cdot\partial\overrightarrow{D}_{j}+\overrightarrow{H}_{i}\cdot\partial\overrightarrow{B}_{j})\label{eq:P3-1-720}
\end{equation}
are all vanish. Hence this formula can also merge into the mutual
energy theorem.The formula is not very important. Even in Maxwell
theory it can only be applied to find a constant. In Maxwell theory 

\begin{equation}
\nabla\times\overrightarrow{E}+\partial\overrightarrow{B}=0\label{eq:P3-1-730}
\end{equation}

\begin{equation}
\nabla\cdot(\nabla\times\overrightarrow{E}+\partial\overrightarrow{B})=0\label{eq:P3-1-740}
\end{equation}
or
\begin{equation}
\nabla\cdot\nabla\times\overrightarrow{E}+\partial\nabla\cdot\overrightarrow{B}=0\label{eq:P3-1-750}
\end{equation}
or
\begin{equation}
\partial\nabla\cdot\overrightarrow{B}=0\label{eq:P3-1-760}
\end{equation}
or

\begin{equation}
\nabla\cdot\overrightarrow{B}=Constant\label{eq:P3-1-770}
\end{equation}
Because we thought the field any way is a problematic, we are not
care a constant field. Hence for this equation we do not need to replace
it.

\section{Reconstruction the electromagnetic field theory from the mutual energy
principle}

\subsection{Define the multiplication of fields}

We do not define the superimposition of the field. We do not know
how to ``add'' is really correct. However we can redefine the ``$\times$'',
``$\cdot$''

\begin{equation}
\overrightarrow{A}\cdot\overrightarrow{B}=\sum_{i=1}^{N}\sum_{j=1,j\neq i}^{N}\overrightarrow{A}_{j}\overrightarrow{B}_{i}\label{eq:P3-1-780}
\end{equation}

\begin{equation}
\overrightarrow{A}\times\overrightarrow{B}=\sum_{i=1}^{N}\sum_{j=1,j\neq i}^{N}\overrightarrow{A}_{j}\overrightarrow{B}_{i}\label{eq:P3-1-790}
\end{equation}
In this way the mutual energy principle 

\[
-\nabla\cdot(\sum_{i}\sum_{j\neq i}\overrightarrow{E}_{i}\times\overrightarrow{H}_{j})
\]
\[
=\sum_{i}\sum_{j\neq i}(\overrightarrow{E}_{i}\cdot\overrightarrow{J}_{j})
\]

\begin{equation}
+\sum_{i}\sum_{j\neq i}(\overrightarrow{E}_{i}\cdot\partial\overrightarrow{D}_{j}+\overrightarrow{H}_{i}\cdot\partial\overrightarrow{B}_{j})\label{eq:P3-1-800}
\end{equation}
can be written as

\begin{equation}
-\nabla\cdot(\overrightarrow{E}\times\overrightarrow{H})=\overrightarrow{E}\cdot\overrightarrow{J}+\overrightarrow{E}\cdot\partial\overrightarrow{D}+\overrightarrow{H}\cdot\partial\overrightarrow{B}\label{eq:P3-1-810}
\end{equation}

This looks exactly to the Poynting theorem. But the above actually
is the mutual energy principle, through redefine the cross multiplication
and point multiplication, it can be written as the exact form of Poynting
theorem. It is clear that
\begin{equation}
\lim_{N\rightarrow\infty}(\times_{new})=\times\label{eq:P3-1-820}
\end{equation}

\begin{equation}
\lim_{N\rightarrow\infty}(\cdot_{new})=\cdot\label{eq:P3-1-830}
\end{equation}
This guarantee that the limit situation of the mutual energy theorem
is the Poynting theorem.

If Poynting theorem is established, then the superimposition of the
field also can be established. Hence if $N\rightarrow\infty$ there
is
\begin{equation}
E=\sum_{j=1}^{N}E_{j}\label{eq:P3-1-840}
\end{equation}
seems can be accept.

\subsection{Linearization of electromagnetic fields}

The fields is not linear, that is not very convenient. In order to
make things simple, We can add self energy current to the mutual energy
theory:

Even we known that self energy current formula 
\begin{equation}
-\nabla\cdot\overrightarrow{E}_{i}\times\overrightarrow{H}_{i}=\overrightarrow{E}_{i}\cdot\overrightarrow{J}_{i}+\overrightarrow{E}_{i}\cdot\partial\overrightarrow{D}_{i}+\overrightarrow{H}_{i}\cdot\partial\overrightarrow{B}_{i}\label{eq:P3-1-860}
\end{equation}
is nonsense in physics, in physics, actually that $\overrightarrow{E}_{i}\times\overrightarrow{H}_{i}$
is 0, $\overrightarrow{E}_{i}\cdot\overrightarrow{J}_{i}$ is $0$,
but the mathematics calculation that is not $0$. We also know that
$\overrightarrow{E}_{i}=\infty$, if the emitter or the absorber is
point, we can just assume it is not a point but the charge of the
electron is inside a small sphere region. Then the above formula can
be a correct formula in mathematics. It is a pseudo self energy current
formula. We can add this pseudo formula to the following mutual energy
theorem,

\[
-\nabla\cdot(\sum_{j}\sum_{i\neq j}\overrightarrow{E}_{i}\times\overrightarrow{H}_{j})
\]
\[
=\sum_{j}\sum_{i\neq j}(\overrightarrow{E}_{i}\cdot\overrightarrow{J}_{j})
\]

\begin{equation}
+\sum_{j}\sum_{i\neq j}(\overrightarrow{E}_{i}\cdot\partial\overrightarrow{D}_{j}+\overrightarrow{H}_{i}\cdot\partial\overrightarrow{B}_{j})\label{eq:P3-1-850}
\end{equation}
We got

\[
-\nabla\cdot(\sum_{j}\sum_{i}\overrightarrow{E}_{i}\times\overrightarrow{H}_{j})
\]
\[
=\sum_{j}\sum_{i}(\overrightarrow{E}_{i}\cdot\overrightarrow{J}_{j})
\]

\begin{equation}
+\sum_{j}\sum_{i}(\overrightarrow{E}_{i}\cdot\partial\overrightarrow{D}_{j}+\overrightarrow{H}_{i}\cdot\partial\overrightarrow{B}_{j})\label{eq:P3-1-870}
\end{equation}
Keep in mind the above formula is only correct in the meaning of mathematics,
not physics. It is a mathematical formula not a physic formula. The
formula can be further written as

\[
-\nabla\cdot(\sum_{i}\overrightarrow{E}_{i})\times(\sum_{j}\overrightarrow{H}_{j})
\]
\[
=(\sum_{i}\overrightarrow{E}_{i})\cdot(\sum_{j}\overrightarrow{J}_{j})
\]

\begin{equation}
+(\sum_{i}\overrightarrow{E}_{i})\cdot(\sum_{j}\partial\overrightarrow{D}_{j})+(\sum_{i}\overrightarrow{H}_{i})\cdot(\sum_{j}\partial\overrightarrow{B}_{j})\label{eq:P3-1-880}
\end{equation}
Write
\begin{equation}
\overrightarrow{E}=\sum_{i}\overrightarrow{E}_{i},\ \ \ \ \overrightarrow{H}=\sum_{i}\overrightarrow{H}_{i}\ \ \ \ \ \overrightarrow{J}=\sum_{i}\overrightarrow{J}_{i}\label{eq:P3-1-890}
\end{equation}
We obtain that,

\[
-\nabla\cdot\overrightarrow{E}\times\overrightarrow{H}
\]
\[
=\overrightarrow{E}\cdot\overrightarrow{J}
\]

\begin{equation}
+\overrightarrow{E}\cdot\partial\overrightarrow{D}+\overrightarrow{H}\partial\overrightarrow{B}\label{eq:P3-1-900}
\end{equation}
Hence we obtained the Poynting theorem. We have obtain the Poynting
theorem. We also obtains the new field definition, which can be superimposed.
Keep in mind this all mathematical result not a results in physics.
It is a result when we add pseudo self-energy current to the mutual
energy principle.

The Poynting theorem can be written as

\[
-(\nabla\times\overrightarrow{E}\cdot\overrightarrow{H}-\nabla\times\overrightarrow{H}\cdot\overrightarrow{E})
\]
\[
=\overrightarrow{E}\cdot\overrightarrow{J}
\]

\begin{equation}
+\overrightarrow{E}\cdot\partial\overrightarrow{D}+\overrightarrow{H}\cdot\partial\overrightarrow{B}\label{eq:P3-1-910}
\end{equation}
or

\begin{equation}
-(\nabla\times\overrightarrow{E}+\partial\overrightarrow{B})\cdot\overrightarrow{H}+(\nabla\times\overrightarrow{H}-\overrightarrow{J}-\partial\overrightarrow{D})\cdot\overrightarrow{E}=0\label{eq:P3-1-920}
\end{equation}
The sufficient condition of the above formula is

\begin{equation}
\nabla\times\overrightarrow{E}+\partial\overrightarrow{B}=0\label{eq:P3-1-930}
\end{equation}
\begin{equation}
\nabla\times\overrightarrow{H}-\overrightarrow{J}-\partial\overrightarrow{D}=0\label{eq:P3-1-940}
\end{equation}
 We got the Maxwell equation. We did not get it as a necessarily condition.
But this is enough. In our electromagnetic theory, Maxwell equation
is not need to be derived, it has problem anyway. The above derivation
of the Maxwell equation is also by dint of the pseudo self-energy
current.

Hence we obtained the results, started from mutual energy energy principle,
by dint of pseudo self-energy current we prove that the Poynting theorem
and Maxwell equation are still correct in the mathematics. It is notice
that it only correct on mathematics not on the physics. 

Understand that we can now further why in quantum physics need a re-normalization
process, when it take away all self energy terms in the formula, they
got correct result. In the correct physics these all self energy terms
actual should be take away from the correct physics or the view of
the mutual energy principle. 

\section{Wireless wave equations}

We have to define what is the wireless wave. We speak the wireless
wave means this wave is send out continuously not as wave package.
We known that the wireless wave frequency is very low the wave length
is wave long. 

In the light situation the assume one photon package has energy
\begin{equation}
E_{engergy}=\omega\hbar\label{eq:P3-1-950}
\end{equation}
The speed of photon is $c$, assume the photon has the life time $\triangle t$
\begin{equation}
pc=fh\label{eq:P3-1-960}
\end{equation}
or
\begin{equation}
\lambda=\frac{c}{f}=\frac{h}{p}\label{eq:P3-1-970}
\end{equation}
If photon become a wave, it need at least one wave length. If the
wireless wave is very long for example one meter. If the wave is even
continues, the surrounding can have thousand environment receiving
device synchronized with this antenna. They all can receive the energy.
This is different to the light situation, in which the wave can only
be received by one absorber. The life time of the wave is
\begin{equation}
T_{life}=\frac{1}{f}\label{eq:P3-1-980}
\end{equation}
Hence if the higher the frequency, the shorter the lift time of the
photon. We known that photon is composed with retarded wave and advanced
wave, the two wave need to be synchronized. The above formula tell
use when the frequency become higher, it is more difficult to make
the two photon synchronized. When frequency is higher energy package
become larger, in the light source there cannot offer continual wave,
hence they can only send wave package. This wave package of retarded
wave looking a synchronized advanced wave. Since the synchronization
become so difficult, it can only find one to be synchronized. Hence
we got photon, it is a synchronized wave from a pair electrons, one
is the emitter another is the absorber. Here we know the synchronization
means the time window, the frequency window and the orientation window
(similar to the receive antenna, which need to adjust its direction
receive more energy).

In the lower frequency, in case one energy package can be received
by many many particle, it be come wireless wave.

We will show that, for photon, there is only wave between two electrons
emitter and the absorber, in this situation the wave do not satisfy
the Maxwell equation in macrocosm. For wireless wave if it satisfy
Maxwell equations in macrocosm, that means it is not send as photon.
The energy of wireless wave is one emitter corresponding to many absorbers.
This kind of wave is not photon wave.

\subsection{Mutual energy principle can be separable}

Here we speak about the separation. That means for example if there
are $N$ charges in a system, we can divided this charges as 2 groups,
then the mutual energy theorem can be established for this two groups.
In this mutual energy theorem for groups, all interaction inside the
groups will not need to be counted. 

The second example for example a charge, we do not need the charge
is with 0 radio. The charge can have radio and can has a few pars.
For example we can divided the a electron's charge to two parts, part
1 and part 2. There is the mutual energy between this two parts. But
this two parts can be see as a group of this two parts. When we calculate
the field can see this group as a whole body. We do not care the mutual
energy or direct interaction between the two parts inside the group.

Hence there is still need some external force to glue the all parts
of the charges inside a electron. But in most situation the electron
can be seen as whole and direct reaction with other electrons. 

\subsection{Why retarded antenna can be see send only retarded waves?}

According to the mutual energy theorem the transmitter antenna send
retarded wave and the receive antenna send advanced wave. We make
a very simple example, in this example the transmitter antenna is
sit at the original point of the world coordinates. Assume there are
infinite more receiving antenna are at the big sphere. The center
of sphere is the transmitting antenna. For the whole system we the
mutual energy theorem,

\[
-\varoiintop_{\Gamma}\cdot(\sum_{i=1}^{N}\ \sum_{j=1,j\neq i}^{N}\overrightarrow{E}_{i}\times\overrightarrow{H}_{j})\cdot\hat{n}d\Gamma
\]
\[
=\iiintop_{V}\sum_{i=1}^{N}\ \sum_{j=1,j\neq i}^{N}(\overrightarrow{E}_{i}\cdot\overrightarrow{J}_{j})dV
\]

\[
+\iiintop_{V}\sum_{i=1}^{N}\ \sum_{j=1,j\neq i}^{N}(\overrightarrow{E}_{i}\cdot\partial\overrightarrow{D}_{j}+\overrightarrow{H}_{i}\cdot\partial\overrightarrow{B}_{j})dV
\]

In the above formula, we assume for example $i=1$ is the transmitter
antenna send retarded wave. All other from $i=2$ to $N$ are all
receive antenna stayed at the big sphere. We know we can add a pseudo
self items to keep the above formula still correct in mathematics,
so we have,

\[
-\varoiintop_{\Gamma}\cdot(\sum_{i=1}^{N}\sum_{j=1}^{N}\overrightarrow{E}_{i}\times\overrightarrow{H}_{j})\cdot\hat{n}d\Gamma
\]
\[
=\iiintop_{V}\sum_{i=1}^{N}\sum_{j=1}^{N}(\overrightarrow{E}_{i}\cdot\overrightarrow{J}_{j})dV
\]

\[
+\iiintop_{V}\sum_{i=1}^{N}\sum_{j=1}^{N}(\overrightarrow{E}_{i}\cdot\partial\overrightarrow{D}_{j}+\overrightarrow{H}_{i}\cdot\partial\overrightarrow{B}_{j})dV
\]
or

\[
-\varoiintop_{\Gamma}\cdot(\sum_{i=1}^{N}\overrightarrow{E}_{i})\times(\sum_{j=1}^{N}\overrightarrow{H}_{j})\cdot\hat{n}d\Gamma
\]
\[
=\iiintop_{V}(\sum_{i=1}^{N}\overrightarrow{E}_{i})\cdot(\sum_{j=1}^{N}\overrightarrow{J}_{j})dV
\]

\[
+\iiintop_{V}(\sum_{i=1}^{N}\overrightarrow{E})_{i}\cdot\sum_{j=1}^{N}\partial\overrightarrow{D}_{j}+(\sum_{i=1}^{N}\overrightarrow{H}_{i})\cdot(\sum_{j=1}^{N}\partial\overrightarrow{B}_{j})dV
\]
or

\[
-\varoiintop_{\Gamma}\cdot\overrightarrow{E}\times\overrightarrow{H}\cdot\hat{n}d\Gamma
\]
\[
=\iiintop_{V}E\cdot\overrightarrow{J}dV
\]

\[
+\iiintop_{V}(\overrightarrow{E}\cdot\partial\overrightarrow{D}+\overrightarrow{H}\cdot\partial\overrightarrow{B})dV
\]
If we take the volume not only for $V_{1}$, the above formula can
be written as,

\[
-\varoiintop_{\Gamma_{1}}\cdot\overrightarrow{E}\times\overrightarrow{H}\cdot\hat{n}d\Gamma
\]
\[
=\iiintop_{V_{1}}E\cdot\overrightarrow{J}_{1}dV
\]

\[
+\iiintop_{V_{1}}(\overrightarrow{E}\cdot\partial\overrightarrow{D}+\overrightarrow{H}\cdot\partial\overrightarrow{B})dV
\]
Here $\Gamma_{1}$ is the surface only includes the antenna $1$ or
the volume $V_{1}$. Here $\overrightarrow{J}_{1}$ is the current
of the transmitter antenna. $\overrightarrow{E}$ and $\overrightarrow{H}$
are the total field include the field of the transmitting antenna
which is the retarded field and all the field of the receiving antenna
which is advanced field. This become all tradition meanings in which
the transmuting antenna only send the retarded wave. 

To all normal people we can just see the whole field is produced by
the current of the transmitting antenna. 

This is the Poynting theorem in our normal sense, in which the field
is all field we can measured which should include the contribution
of the retarded fields and advanced fields, and also the pseudo fields.
$\overrightarrow{J}_{1}$ is only the current of transmitting antenna,
all the currents of receiving antenna is not list out, same as our
traditional way to work with electromagnetic field. In this situation
we can think our the field is produced by current of the emitter.
In this way we have the felling that all the field si produced by
the antenna 1 and all the field is retarded fields.

The field satisfy the above Poynting theorem guarantees it also satisfy
the Maxwell equations. We have said that any field satisfy Poynting
theorem, it will also satisfy the reciprocity theorem and also Green
function theory, which gives all solution of the Poynting theorem,
from that we obtains the Maxwell equation by induction. Hence the
above field also satisfies,

\[
\begin{cases}
\nabla\times\overrightarrow{E}=-\partial\overrightarrow{B}\\
\nabla\times\overrightarrow{H}=\overrightarrow{J}+\partial\overrightarrow{D}
\end{cases}
\]
Where 
\[
\overrightarrow{E}=\overrightarrow{E}_{1}+\cdots\overrightarrow{E}_{i}\cdots+\overrightarrow{E}_{N}
\]
\[
\overrightarrow{H}=\overrightarrow{H}_{1}+\cdots\overrightarrow{H}_{i}\cdots+\overrightarrow{H}_{N}
\]
Please keep mind inside above field, there is pseudo self energy current
terms. Hence even we obtained the Maxwell equation, it is still only
correct in some mathematics meaning. 

In the case need to consider the question what is the self action
\[
\iiintop_{V}\overrightarrow{E}_{i}\cdot\overrightarrow{J}_{i}dV
\]
self energy current
\[
\varoiintop_{\Gamma_{1}}\cdot\overrightarrow{E}_{i}\times\overrightarrow{H}_{i}\cdot\hat{n}d\Gamma
\]
 and what is self energy
\[
\iiintop_{V}(\overrightarrow{E}_{i}\cdot\partial\overrightarrow{D}_{i}+\overrightarrow{H}_{i}\cdot\partial\overrightarrow{B}_{i})\ dV
\]
The solution of Maxwell equation will meet big problems. We should
be clear that the above all self energy terms are all not exist in
the physics. They all illusive.

When I was study electromagnetic field theory. One of my teacher tell
me that the reciprocity theorem is very strong. It can solve nearly
all electromagnetic problems. In that time I asked myself that is
the reciprocity theorem can replace the Maxwell equations? Now I know
the reason. The reciprocity theorem is a transform of the mutual energy
theorem or the mutual energy principle, which can solve all problems
of electromagnetic fields.

\subsection{The influence of the environment to the transmitting antenna}

We have noticed that the field of the transmitting antenna has also
the contribution from all receiving antenna which is actually the
environment of the transmuting antenna. It is clear if this environment
is not equal in all direction our calculated field will have big difference
with the situation when we calculate it as retarded wave alone. As
a antenna engineer I know the directivity diagram of the transmitting
antenna is always deviate from the calculation, we often interpreted
this is because the refection of the environment. From the discussion
of this section we known that this deviation perhaps is because the
influence of the advanced wave from the environment. 

In the next section we will assume the transmitting antenna only send
the the retarded wave and we do not need to take the contribution
of the advanced wave of the receiving antenna, which offers us a field
of from the environment. 

\subsection{Mutual energy theorem can be applied to a part of system}

For example we have a antenna is contains $N$ electrons running inside
a wire. Assume this $N$ electron send retarded potential to outside.
We would like to calculate the whole energy of this system to the
outside. We know that the power of the system is

\begin{equation}
\sum_{i=1}^{N}\ \sum_{j=1,j\neq i}^{N}(\overrightarrow{E}_{i}\cdot\overrightarrow{J}_{j})\label{eq:P3-2-10}
\end{equation}
we can calculate all the energy current go to outside space while
is
\begin{equation}
-\nabla\cdot(\sum_{i=1}^{N}\ \sum_{j=1,j\neq i}^{N}\overrightarrow{E}_{i}\times\overrightarrow{H}_{j})\label{eq:P3-2-20}
\end{equation}
The energy saved to the space is 

\begin{equation}
\sum_{i}\sum_{j\neq i}(\overrightarrow{E}_{i}\cdot\partial\overrightarrow{D}_{j}+\overrightarrow{H}_{i}\cdot\partial\overrightarrow{B}_{j})\label{eq:P3-2-30}
\end{equation}

When $N$ is very big, i.e.,$N\rightarrow\infty$ the above energy
go to out side have very small difference to 
\[
-\nabla\cdot(\sum_{i=1}^{N}\ \sum_{j=1}^{N}\overrightarrow{E}_{i}\times\overrightarrow{H}_{j})
\]

\[
=\sum_{i=1}^{N}\sum_{j=1}^{N}(\overrightarrow{E}_{i}\cdot\overrightarrow{J}_{j})
\]

\begin{equation}
+\sum_{i=1}^{N}\sum_{j=1}^{N}(\overrightarrow{E}_{i}\cdot\partial\overrightarrow{D}_{j}+\overrightarrow{H}_{i}\cdot\partial\overrightarrow{B}_{j})\label{eq:P3-2-40}
\end{equation}
or

\[
-\nabla\cdot(\sum_{i=1}^{N}\ \overrightarrow{E}_{i})\times(\sum_{j=1}^{N}\overrightarrow{H}_{j})
\]

\[
=\ (\sum_{i=1}^{N}\overrightarrow{E}_{i})\cdot(\sum_{j=1}^{N}\overrightarrow{J}_{j})
\]

\begin{equation}
+(\sum_{i=1}^{N}\overrightarrow{E}_{i})\cdot(\sum_{j=1}^{N}\partial\overrightarrow{D}_{j})+(\sum_{i=1}^{N}\overrightarrow{H}_{i})\cdot(\sum_{j=1}^{N}\partial\overrightarrow{B}_{j})\label{eq:P3-2-50}
\end{equation}
or

\[
-\nabla\cdot\overrightarrow{E}\times\overrightarrow{H}
\]

\[
=\ \overrightarrow{E}\cdot\overrightarrow{H}
\]

\begin{equation}
+\overrightarrow{E}\cdot\partial\overrightarrow{D}+\overrightarrow{H}\cdot\partial\overrightarrow{B}\label{eq:P3-2-60}
\end{equation}
This is just the Poynting theorem, hence Poynting theorem still can
be applied approximately.

\subsection{Example of 3 charges or 3 antennas}

Think a example of antenna system there are one transmitting antenna
1, two receiving antenna 2 and 3 in the whole space. Assume the two
receive antenna is very close to each other.

The energy sent out from the transmitting antenna 1 is received by
antenna 2 and 3. The transmitting antenna sent retarded potential,
the two receiving antenna sent advanced potentials. 

From the mutual energy principle, it tells us that

\[
-\varoiintop_{\Gamma}\cdot(\sum_{i=1}^{3}\ \sum_{j=1,j\neq i}^{3}\overrightarrow{E}_{i}\times\overrightarrow{H}_{j})\cdot\hat{n}d\Gamma
\]
\[
=\iiintop_{V}\sum_{i=1}^{3}\ \sum_{j=1,j\neq i}^{3}(\overrightarrow{E}_{i}\cdot\overrightarrow{J}_{j})dV
\]

\begin{equation}
+\iiintop_{V}\sum_{i=1}^{3}\ \sum_{j=1,j\neq i}^{3}(\overrightarrow{E}_{i}\cdot\partial\overrightarrow{D}_{j}+\overrightarrow{H}_{i}\cdot\partial\overrightarrow{B}_{j})dV\label{eq:P3-2-70}
\end{equation}
In another side, the mutual energy theorem is also true to each pair
of two antenna. That means there is also, for example the antenna
1 and 2 have the following mutual energy theorem,

\[
-\varoiintop_{\Gamma}(\overrightarrow{E}_{1}\times\overrightarrow{H}_{2}+\overrightarrow{E}_{2}\times\overrightarrow{H}_{1})\cdot\hat{n}d\Gamma
\]

\[
=\iiintop_{V}(\overrightarrow{E}_{1}\cdot\overrightarrow{J}_{2}+\overrightarrow{E}_{1}\cdot\overrightarrow{J}_{2})dV
\]
\begin{equation}
\iiintop_{V}(\overrightarrow{E}_{1}\cdot\partial\overrightarrow{D}_{2}+\overrightarrow{H}_{1}\cdot\partial\overrightarrow{B}_{2}+\overrightarrow{E}_{2}\cdot\partial\overrightarrow{D}_{1}+\overrightarrow{H}_{2}\cdot\partial\overrightarrow{B}_{1})\ dV\label{eq:P3-2-80}
\end{equation}
Considering antenna 1 send retarded wave and antenna 2 send the advanced
wave. Hence, in the big sphere surface $\Gamma$, the field $\zeta_{1}$
and field $\zeta_{2}$ are not nonzero in the same time hence
\begin{equation}
\varoiintop_{\Gamma}(\overrightarrow{E}_{1}\times\overrightarrow{H}_{2}+\overrightarrow{E}_{2}\times\overrightarrow{H}_{1})\cdot\hat{n}d\Gamma=0\label{eq:P3-2-90}
\end{equation}
Here $\Gamma$ is the surface contained the antenna 1 and antenna
2.

If considering a integral with a time

\begin{equation}
\intop_{-\infty}^{\infty}\iiintop_{V}(\overrightarrow{E}_{1}\cdot\partial\overrightarrow{D}_{2}+\overrightarrow{H}_{1}\cdot\partial\overrightarrow{B}_{2}+\overrightarrow{E}_{2}\cdot\partial\overrightarrow{D}_{1}+\overrightarrow{H}_{2}\cdot\partial\overrightarrow{B}_{1})dVdt=0\label{eq:P3-2-100}
\end{equation}
hence we have

\begin{equation}
\intop_{-\infty}^{\infty}\iiintop_{V}(\overrightarrow{E}_{1}\cdot\overrightarrow{J}_{2}+\overrightarrow{E}_{1}\cdot\overrightarrow{J}_{2})dVdt=0\label{eq:P3-2-110}
\end{equation}
or
\begin{equation}
-\intop_{-\infty}^{\infty}\iiintop_{V}\overrightarrow{E}_{1}\cdot\overrightarrow{J}_{2}dVdt=\intop_{-\infty}^{\infty}\iiintop_{V}\overrightarrow{E}_{1}\cdot\overrightarrow{J}_{2}dVdt\label{eq:P3-2-120}
\end{equation}
The left side is the energy sent by the transmitting antenna. The
right side is the energy received by the second antenna.

In the same way we have
\begin{equation}
-\intop_{-\infty}^{\infty}\iiintop_{V}\overrightarrow{E}_{1}\cdot\overrightarrow{J}_{3}dVdt=\intop_{-\infty}^{\infty}\iiintop_{V}\overrightarrow{E}_{1}\cdot\overrightarrow{J}_{3}dVdt\label{eq:P3-2-130}
\end{equation}

The left is the energy sent by the first antenna to the second antenna.
The second is the energy received by the third antenna. 

If the antenna 2 and 3 is very close, they can have influence by each
other. This can be described by the mutual energy theorem between
2 and 3. Since these two antenna are all receiving antenna. They sent
all advanced wave. in this situation the surface integral is nonzero,
hence we have

\begin{equation}
\intop_{-\infty}^{\infty}\iiintop_{V}(\overrightarrow{E}_{2}\cdot\overrightarrow{J}_{3}+\overrightarrow{E}_{2}\cdot\overrightarrow{J}_{3})dVdt=\intop_{-\infty}^{\infty}\varoiintop_{\Gamma}(\overrightarrow{E}_{2}\times\overrightarrow{H}_{3}+\overrightarrow{E}_{3}\times\overrightarrow{H}_{2})\cdot\hat{n}d\Gamma\label{eq:P3-2-140}
\end{equation}
The two antenna will enforce the receiving strength make two receiving
antenna receive power more than the summation of the two antenna worked
alone. This become more complicated we do not continually discuss
here.

It is same if there two transmitting antenna close to each other,
they can send energy more than two 2 times as the antenna worked alone.

We can add the pseudo self energy terms to the mutual energy theorem,
hence we obtained,

\[
-\varoiintop_{\Gamma}\cdot(\sum_{i}^{3}\sum_{j=1}^{3}\overrightarrow{E}_{i}\times\overrightarrow{H}_{j})\cdot\hat{n}d\Gamma
\]
\[
=\iiintop_{V}\sum_{i=1}^{3}\sum_{j=1}^{3}(\overrightarrow{E}_{i}\cdot\overrightarrow{J}_{j})dV
\]

\begin{equation}
+\iiintop_{V}\sum_{i=1}^{3}\sum_{j=1}^{3}(\overrightarrow{E}_{i}\cdot\partial\overrightarrow{D}_{j}+\overrightarrow{H}_{i}\cdot\partial\overrightarrow{B}_{j})dV\label{eq:P3-2-150}
\end{equation}
or

\[
-\varoiintop_{\Gamma}\cdot(\sum_{i}^{3}\overrightarrow{E}_{i})\times(\sum_{j=1}^{3}\overrightarrow{H}_{j})\cdot\hat{n}d\Gamma
\]
\[
=\iiintop_{V}(\sum_{i=1}^{3}\overrightarrow{E}_{i})\cdot(\sum_{j=1}^{3}\overrightarrow{J}_{j})dV
\]

\begin{equation}
+\iiintop_{V}(\sum_{i=1}^{3}\overrightarrow{E})_{i}\cdot\sum_{j=1}^{3}\partial\overrightarrow{D}_{j}+(\sum_{i=1}^{3}\overrightarrow{H}_{i})\cdot(\sum_{j=1}^{3}\partial\overrightarrow{B}_{j})dV\label{eq:P3-2-160}
\end{equation}
or

\[
-\varoiintop_{\Gamma}\cdot\overrightarrow{E}\times\overrightarrow{H}\cdot\hat{n}d\Gamma
\]
\[
=\iiintop_{V}E\cdot\overrightarrow{J}dV
\]

\begin{equation}
+\iiintop_{V}(\overrightarrow{E}\cdot\partial\overrightarrow{D}+\overrightarrow{H}\cdot\partial\overrightarrow{B})dV\label{eq:P3-2-170}
\end{equation}
if we take the volume not only for $V_{1}$, the above formula can
be written as,

\[
-\varoiintop_{\Gamma_{1}}\cdot\overrightarrow{E}\times\overrightarrow{H}\cdot\hat{n}d\Gamma
\]
\[
=\iiintop_{V_{1}}E\cdot\overrightarrow{J}_{1}dV
\]

\begin{equation}
+\iiintop_{V_{1}}(\overrightarrow{E}\cdot\partial\overrightarrow{D}+\overrightarrow{H}\cdot\partial\overrightarrow{B})dV\label{eq:P3-2180}
\end{equation}
This is the Poynting theorem in our normal sense, in which the field
is all field we can measured which should include the contribution
of the retarded fields and advanced fields, and also the pseudo fields.
$\overrightarrow{J}_{1}$ is only the current of transmitting antenna,
all the currents of receiving antenna is not list out, same as our
traditional way to work with electromagnetic field. In this situation
we can think our the field is produced by current of the emitter.
In this way we have the felling that all the field si produced by
the antenna 1 and all the field is retarded fields.

The field satisfy the above Poynting theorem guarantees it also satisfy
the Maxwell equations. We have said that any field satisfy Poynting
theorem, it will also satisfy the reciprocity theorem and also Green
function theory, which gives all solution of the Poynting theorem,
from that we obtains the Maxwell equation by induction. Hence the
above field also satisfies,

\begin{equation}
\begin{cases}
\nabla\times\overrightarrow{E}=-\partial\overrightarrow{B}\\
\nabla\times\overrightarrow{H}=+\overrightarrow{J}+\partial\overrightarrow{D}
\end{cases}\label{eq:P3-2-190}
\end{equation}
Where 
\[
\overrightarrow{E}=\overrightarrow{E}_{1}+\overrightarrow{E}_{2}+\overrightarrow{E}_{3}
\]
\begin{equation}
\overrightarrow{H}=\overrightarrow{H}_{1}+\overrightarrow{H}_{2}+\overrightarrow{H}_{3}\label{eq:P3-2-200}
\end{equation}
Please keep mind inside above field, there is pseudo self energy current
terms. Hence even we obtained the Maxwell equation, it is still only
correct in some mathematics meaning. 

In the case need to consider the question what is the self action
\begin{equation}
\iiintop_{V}\overrightarrow{E}_{i}\cdot\overrightarrow{J}_{i}dV\label{eq:P3-2-210}
\end{equation}
, self energy current
\begin{equation}
\varoiintop_{\Gamma_{1}}\cdot\overrightarrow{E}_{i}\times\overrightarrow{H}_{i}\cdot\hat{n}d\Gamma\label{eq:P3-2-220}
\end{equation}
 and what is self energy
\begin{equation}
\iiintop_{V}(\overrightarrow{E}_{i}\cdot\partial\overrightarrow{D}_{i}+\overrightarrow{H}_{i}\cdot\partial\overrightarrow{B}_{i})\ dV\label{eq:P3-2-230}
\end{equation}
The solution of Maxwell equation will meet big problems. We should
be clear that the above all self energy tems are all not exist in
the physics. They all illusive.

When I was study electromagnetic field theory. One of my teacher tell
me that the reciprocity theorem is very strong. It can solve nearly
all electromagnetic problems. In that time I asked myself that is
the reciprocity theorem can replace the Maxwell equations? Now I know
the reason. The reciprocity theorem is a transform of the mutual energy
theorem or the mutual energy principle, which can solve all problems
of electromagnetic fields.

\section{Photon equations}

\subsection{The mutual energy theorem}

Now I am cleared that I am actually do not need to calculate the self
energy current, which doesn't exist. 

The reason, the self energy current doesn't exist is because, the
concept electromagnetic field is wrong. The electromagnetic field
need a test charge in static field situation or an absorber in light
wave situation to measure the field or absorb the field. However this
charge or absorber actually joined the creation of the action or reaction.
We have measured the field we cannot show whether or not if the test
charges or the absorber is removed from the system the electromagnetic
field still exist. According to the direct interaction principle,
this action or reaction is only exist in the case there are two electrons,
the emitter and the absorber.

This problem cannot solved through the concept field. But we have
successfully solved it through the energy. From the Poynting theorem
we remove all self items include self energy current, self energy
increase and self reaction ($\overrightarrow{J}_{i}\cdot\overrightarrow{E}_{i})$.
After this removal, the Poynting theorem is changed to the mutual
energy theorem. 

If self energy current doesn't not exist. The Maxwell equation is
clear wrong, because from Maxwell equations we can got a solution
with self energy current. It even worse, we can not got a solution
with Maxwell equation that self energy vanishes. Hence, up to now
the only way is to abandon the Maxwell equations. Take self energy
current away and Maxwell equation away, the left is only the mutual
energy theorem and we can call it as mutual energy energy principle.
The above few sections we have shown that the mutual energy can be
used to replace Poynting theorem, after this replacement, we got a
new ``Poynting theorem'' which is actually the mutual energy theorem.
We also shows the Poynting theorem is equivalent in principle to the
Maxwell equations. Hence when we can replace the Poynting theorem
with the mutual energy theorem, the mutual energy theorem actually
can also replace the Maxwell equations. After we also shows that the
Gauss law can also merged to the mutual energy theorem, Hence we can
use mutual energy theorem only one formula to replace all 4 formula
of the Maxwell equations. This replacement is not because 1 formula
is simpler than 4 formula, it is because this one formula is correct
and the 4 formula is not correct.

We also shows the problem of the Maxwell equation in macrocosm, if
we add a pseudo field to the two sides of the formula of the mutual
energy theorem, it be come the Poynting theorem, and the Poynting
theorem is still correct in mathematics. Hence we also can show the
Maxwell equations are also correct in mathematics. It is not correct
in physics, we have notice there are pseudo field items. However,
most wireless problem which still can be solved with Maxwell equations.
Here the only need to notice is that the concept of the field is only
correct at adding the pseudo items, which is all self energy items.
For most wireless situation we still can apply all engineer problem
with Maxwell equations. 

In the case of light, there is only two charges, one is emitter and
one is absorber. This is the place we really need to deal the problem
of self energy items. We have calculated that if it exist, it will
contribute half energy transfer from emitter to absrober. In this
situation the self energy items cannot be omitted. We endorse the
direct interaction principle, which leads us to denies all existent
of the self energy items. After removal of all self energy items,
we obtained the mutual energy theorem or the mutual energy principle.
Now we need to looking the solution from the mutual energy principle. 

Hence, this means that we only need to find a solution which satisfy
the mutual energy principle. For the photon situation there is only
the emitter and absorber two electron, we assume the index of the
emitter is $1$ and the index of the absorber is $2$, the mutual
energy theorem is list as following, 
\[
-\nabla\cdot(\overrightarrow{E}_{1}\times\overrightarrow{H}_{2}+\overrightarrow{E}_{2}\times\overrightarrow{H}_{1})
\]
\[
=\overrightarrow{E}_{2}\cdot\overrightarrow{J}_{1}+\overrightarrow{E}_{1}\cdot\overrightarrow{J}_{2}
\]

\begin{equation}
+\overrightarrow{E}_{1}\cdot\partial\overrightarrow{D}_{2}+\overrightarrow{E}_{2}\cdot\partial\overrightarrow{D}_{1}+\overrightarrow{H}_{1}\cdot\partial\overrightarrow{B}_{2}+\overrightarrow{H}_{2}\cdot\partial\overrightarrow{B}_{1}\label{eq:P3-4-10}
\end{equation}

\subsection{The solution of photon equations 1 }

We know the Maxwell equations are the sufficiant condition of the
mutual energy theorem, hence we can got the solution of the mutual
energy theorem by solving the Maxwell equations. One of the solution
of the above photon equation is Maxwell equation solutions which is,
\begin{equation}
\begin{cases}
\nabla\times\overrightarrow{E}_{1}=-\partial\overrightarrow{B}_{1}\\
\nabla\times\overrightarrow{H}_{1}=+\overrightarrow{J}_{1}+\partial\overrightarrow{D}_{1}
\end{cases}\label{eq:P3-4-20}
\end{equation}
and

\begin{equation}
\begin{cases}
\nabla\times\overrightarrow{E}_{2}=-\partial\overrightarrow{B}_{2}\\
\nabla\times\overrightarrow{H}_{2}=+\overrightarrow{J}_{2}+\partial\overrightarrow{D}_{2}
\end{cases}\label{eq:P3-4-30}
\end{equation}
It must notice that (a) we are looking the solutions $\zeta_{1}=[\overrightarrow{E}_{1},\overrightarrow{H}_{1},J_{1}]$,
$\zeta_{2}=[\overrightarrow{E}_{2},\overrightarrow{H}_{2},J_{2}]$
nonzero simultaneously. If $\xi_{1}=[\overrightarrow{E}_{1},\overrightarrow{H}_{1}]$=0,
$\xi_{2}=[\overrightarrow{E}_{2},\overrightarrow{H}_{2}]\neq0$, this
become the $0$ solution of the mutual energy principle which is not
what we are looking for. Hence the above simultaneously nonzero solution
of Maxwell equations is not just the solution of the Maxwell equations
but is the solution of the mutual energy principle.

\begin{figure}
\includegraphics[scale=0.5]{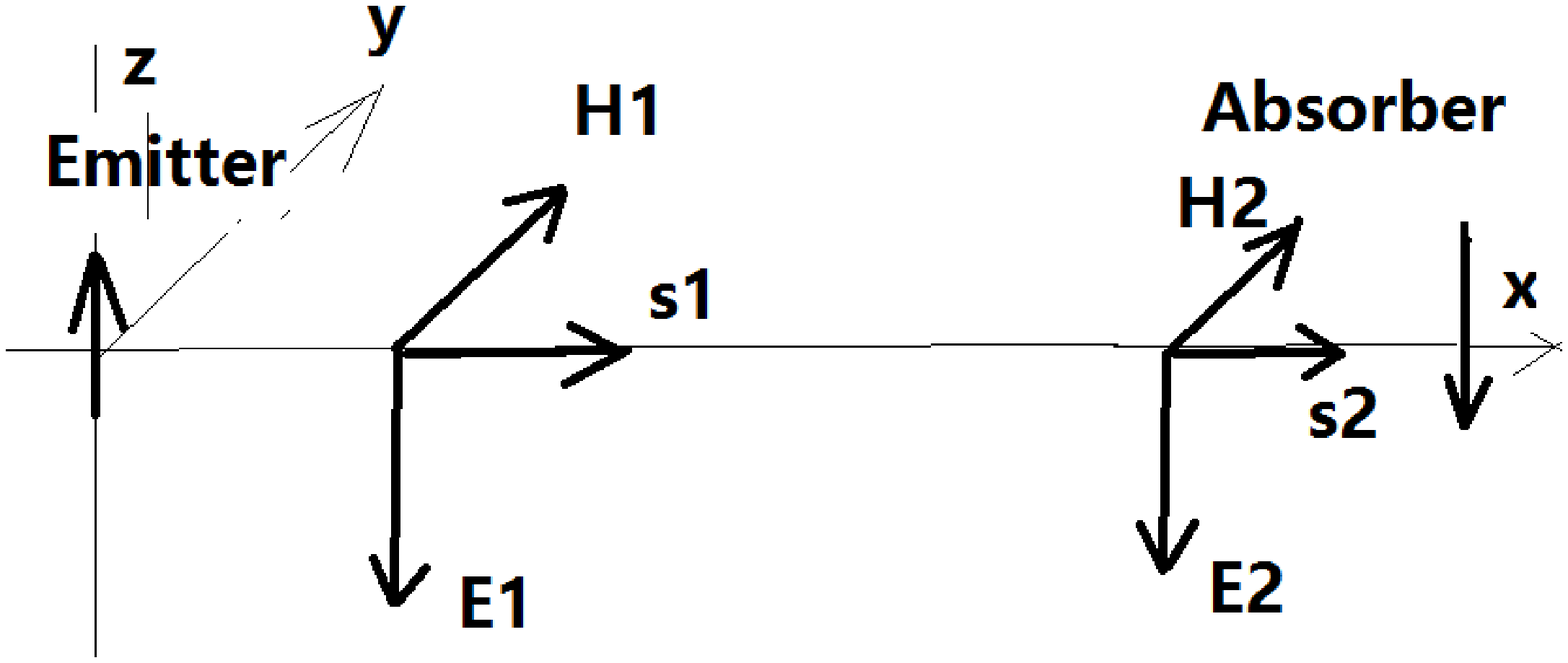}

\caption{photon model, in this model the field $\zeta_{1}=[\protect\overrightarrow{E}_{1},\protect\overrightarrow{H}_{1},\protect\overrightarrow{J}_{1}]$,$\zeta_{2}=[\protect\overrightarrow{E}_{2},\protect\overrightarrow{H}_{2},\protect\overrightarrow{J}_{2}]$
all satisfy Maxwell equations. \label{fig:photon-model_1}}
\end{figure}

Figure \ref{fig:photon-model_1} shows the photon model of this kind
solution.

Assume we have put a metal place between the emitter and the absorber.
We make a hole to allow the light can go through it from the emitter
to the absorber. The mutual energy is exist only on the overlap of
the two field $\zeta_{1}=[\overrightarrow{E}_{1},\overrightarrow{H}_{1}]$
and $\zeta_{1}=[\overrightarrow{E}_{2},\overrightarrow{H}_{2}]$,
see Figure.
\begin{figure}
\includegraphics[scale=0.5]{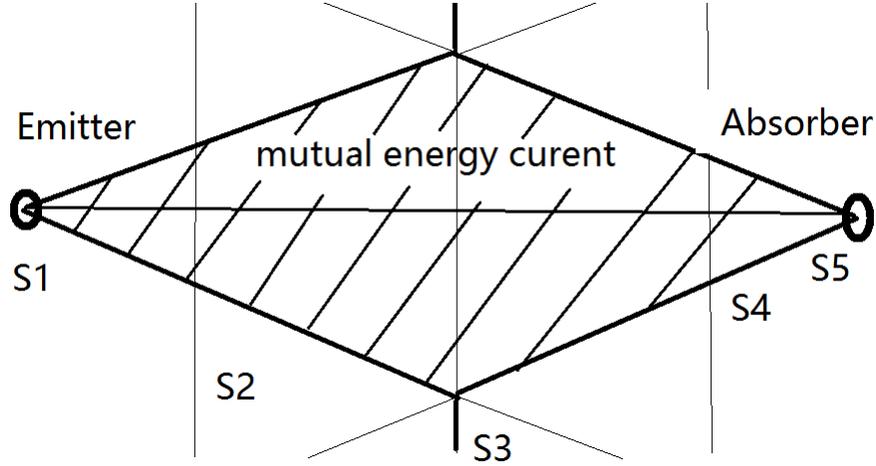}

\caption{The mutual energy current is only existent at the overlap place of
the the two solutions of the Maxwell equation. The field of the emitter
is retarded wave. The wave of the absorber is advanced wave. \label{fig:photon-model_1b}}
\end{figure}

The disadvantage of this photon model is that it can only send the
wave with linear polarization. If we need the photon send circular
polarized field, we have to make the current $\overrightarrow{J}_{1}$
and $\overrightarrow{J}_{2}$ all have two components for example
along $y$ and $z$, or to make the currents rotating along $x$ axis.
This is perhaps possible, because the electron is at spin, there current
is also possible to spin. In this wave the radiate wave become circular
rotated. 

We can take the volume only includes the emitter or only includes
only the absorber, this way we can prove the mutual energy current
go through each surface $S_{1}$, $S_{2}$, $S_{3}$, $S_{4}$ and
$S_{5}$ are equal that is
\[
-\intop_{t=-\infty}^{\infty}\iiintop_{V_{1}}(\overrightarrow{E}_{2}\cdot\overrightarrow{J}_{1})
\]
\[
=Q_{1}=Q_{2}=Q_{3}=Q_{4}=Q_{5}
\]
\begin{equation}
\intop_{t=-\infty}^{\infty}\iiintop_{V_{2}}(\overrightarrow{E}_{1}\cdot\overrightarrow{J}_{2})dV\label{eq:P3-4-40}
\end{equation}
where

\begin{equation}
Q_{i}=\intop_{t=-\infty}^{\infty}\varoiintop_{S_{1}}\cdot(\overrightarrow{E}_{1}\times\overrightarrow{H}_{2}+\overrightarrow{E}_{2}\times\overrightarrow{H}_{1})\cdot d\Gamma dt\ \ \ \ \ \ i=1,2,3,4,5\label{eq:P3-4-50}
\end{equation}

This formula clear tell us the photon's energy is just the mutual
energy current. The mutual energy current is equal at the 5 different
surface. We know that the surface $S_{1}$ and $S_{2}$ are very close
to the emitter or absorber. This time the surface be come so small
hence the wave beam is concentrated to a point which look very like
a particle. In the middle, the wave beam is very thick. We can put
other kind plate for example the metal plate with two slits. In this
case the wave will produce interference patterns. This will explain
the duality character of the photon. 

For the above solution, we have use the the sufficient condition of
the mutual energy theorem. It is not the necessary condition. The
mutual energy principle perhaps has some other solution which do not
satisfy Maxwell equations, which will be discussed in next section.

\subsection{The solution of photon equations 2}

Hence this is not solution we are looking for. However, we can just
rotated the 90 degree hence it will parallel to the The above self-energy
current vanishes. This means $\overrightarrow{E}\times\overrightarrow{H}$
must parallel to each other.

In this situation the perhaps we can find a solution which is not
the solution of Maxwell equation but it still satisfy the mutual energy
principle.

If it is true this solution is also possible the solution for electromagnetic
fields, that means we have found other photon model.
\begin{figure}
\includegraphics[scale=0.5]{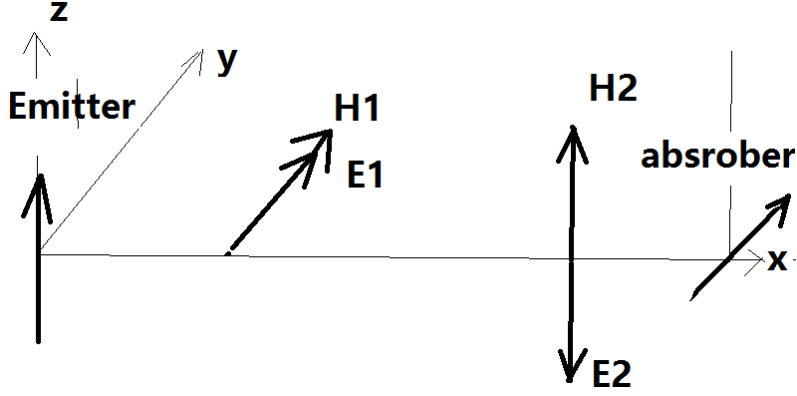}

\caption{In this photon model, the fields $\zeta_{1}=[\protect\overrightarrow{E}_{1},\protect\overrightarrow{H}_{1},\protect\overrightarrow{J}_{1}]$,$\zeta_{2}=[\protect\overrightarrow{E}_{2},\protect\overrightarrow{H}_{2},\protect\overrightarrow{J}_{2}]$
do not satisfy the Maxwell equations. However they still satisfy the
mutual energy theorem.\label{fig:In-this-photon_2}}
\end{figure}

We can easy to prove the above Figure \ref{fig:In-this-photon_2}
satisfy mutual energy theorem even it is not satisfy Maxwell equations.
We can see that this model is just a rotation from the Figure \ref{fig:photon-model_1}.
We know that the photon model in Figure \ref{fig:photon-model_1}
satisfy Maxwell equations, that guarantees it also satisfies the mutual
energy theorem. Know we have rotated $\overrightarrow{J}_{2}$, $\overrightarrow{H}_{2}$
and $\overrightarrow{E}_{1}$ along the $x$ axis 90 degree. Hence
$\overrightarrow{E}_{1}\times\overrightarrow{H}_{2}$ and $\overrightarrow{E}_{1}\cdot\overrightarrow{J}_{2}$
don't change. This guaranties all the items in the mutual energy theorem
do not change, if it is satisfied before the rotation, after the rotation
the equation should still satisfied. 

Even this model do not satisfy the Maxwell equations, it has 3 advantages.

(1) Since field $\overrightarrow{E}_{1}||\overrightarrow{H}_{1}$
and $\overrightarrow{E}_{2}||\overrightarrow{H}_{2}$ the self energy
current vanish automatically. 

\begin{equation}
\overrightarrow{E}_{1}||\overrightarrow{H}_{1},\ \ \ \ \ \ \ \ \overrightarrow{E}_{2}||\overrightarrow{H}_{2}\label{eq:P3-4-60}
\end{equation}
or

\begin{equation}
\varoiintop_{\Gamma}\cdot(\overrightarrow{E}_{1}\times\overrightarrow{H}_{1})\cdot d\Gamma=0,\ \ \ \ \ \ \ \ \varoiintop_{\Gamma}\cdot(\overrightarrow{E}_{2}\times\overrightarrow{H}_{2})\cdot d\Gamma=0\label{eq:P3-4-70}
\end{equation}

(2) we can see that, $\overrightarrow{E}_{1}\perp\overrightarrow{J}_{1}$
and $\overrightarrow{E}_{2}\perp\overrightarrow{J}_{2}$ we have clear
that the self energy action vanish automatically. 

\begin{equation}
\iiintop_{V}(\overrightarrow{E}_{1}\cdot\overrightarrow{J}_{1})dV=0\label{eq:P3-4-80}
\end{equation}
\begin{equation}
\iiintop_{V}(\overrightarrow{E}_{2}\cdot\overrightarrow{J}_{2})dV=0\label{eq:P3-4-90}
\end{equation}

(3) this model are easy to support circular polarization. Because
not $\overrightarrow{E}_{1}$ and $\overrightarrow{E}_{2}$ also perpendicular.
We know it is difficult to explain that why electron is so small it
still can send circular polarization waves. In our antenna experience
we need very complicated antenna to send circular polarized waves.
When two electric field are perpendicular if the two fields have 90
degree phase difference, we have obtained circular polarization. 

Even this solution do not satisfy Maxwell equation, it has $0$ self
energy current and $0$ self action and can offer a simple current
to support the circular polarization. It looks as a miracle.

Even this is not a physic solution of photon, using the mutual energy
theorem as principle still much better than to take the Maxwell equations
as the principle. 

\subsection{Is light satisfy Maxwell equation?}

Many people ask in the internet the question whether or not the field
of light satisfy Maxwell equation? We need to distinguish the problem
in microcosm and in macrocosm. 

In microcosm deal the problem of photon, the photon is a small system
only with two electrons one is the emitter, the other is the absorber.
From above discussion we know that photon satisfies the mutual energy
principle. We also know that the Maxwell equation in microcosm is
the sufficient condition of the mutual energy principle, hence, it
is possible that photon satisfies the Maxwell equations. But it is
also possible photon satisfies the mutual energy principle but do
not satisfy the Maxwell equations.

In macrocosm, many photo do not satisfy the Maxwell equations in macrocosm,
the reason is photon is energy package, here, the energy is linear.
For example one photon the energy is 1, 2 photon the energy become
2. If energy is linear, its field can not be linear. Hence in macrocosm
the many photon is also do not satisfy the Maxwell equations. This
result is different with wireless wave. In case of wireless wave,
the wireless wave satisfy linear (do not consider the effect of pseudo
field) It satisfy Maxwell equations.

According discussion that field of emitter and absorber of one photon
is possible do not satisfy Maxwell equations. There is the possibility
it do not satisfy Maxwell equation but it still satisfy mutual energy
principle. 

However to make things simple we still assume for each photon, the
emitter and the absorber still satisfy the Maxwell equations. For
a system with big number of photon $N$, there is $2N$ emitters and
absorbers. If the field of the light can be superimposed, then it
is clear the $2N$ system also satisfy Maxwell equations. 

If 

\begin{equation}
\begin{cases}
\nabla\times\overrightarrow{E}_{i}=-\partial\overrightarrow{B}_{i}\\
\nabla\times\overrightarrow{H}_{i}=+\overrightarrow{J}_{i}+\partial\overrightarrow{D}_{i}
\end{cases}\label{eq:P3-4-100}
\end{equation}

is satisfied it is clear that
\begin{equation}
\begin{cases}
\nabla\times\sum_{i=1}^{2N}\overrightarrow{E}_{i}=-\partial\sum_{i=1}^{2N}\overrightarrow{B}_{i}\\
\nabla\times\sum_{i=1}^{2N}\overrightarrow{H}_{i}=+\overrightarrow{J}_{i}+\partial\sum_{i=1}^{2N}\overrightarrow{D}_{i}
\end{cases}\label{eq:P3-4-110}
\end{equation}

However the field of photon is not linear and cannot superimposed.
There are two reasons we speak the field of light is not linear. The
first reason is we have mentioned before in last few sections. The
second reason is photon as energy package and hence its energy is
linear. $N$ photons energy has $N$ times of energy of one photon.
Energy is linear tell us that the field is not linear. if the field
is linear the energy must quadratic. 

Hence in general that light's field do not satisfy Maxwell equations
in macrocosm.

\subsection{Is the wireless wave are composed as photons?}

Some people ask in the internet that is the wireless wave also composed
as many photons or there is frequency limit beyond that frequency
all energy be come packaged as photon. We thought this is a very good
question and would like make it clear here. 

In the antenna system one transmitting antenna can send the energy
to many receiving antennas. It is not send a package of the energy
to only one antenna in a time and randomly send the the energy to
another antenna. It send the energy to all space if any antenna can
receiving this energy. 

This is a system with charges more than two. All these receiving antennas
can synchronized with the transmitting antenna with frequency and
orientation. This kind of energy is not a energy package. Hence wireless
wave is not composed as photons. Actually the high frequency wave
become photon is only because the field of the emitter and absorber
is difficult the synchronized.

\subsection{Is all current happened in the emitter or absorber will create any
action or reaction?}

Assume in the emitter it has a current change for example a electron
has from a higher energy level jump to a lower, is this current must
send a photon? We think the answer is no. This current change will
send the retarded potential but if there can not found a absorber
and the field of the absorber must just synchronized to the field
of the emitter. If this kind of absorber is not happened. The energy
is not send out. The electron has energy perhaps it will return to
the higher level. And next time it will spring to low level again
to found the matched advance wave.

In the case only have a emitter and a absorber, the current of the
emitter produced a retarded wave and advanced wave. The absorber as
produced a retarded wave and advanced wave. We know that the retarded
wave or the emitter has synchronized with the advanced wave of the
absorber. But the advanced wave of the emitter and the retarded wave
of the absorber has been sent out but doesn't find a corresponding
matured wave. 

Maxwell equation cannot tell us if there is a current advanced wave
or retarded wave should be associated to this current. Hence Wheeler
and Feynman assume there always half advanced wave and half retarded
wave associated to the current.

From my experience I would thought only one kind wave can associated
to the emitter and absorber. We can not obtained which wave this current
can produce from Maxwell's theorem, but perhaps we can obtained some
information from the mutual theorem.

If the charge jump from the high energy to lower energy, we can assume
there are many advanced wave have been at the place of the emitter,
if it just run against the advanced wave it will gave this wave as
retarded potential. When this retarded potential reached to the absorber.
The electron in the absorber go along the retarded wave hence it will
send the advanced wave. From this reaction who should send retarded
wave and who should send advance wave is decided. 

The retarded wave is not real wave, it is only offers the possibility
to support a absorber to receive it. If this absorber appear by produce
a current synchronized with the emitter. The energy will send from
the emitter to the absorber. If the no absorber to receive this energy.
The energy of the emitter doesn't loss any energy to the empty space.
There is no any energy is lost to the space and move in the space
without the emitter and the absorber. 

\subsection{Retarded potential}

We know that the mutual energy theorem endorse the direct interact
theory, hence the retarded wave is not a real thing. If there is no
absorber or the advance wave, the retarded wave cannot send the the
energy out. This tell us the retarded wave is only offers the ability
to send out the energy but doesn't really offers the energy. This
is real reason of the probability interpretation of Copenhagen. However
now I know that the probability interpretation of Copenhagen is at
least correct at about the retarded wave. The retarded wave is only
offers the ability to do some thing. It is not send real energy, the
real energy is sent only in case there is a advanced wave. 

In this way we actually endorse both the Copenhagen interpenetration
of the quantum physics and also the transaction interpretation of
the quantum physics in which allows the retarded wave and the advanced
wave.

When I speak that the retarded wave is not real, actually we also
means the advanced wave is also not real, not energy is send by advanced
wave, only if the retarded wave and the advanced wave meet together,
the energy current is send from emitter to the absorber. 

\section{the story about this discovery}

1984 I have enter Xidian University in China to learn electromagnetic
field and microwave technology as a master degree graduate student.
1985 I begin to learn the modified reciprocity theorem from the book
about electromagnetic wave of J.A Kong. In his book he has talked
the modified Lorentz theorem and electromagnetic field transform,
in another different section. As application I applied the transform
to the reciprocity theorem I got a new formula, when I looked the
new formula it is much meaningful compare to the Lorentz reciprocity
theorem. It clear tells the energy between two the receiving and transmitting
antennas. I call it ``mutual energy theorem'' and published 3 papers
about that\cite{IEEEexample:shrzhao1,IEEEexample:shrzhao2,IEEEexample:shrzhao3}
in 1987-1989. In the end of 1986, my master degree defense is with
the ``mutual energy theorem'' which is failed. The professors said:
``make a transform to the reciprocity theorem and then call it the
mutual energy theorem that is not worthy of the name''.

Later I worked as antenna designer for 3 years in China. Then go to
Germany working at CT and MEG medical image in Julich research center
Germany for 7 years as scientist. This work allow me to become a Ph.D
in Xi'an Jiaotong University in 1998 in the field CT image reconstructions. 

Later I Changed more than 10 different companies and institute in
Germany, Canada and USA, most working as C++ software research and
development. My publication most is about CT image reconstructions. 

From 2014 I came back to the mutual energy theorem, I have long the
felling this topic I did not finish, there is some important thing
I need to do. First I wrote the ``Concept of mutual energy''. This
time I can search all reference free, because I am working at Cimtec
Inc which belongs to Western University of Canada. The most important
reference are two I found, one is from W.J. Welch \cite{IEEEexample:Welch,IEEEexample:Welch-2}.
He seems first derived the embryonic form of the mutual energy theorem
and introduced the concept of advanced potential. This is the first
time I touched the concept of advanced potential. The second reference
is the time domain mutual energy theorem published by Adrianus T.
de Hoop\cite{IEEEexample:Adrianus2}. The difference of this publication
with the mutual energy theorem is only a Fourier transform. It is
referred as time-domain correlation reciprocity theorem published
also in 1987 seem year as the mutual energy theorem published. It
is lucky for me that the publication is half years earlier than his.
This allow me still can call it as the mutual energy theorem. I submitted
8 manuscripts about this mutual energy concepts to IEEE Transactions
on Microwave Theory and Techniques. It is nearly go through the peer
review process. They not reject it immediately but ask me to make
corrections. After 3 time corrections, but finally they rejected them.
I resent it to IEEE Transactions on Antennas and Propagation, they
are rejected again.

In the end of 2016 I begin to realized that the mutual energy theorem
is not only a theorem but is a principle. Here I did not mean it is
a principle can replace Maxwell equations. I think if electromagnetic
field transfers the energy by the mutual energy theorem, then other
particles, for example electron will do also energy transfer by the
mutual energy of the electron. In this way it is a principle.

If in wireless wave, transfer energy between transmitting antenna
and receiving antenna is a combination of retarded potential and advanced
potential, then the energy transfer between the emitter of the electron
and the absorber of the electron also should through the mutual energy
current of the electrons. In this way, the mutual energy theorem is
a principle. I begin to work at to prove photon transfer the energy
by the mutual energy current. I prove in a lossy media the mutual
energy theorem offers correct result but the reciprocity theorem cannot.
I have tried to explain the duality of the photon with the mutual
energy theorem.

In the December of 2016 I begin to work at the theory of time domain
mutual energy current. I begin to interesting to know whether or not
that the self-energy current do transfer the energy. In this time
I am sure that the mutual energy current can transfers energy, but
I am not clear what about the self-energy current. Is self energy
current also transfer energy? If the self energy current transfer
energy I calculated, it can have half contribution to the whole energy
transfer for a system with an emitter and an absorber. Hence for photon
the self energy cannot be omitted. In case $N$ is large, the self
energy have $N$ items. The mutual energy current has $N^{2}-N$ items.
Each items has same energy, hence if $N$ is very large, the contribution
of self-energy current can be omitted. For a photon there is only
$N=2$, one emitter and one absorber. $2^{2}-2=2$. Hence the self
energy current cannot omit in the case of photon.

In this time I still believe Maxwell equations. According to the Maxwell
equations, if self energy current vanishes, I found that the field
also will vanish. 

If
\begin{equation}
self\ energy\ current\equiv\varoiintop_{\Gamma}(\overrightarrow{E}\times\overrightarrow{H})\hat{n}d\Gamma=0\label{eq:P3-5-10}
\end{equation}
means that 
\begin{equation}
\iiintop_{V}(\overrightarrow{E}\cdot\partial\overrightarrow{E}+\overrightarrow{E}\cdot\partial\overrightarrow{E})=0\label{eq:P3-5-20}
\end{equation}
Here $V$ is at the outside of the surface $\Gamma$ that also further
means
\[
\overrightarrow{H}=\overrightarrow{E}=0
\]
However if the electromagnetic field vanishes, the mutual energy current
also will vanish. This is really confused me.

In other hand, I cannot accept the concept of self energy current
collapsed to absorber. Collapse is not a physics concept, if you accept
the concept could you offer me the equation of the collapse process?

Assume the self energy current doesn't vanishes. I have designed a
few possibility for that. The emitter's self energy current sends
to future and infinity. The absorber's self energy current sends to
past and infinity. The emitter lost some energy, the absorber obtained
some energy hence, the self energy is transferred from emitter to
the absorber. 

Another possibility is that the emitter sends the retarded self energy
current to the future and infinity, in the same time it sends advanced
self energy current to the past and infinity, hence there is no pure
energy gain for the emitter. It is same to the absorber theory. 

I noticed that there is problem for these two possibility. If the
source of light is put inside of the metal container, and there is
only a small hole for light, how can allow the self energy go through
the hole and got to infinity? Hence I believe the self energy current
exist is still a very bed idea.

I begin to look the possibility the self energy current helps the
mutual energy to send energy from emitter to absorber, then it returned
to it's sender. The emitter's self-energy return to the emitter, the
absorber's energy current returns to the absorber. However the electromagnetic
theory do not support time-reversed wave. From Maxwell equations we
can obtained retarded wave which is sent to future and infinity and
advanced wave which is sent to the past and infinity. The returned
wave for the retarded wave needs to go from future and infinity come
back to the emitter. This wave actually obeys so called time-reversed
Maxwell equations which is
\begin{equation}
\nabla\times\overrightarrow{E}=\partial\overrightarrow{B}\label{eq:P3-5-30}
\end{equation}

\begin{equation}
\nabla\times\overrightarrow{H}=-J-\partial\overrightarrow{D}\label{eq:P3-5-40}
\end{equation}
I have to introduce the new questions that is also not a good idea.
There is also other problem, I have to ask the return process is not
done immediately otherwise the returned field will cancels the original
self energy current. The returned wave must stayed at infinity for
a moment then returns. This is even more strange. How the wave can
stay at infinity for a moment and then returns? 

Hence the emitter and absorber send self-energy currents become very
confuse to me. I come bake to the idea the self current should vanish
hence
\begin{equation}
\varoiintop_{\Gamma}(\overrightarrow{E}\times\overrightarrow{H})\hat{n}d\Gamma=0\label{eq:P3-5-70}
\end{equation}
if
\begin{equation}
\overrightarrow{E}\times\overrightarrow{H}=0\label{eq:P3-5-80}
\end{equation}
The above self-energy current vanishes. This means $\overrightarrow{E}\times\overrightarrow{H}$
must parallel to each other.
\begin{equation}
\overrightarrow{E}||\overrightarrow{H}\label{eq:P3-5-90}
\end{equation}
However, Maxwell equation is clear do not support this kind of field.
This parallel field cannot be the solution of the Maxwell equations.
I begin to think perhaps we do not need to satisfy the Maxwell equations,
we only need to find solution to satisfy

\begin{equation}
\nabla\times(\overrightarrow{E}_{1}+\overrightarrow{E}_{1})=-\partial(\overrightarrow{B}_{1}+\overrightarrow{B}_{2})\label{eq:P3-5-105}
\end{equation}

\begin{equation}
\nabla\times(\overrightarrow{H}_{1}+\overrightarrow{H}_{2})=(J_{1}+J_{2})+\partial(\overrightarrow{D}_{1}+\overrightarrow{D}_{2})\label{eq:P3-5-110}
\end{equation}
I have drew the picture about the field possibility. But I still could
not prove this kind of field satisfy the above equations. From my
experience, the new equation has doubled the variables compared to
the original Maxwell equations, hence has more solutions. And should
be easy to find solutions. But I still cannot cannot prove this solution
satisfy this relaxed Maxwell equations. And also I cannot prove this
relaxed is reasonable. 

In beginning of the March in 2017, I begin to think what is correct
way to define the electromagnetic fields and the magnetic field. It
lead me to have 3 ways to define the field, which is the correct way?
After I notice the correct power is
\begin{equation}
P=\sum_{i=0}^{N}\sum_{j=1,j\neq i}^{N}\overrightarrow{E}(x_{j},x_{i})\cdot\overrightarrow{J}_{i}\label{eq:P4-6-120}
\end{equation}
 I begin to realized, in the total power, there is no any thing called
self action, which is
\begin{equation}
\overrightarrow{E}(x_{i},x_{i})\cdot\overrightarrow{J}_{i}\label{eq:P3-5-130}
\end{equation}
We do not need to calculate $\overrightarrow{E}(x_{i},x_{i})$ that
is $\infty$. Quantum physics try to solve this zero infinity by re-normalization.
But this formula clear tell us the self-energy action $\overrightarrow{E}(x_{i},x_{i})\cdot\overrightarrow{J}_{i}$
is a concept not necessary. If self energy current is take away from
current calculation, it should be also take away from all other corresponding
energy calculation. Hence all energy related value should be defined
like following,
\begin{equation}
\overrightarrow{E}(x)\times\overrightarrow{H}(x)=\sum_{i=1}^{N}\sum_{j=1,j\neq i}^{N}\overrightarrow{E}(x_{j},x)\times\overrightarrow{H}(x_{i},x)\label{eq:P3-5-140}
\end{equation}

\begin{equation}
\overrightarrow{E}(x)\cdot\overrightarrow{E}(x)=\sum_{i=1}^{N}\sum_{j=1,j\neq i}^{N}\overrightarrow{E}(x_{j},x)\cdot\overrightarrow{E}(x_{i},x)\label{eq:P3-5-150}
\end{equation}

\begin{equation}
\overrightarrow{H}(x)\cdot\overrightarrow{H}(x)=\sum_{i=1}^{N}\sum_{j=1,j\neq i}^{N}\overrightarrow{H}(x_{j},x)\cdot\overrightarrow{H}(x_{i},x)\label{eq:P3-5-160}
\end{equation}
The field superimposition is a confused concept and hence can be removed.
We define the field just as the collection of all contributions,
\begin{equation}
\overrightarrow{E}(x)=\{\overrightarrow{H}(x_{1},x),\overrightarrow{H}(x_{2},x),\cdots\overrightarrow{H}(x_{i},x)\cdots\}\label{eq:P3-5-170}
\end{equation}
I am really so glad to found out the problem is at the superimposition
of the field. The Maxwell equation in the microcosm is perhaps still
correct, but it can also be replaced with the mutual energy principle.
In the macrocosm, since the field concept is not correct, we cannot
superimpose the field. Even if Maxwell equations are correct in microcosm,
we still cannot prove the Maxwell equations in macrocosm is correct.

In the end I found if add the pseudo self items for the field, the
Poynting theorem still correct in the meaning of the mathematics.
Maxwell equations can be done the same in macrocosm. Hence all traditional
way to solve the classical problem still can work with all Maxwell's
theory. The only things need to be notice is that the Poynting theorem
and Maxwell's equation have added the pseudo self field, which can
cause problem especially in the situation of quantum physics. If we
use the mutual energy theorem as a principle, all normalization process
in quantum physics become clear correct.

\section{Conclusion}

In photon situation, Maxwell equation for a emitter or absorber, only
offers a illusive solution which are nonzero solution of self energy
terms. For a system of a photon with a emitter and absorber, the solution
of simultaneously nonzero solution for the emitter and absorber is
possible to be obtained from the mutual energy principle.

Even if we assume all the emitters and absorbers satisfy Maxwell equations
in microcosm, we cannot prove the Maxwell equation in in macrocosm
situation. for example wireless wave case.

That can only achieve if the field can be superimposed. However we
have found in general the field cannot be superimposed. The field
can be linearized, but that need to add all pseudo self energy terms.
The macrocosm Maxwell equation need to be proved with mutual energy
principle together the pseudo self energy terms.

In other side the mutual energy principle is established both in photon
situation there is only very few source and the macrocosm situation
there are infinite charges all synchronized. From mutual energy principle
we can prove the Poynting theorem and Maxwell equations in macrocosm(even
they are only correct in mathematics). 

Hence we obtained the conclusion it should not necessary to offer
the Maxwell equations as a principle. It is only a mathematics method
can be used to found the solution of the mutual energy principle. 

Mutual energy principle ask all field cannot vanishes. Hence the solution
with mutual energy theorem will be nonzero solution for all fields.
Maxwell equations can obtained the solution only for emitter or for
absorber which is not the solution we are looking for. From Maxwell
equations we can obtained the pseudo self energy terms which are very
confuse. 

Hence we have to put the mutual energy principle as the starting point
of all electromagnetic theory. It is the real principle in the theory
of the electromagnetic fields.

\end{document}